# GW Notes

July to September 2009

Notes & News for GW science

Editors:

P. Amaro-Seoane and B. F. Schutz

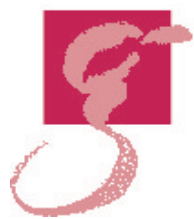

GW Notes was born from the need for a journal where the distinct communities involved in gravitation wave research might gather. While these three communities - Astrophysics, General Relativity and Data Analysis - have made significant collaborative progress over recent years, we believe that it is indispensable to future advancement that they draw closer, and that they speak a common idiom.

> *Editorial*
>
> A powerful camera: Snapshotting black holes

Close monitoring of the Keplerian orbits of the so-called "S-stars" or "SO-stars"[1] has led to the revolutionary discovery that a massive black hole is lurking in the centre of our own galaxy.

These S-stars are located in very tight orbits around a very compact astronomical radio source at the Galactic Centre, known as Sagittarius A$^*$. After observing these stars for some ten years, a group led by Reinhard Genzel of the Max Planck Institute for Extraterrestrial Physics reported in 2002 that the star called S-2 was moving on a Keplerian orbit around Sagittarius A$^*$. The radio source therefore had to contain a dark massive object confined in a volume with a radius of some 120AU. Later observations by Genzel's group and one at UCLA led by Andrea Ghez determined that the mass is about $4 \times 10^6 \, M_\odot$, which is a clear –though indirect– proof that it is a massive black hole.

From a theoretical point of view, we know that ordinary stars feed the central massive black hole via accretion of gases produced by tidal disruptions. However, if they are compact enough, stars can be swallowed whole when they reach the horizon after gradually spiralling in due to the emission of GWs. This process is known as an "*Extreme Mass Ratio Inspiral*" and it is one of the most important objectives of LISA.

The "doomed" compact object spends many orbits around the massive black hole before it is swallowed. While doing so, it radiates energy which can be thought of as a snapshot containing detailed information about the masses and spins of black holes to an accuracy which is beyond that of any other astrophysical technique. Such information will tell us about cosmic evolution and the history and growth of massive black holes in the nearby universe. Therefore, an early understanding of how we can model these waveforms is of prime importance to extract the signal from the LISA datastream. For this GW Notes issue we have approached Nicolás Yunes (Princeton University) to extend in high detail his recent work on EMRI waveforms for our highlight article.

Pau Amaro-Seoane & Bernard F. Schutz, editors

---

[1] We note that the S stands simply for "source"; they should not to be confused with spectral type S stars, which are late- type giant stars (such as class K5-M)

*GW Notes highlight article*

*EMRIs*

# GRAVITATIONAL WAVE MODELLING OF EXTREME MASS RATIO INSPIRALS AND THE EFFECTIVE-ONE-BODY APPROACH

Nicolás Yunes
Department of Physics
Princeton University
Princeton, NJ 08544, USA
e-mail: nyunes@princeton.edu

**Abstract**

Accurate and efficient models to calculate the gravitational wave response of the proposed *Laser Interferometer Space Antenna* are crucial for the accurate extraction of physical parameters from noisy data, especially for moderate signal-to-noise ratio events. One of the most challenging and interesting sources of such waves is *extreme-mass ratio inspirals*, where a small compact object winds into a supermassive black hole in a generic orbit. The interest in these sources stems from their ability to accurately map the spacetime around supermassive black holes, thus revealing otherwise inaccessible astrophysical information and allowing for exquisite tests of general relativity. The difficulty in modelling gravitational waves produced in such inspirals is two-fold. First, extreme-mass ratio orbits are generic, including zoom-whirl episodes where the small object pirouettes in the strong gravitational field of the supermassive black hole at large velocities. Second, gravitational waves generated by these sources can contain millions of cycles in the detector's sensitivity band, and thus a small error in the modelling can lead to a large accumulated error in the template after a one-year observation. For these reasons, one must develop sophisticated techniques to approximate these waves as accurately and efficiently as possible. This article focuses on these techniques, explaining the difficulty in the modelling and suggesting possible routes to their resolution. We first set the stage through a brief summary of some of the current models available for constructing approximate extreme-mass ratio inspiral templates. We then introduce in detail a new scheme that combines ingredients from both black hole perturbation theory and the effective-one-body approach. We conclude with comparisons between this new scheme and Teukolsky-based waveforms for quasi-circular inspirals into non-spinning supermassive black holes.

A probe of spacetime and astrophysics: EMRIs

# Contents

# 1 Introduction

An "extreme mass ratio inspiral" (EMRI) is the name given to the gravitational wave (GW) driven coalescence of stellar mass compact objects into supermassive black holes (SMBHs). Astrophysical evidence supports the existence of $(10^6, 10^9)M_\odot$ SMBHs at the center of galaxies (Amaro-Seoane et al., 2007), while several scenarios have been suggested (tidal separation Miller et al., 2005, accretion disk capture Syer et al., 1991, etc.), where small BHs or NSs in the mass range $(10, 10^2)M_\odot$ could be driven close enough to inspiral via GW emission. We shall not review the astrophysics of EMRIs here, but instead satisfy ourselves with the understanding that astrophysically they are expected to be abundant (Gair et al., 2004) (the interested reader can refer to the review in Amaro-Seoane et al., 2007 for more astrophysical details).

Although EMRIs can be classified as coalescences with mass ratios $(10^{-4}, 10^{-8})$ and total masses $\sim (10^6, 10^9)M_\odot$, only a subset of these are of interest to the planned *Laser Interferometer Space Antenna* (LISA). The low-frequency noise of this detector raises sharply for frequencies $\lesssim 10^{-4}$Hz due to white dwarf confusion noise (Farmer and Phinney, 2003), thus excluding EMRIs with total mass $\gtrsim 10^8 M_\odot$. On the other hand, EMRIs with total mass $\lesssim 10^6 M_\odot$ can be seen by LISA only when the inspiral is widely separated, since for smaller separations (eg. separations smaller than ten times the SMBH mass) the GW frequency is outside the detector's optimal sensitivity band (LISA's noise raises sharply above $\sim 10^{-2}$ Hz). Astrophysically realistic EMRIs of interest to LISA and that sample the strong gravitational field of the SMBH are those with total mass in the range $(10^6, 10^7)M_\odot$ and mass ratio $(10^{-4}, 10^{-6})$.

EMRIs are one of the main targets for LISA both because of their expected abundance and because their waves encode a detailed map of the strong gravitational field of SMBHs (Hughes, 2006). Such information includes the mass and spin of SMBHs, and the location of their horizon and innermost stable circular orbit (ISCO), all of which are critical for the understanding of certain astrophysical phenomena such as BH growth scenarios and accretion disk physics. Perhaps more importantly still, GWs from EMRIs should allow for exquisite tests of general relativity (GR) (see eg. Ryan, 1995, 1997, Poisson, 1996, Collins and Hughes, 2004, Glampedakis, 2005, Glampedakis and Babak, 2006, Barack and Cutler, 2007, Barausse et al., 2007, Gair et al., 2008, Alexander et al., 2008b, Yunes and Finn, 2009, Sopuerta and Yunes, 2009, Yunes and Pretorius, 2009b, Yunes and Sopuerta, 2009, Apostolatos et al., 2009 and Schutz et al., 2009).

The characterisation of EMRI GWs, unfortunately, is a tremendously difficult problem because of the intrinsic feebleness of these waves. Generically, one expects the detector's noise curve to be louder than the GW signal from EMRIs, which is why all detection strategies must rely heavily on the technique of matched filtering (Porter, 2009). This technique consists in maximising the cross-correlation between the product of data and certain GW templates (weighted by a measure of the detector noise) over the template parameters. We shall not review matched filtering or data analysis here, but we refer the interested reader to the second GW Notes article (Porter, 2009).

Even without a more profound understanding of matched filtering, it is painstakingly clear that parameter estimation for EMRIs will necessarily be extremely sensitive to the construction of accurate templates. Template errors, due to the use of approximation schemes to solve the Einstein equations, can lead to large systematic errors (Cutler and Vallisneri, 2007 and Yunes and Pretorius, 2009b). Recently, Lindblom et al., 2008 estimated that an accuracy of 0.07 ($10^{-4}$) radians in the phase is required for detection (parameter estimation) with LISA (for a signal-to-noise ratio (SNR) of about four thousand). Of course, EMRI signals are expected to have more moderate SNRs, and thus the accuracy requirements for parameter estimation are less stringent. Not only should templates be accurate, but their construction should be fast and efficient, as the maximisation of the matched filtering cross-correlation requires the repeated evaluation of a certain integral while varying template parameters.

The above problem worsens once one realises that the usual approximation schemes commonly employed to build waveforms for equal-mass systems do not apply to EMRIs. More precisely, the traditional post-Newtonian (PN) approximation relies on expansions in powers of the orbital velocity, which for EMRIs can be in the range $v/c \sim (0.1, 0.5)$ during most of the coalescence. Black hole (BH) perturbation theory schemes are applicable, where one relies on expansions in the mass ratio, but these are necessarily numeric in nature and their computational cost to cover the full EMRI parameter space is likely to be prohibitive (Gair et al., 2004). Fully numerical simulations could also be employed for template construction, but these become computationally impractical in the small mass ratio limit.

Such considerations have motivated many researchers to rely on some combination of the above approximation schemes to construct quasi-analytic templates. Until now, there has not been a comprehensive summary of the different models employed to construct such templates. The present article begins to fill this void, while setting the stage for its primary goal: to introduce in detail a new model that combines BH perturbation theory with the effective-one-body approach (EOB). This model was first introduced in Yunes et al., 2009, and here we fill in some of the missing details left out from that publication due to space limitations. We conclude by comparing in detail this new approach to Teukolsky-based waveforms (Hughes, 2000, 2001) for a certain class of orbits.

This article is divided as follows:

**Section 2** (**page 7**) introduces the basics of modelling; **Section 3** (**page 10**) summarises some of the available models to construct EMRI waveforms; **Section 4** (**page 17**) describes in detail a new model for EMRI GW construction; **Section 5** (**page 29**) compares the new EMRI model to numerically computed templates; **Section 7** (**page 39**) concludes and points to future work.

We employ the following conventions: we work exclusively in four spacetime dimensions with signature $(-,+,+,+)$ (Misner et al., 1973), with Greek letters ranging over all spacetime indices; round and square brackets around indices denote symmetrisation and anti-symmetrisation respectively, namely $T_{(\mu\nu)} = \frac{1}{2}(T_{\mu\nu} + T_{\nu\mu})$ and $T_{[\mu\nu]} = \frac{1}{2}(T_{\mu\nu} - T_{\nu\mu})$; partial derivatives are sometimes denoted by commas (e.g. $\partial\theta/\partial r = \partial_r \theta = \theta_{,r}$). The Einstein summation convention is employed unless otherwise specified, and we use geometrised units where $G = c = 1$.

## 2 Gravitational Wave Modelling of Extreme Mass Ratio Inspirals

The present article is not intended only for experts in the field of GW modelling, and as such, a certain degree of introduction is required. This sections serves this purpose. We shall summarise the main points related to EMRI modelling, following mostly the review papers Poisson, 2004a and Barack, 2009.

**2.1 Exact Schemes**

Waveform modelling reduces to solving the Einstein equations for a given system under consideration. Of course, this is a tremendously difficult task due to the non-linearities of the field equations. In fact, contrary to Newtonian expectation, a solution to the two-body problem in GR is still lacking. One can argue that this is because by construction GR replaces the two-body problem by a one-spacetime problem, where one must solve both for the motion of the particles and the evolution of the fields *simultaneously* (Hughes, ). Let us see how this comes about.

Perhaps contrary to the GR spirit, solutions to the Einstein equations are usually sought after by decoupling the equations of motion for matter from the evolution equations for the gravitational field. By the Strong Equivalence Principle (SEP), the former are given by $\nabla^\alpha T_{\alpha\beta} = 0$, where $\nabla^\alpha$ is the covariant derivative associated with the full metric $g_{\mu\nu}$ and $T_{\alpha\beta}$ is the matter stress-energy tensor. One would like to think of this equation as a differential relation for the behaviour of matter, which one immediately sees is controlled by the metric tensor through the Christoffel connection entering the covariant derivative and also the stress-energy tensor itself. One then requires knowledge of the gravitational field to compute the trajectories, but these fields are controlled by the Einstein equations $G_{\mu\nu} = 8\pi T_{\mu\nu}$, where $G_{\mu\nu}$ is the Einstein tensor. One would like to think of this as a differential equation for $g_{\mu\nu}$ only, but one immediately sees that these fields are sourced by matter. In this manner, one rediscovers that an exact decoupling is impossible in GR as *geometry tells matter how to move and matter tells geometry how to curve.*

Without a decoupling of the field equations, it becomes extraordinarily difficult to find an analytic solution, which is why one sometimes relies on full numerics. Numerical relativity has recently undergone a revolution, where the equal-mass (and comparable-mass) binary problem has been essentially solved (Pretorius, 2005, Baker et al., 2006 and Campanelli et al., 2006). The EMRI problem, however, poses a new set of difficulties associated with the additional scale introduced by the mass of the small object. The minimum spatial and temporal discretisation allowed is typically determined by the smallest scale in the problem, now proportional to the mass ratio. Moreover, resolving the GW content of the simulations requires evaluation of quantities in the so-called wave or radiation zone, which is set by the GW wavelength, in turn determined by the orbital separation and the mass scale of the SMBH. Finally, the number of cycles contained in EMRI GWs scales inversely with the mass ratio, and thus, one requires extremely long evolutions. All of these concerns make numerical relativistic evolutions of EMRIs extremely difficult and probably unlikely, without the implementation of new methods and techniques.

## 2.2 Approximation Schemes

For certain systems, such as EMRIs, an *approximate* decoupling of the equations of motion for matter and field equations for the gravitational field is indeed possible. EMRIs possess a natural (and time-independent) small parameter, the mass ratio of the system $q := m_s/m_L \ll 1$, where $m_s$ is the mass of the small compact object and $m_L$ is the SMBH mass. Notice that this parameter remains small all throughout the coalescence, a non-trivial observation that does not necessarily hold in other perturbative scheme (eg. in the PN approximation one expands in powers of the orbital velocity, which varies throughout the trajectory). Expanding in this small parameter allows one to decouple the above system of equations by proposing the following perturbative decomposition of the fields of interest:

$$g_{\mu\nu} = \bar{g}_{\mu\nu} + q\, h_{\mu\nu},$$
$$z^\mu = z^\mu_{geod} + q\, \delta z^\mu,$$

where $\bar{g}_{\mu\nu}$ is the background SMBH metric, $h_{\mu\nu}$ is a perturbation, $z^\mu_{geod}$ are geodesic trajectories in the background of $\bar{g}_{\mu\nu}$ and $\delta z^\mu$ is a perturbation. One can check that to leading order in $q$, the SEP requires $z^\mu_{geod}$ to be geodesics of $\bar{g}_{\mu\nu}$. To leading order in $q$, the linearised Einstein equations describe the evolution of $h_{\mu\nu}$ as sourced by $z^\mu_{geod}$. This boot-strapping scheme can then be repeated to obtain the correction in the trajectories as induced by $h_{\mu\nu}$ to next order in $q$.

Let us now examine in more detail the schematic form of the decoupled evolution equations. To zeroth order, the SEP requires that one solve the evolution equation for the trajectories in a fixed background geometry, ie. the geodesic equations

$$\ddot{z}^\alpha_{geod} = F^\alpha_{BG} := \Gamma^\alpha_{\mu\nu}[\bar{g}] u^\mu_{geod} u^\nu_{geod}, \tag{1}$$

where overhead dots stand for differentiation with respect to proper time $\tau$ and $u^\mu_{geod} = \dot{z}^\mu_{geod}$ is the small body's geodesic 4-velocity. The quantity $F^\alpha_{BG}$ can be thought of as a "forces" exerted by the background geometry, with $\Gamma^\alpha_{\mu\nu}[\bar{g}]$ the Christoffel connection associated with $\bar{g}_{\alpha\beta}$. Once the background trajectory has been found, one can solve the linearised Einstein equations

$$\Box \bar{h}_{\alpha\beta} + 2\bar{R}^\mu{}_\alpha{}^\nu{}_\beta \bar{h}_{\mu\nu} = -16\pi m_s \int \frac{\delta^4\left[x^\mu - z^\mu_{geod}\right]}{\sqrt{-\bar{g}}} u_\alpha u_\beta d\tau, \quad (2)$$

where $\bar{h}_{\alpha\beta}$ is some regularised part of the metric perturbation $h_{\alpha\beta}$, $\bar{R}^\mu{}_{\alpha\nu\beta}$ is the Riemann tensor of the background and $\bar{g}$ is the determinant of the background metric. Once the metric perturbation is calculated, one can compute the deviation vector $\delta z^\mu$ via the equation

$$\delta \ddot{z}^\alpha = F^\alpha_{SF} := \nabla^{\alpha\beta\gamma} h_{\beta\gamma}, \quad (3)$$

where $F_{SF}$ is the self-force induced by the particle on itself, with $\nabla^{\alpha\beta\gamma}$ some derivative operator defined in Barack, 2009. In principle, one can then compute second-order corrections to the gravitational field due to the deviation vector, but in practice such second order computations are quite difficult.

The correctness of such a decoupling scheme hinges on a fundamental assumption in EMRI modelling: *the adiabatic approximation*. This approximation refers to the assumption that the deviation vector $\delta z^\mu$ is small relative to $z^\mu_{geod}$ in that the former's rate of change is much smaller than the latter's orbital period. As first pointed out by Gralla and Wald, 2008, this smallness assumption must only hold quasi-locally for some small proper time, as otherwise such a linear approximation would lead to a linear drift in the solution to the equations of motion. Their proposed solution was to self-consistently recalculate the linear correction dynamically while the trajectory evolves, a proposal later formalised by Pound and Poisson, 2008 and Pound, 2009 in the method of *osculating orbits*.

This method refers to the idea that at each point of the small body's world-line there exists a geodesic that lies tangent to it. The exact world-line is then nothing more than an interpolation between these different tangent geodesics. From such a standpoint, EMRI modelling reduces to understanding geodesics with varying orbital elements $\mathcal{I}^A := (E, L, Q)$, where $E$, $L$ and $Q$ are the geodesic energy per unit mass, angular momentum per unit mass and Carter constant, which are not constant anymore but assumed to vary adiabatically. Moreover, one must also account for conservative corrections to the initial conditions of this sequence of geodesics. One practical implementation of this idea is to enhance the geodesic equations with new evolution equations for the orbital elements:

$$\dot{z}^\mu = F(x^\mu; \mathcal{I}^A), \qquad \dot{\mathcal{I}}^A = G(x^\mu; z^\mu; \mathcal{I}^A) \quad (4)$$

where $F(\cdot)$ is some function of the background coordinates $x^\mu$ and the orbital elements, determined by the geodesic equations, and $G(\cdot)$ is some other function determined by the self-force.

The implementation of this approximation can take different forms, depending on how the evolution equation for the orbital elements is handled (Pound and Poisson, 2008). The *adiabatic secular approximation* replaces this evolution by an averaged version that removes all oscillatory terms. This approximation neglects both conservative corrections (time-symmetric corrections to the background geometry or the effective potential), as well as oscillatory corrections to the dissipative dynamics. On the other hand, the *adiabatic radiative approximation* constructs the evolution of $\mathcal{I}^A$ through the time-asymmetric part of the radiation field (ie. the half-retarded minus half-advanced solution of Mino, 2003). This approximation then only neglects conservative corrections to the self-force. Whether these conservative terms are required to obtain an accurate representation of the waveform will depend on the type of orbit considered (Pound and Poisson, 2008). In the next section, we summarise the established models for the calculation of EMRI templates and classify them based on the definitions introduced here.

## 3  Some EMRI Models

In this section we summarise some proposed models to construct EMRI waveforms. This summary is essential to set the stage for the introduction of a new method, allowing us to compare the new scheme to previously developed ones. We shall not attempt to make a comprehensive review of these models here.

### 3.1  Black Hole Perturbation Theory Approach

BH perturbation theory refers to a scheme where the full metric is decomposed into a background plus a perturbation, with the latter assumed much smaller than the former. For EMRIs, such a perturbation is proportional to the mass ratio, which is always much smaller than unity, and thus, the above decomposition is valid. As for the background, one does not assume that it is Minkowski, but instead it is usually assumed to be some non-linear, exact BH metric, such as Schwarzschild or Kerr. The avoidance of simultaneously employing a small-mass ratio approximation and some other approximation renders this mono-variate expansion incredibly difficult to implement in practice.

Most BH perturbation theory calculations are carried out in the *Lorenz gauge*. In this context, a gauge condition is nothing but a differential constraint on the metric perturbation, which thus restricts some of the gauge freedom inherent in GR. In the Lorenz case, this differential condition is $\bar{\nabla}^\alpha \bar{h}_{\alpha\beta} = 0$, where $\bar{\nabla}$ is the covariant derivative operator associated with the background metric and $\bar{h}_{\alpha\beta}$ is the trace-reversed metric perturbation. This gauge is different from the harmonic gauge condition, usually employed in PN theory, due to the non-linearities of the Christoffel symbols in the covariant derivative (for a detailed analysis of these differences see Damour, 2009). A final noteworthy point is that no restriction is in principle imposed on the coordinate system of the background or the metric perturbation.

As described in the previous section, the calculation of the metric perturbation to first order hinges on the computation of the self-force that allows the small object to inspiral and deviate from a geodesic. If the self-force were to be neglected, then the small object would never inspiral (there would not be radiation-reaction in the trajectories) and the waveform would have a constant amplitude (there would not be any chirping). An extensive program for the calculation of the self-force is currently underway, but we shall not summarise this here, instead referring the interested reader to Barack, 2009. Suffice it to say, this endeavour is an extremely difficult one that has not yet been solved in full generality. However, there are recent promising advances that suggest self-force calculations on a Schwarzschild background are finally under control, even if waveform production has not yet been fully investigated (Sago, 2009 and Detweiler, 2009).

Based on the classification discussed in the previous section, one would agree that BH perturbation theory relies only on the adiabatic approximation. That is, the self-force that one is attempting to compute is the mfull force (exact to linear order in the mass ratio), and not its averaged form, time-asymmetric form or a slow-velocity approximation. This is one of the main reasons why the calculation of the self-force is extremely difficult and only brave souls have embarked on it.

**3.2 Teukolsky Approach**

The Teukolsky approach is a scheme where one computes the waveform assuming one can interpolate over a sequence of geodesics as guided by the amount of radiation emitted in GWs at infinity and down the BH horizon. This method was initially proposed in Poisson, 1993a and then implemented in Hughes, 2000, 2001, with follow-up refinements in the frequency-domain by Glampedakis and Kennefick, 2002, Drasco and Hughes, 2006, Drasco et al., 2005 and Hughes et al., 2005 and in the time-domain by Lopez-Aleman et al., 2003, Khanna, 2004, Burko and Khanna, 2007 and Sundararajan et al., 2007, 2008. We shall discuss this approach in more detail than any other one, as these are the waveforms we employ to compare against the new EOB approach in **Sec. 4**.

The waveform construction proceeds as follows. First, one chooses a set of orbital elements and initial conditions to define a geodesic and solves the geodesic equations to obtain the bounded trajectories of the small object in the SMBH background. Second, these trajectories are used to solve the linearised Einstein equations, from which one can compute the rates of change of the orbital elements. Third, these rates of change are used to infer some new orbital elements and repeat the algorithm. One can think of this scheme as obtaining snapshots of the waveforms for each set of orbital elements and then interpolating to obtain the entire waveform as a function of time (Drasco and Hughes, 2006).

Let us now discuss each step in more detail. The first step requires the time-domain, numerical solution to the geodesic equation. After carefully dealing with turning points in the orbit (see eg. Hughes, 2000), one obtains the temporal evolution of the background geodesic trajectories. Given these trajectories, the second step requires the solution to the linearised Einstein equations, a dramatically more difficult endeavour. Teukolsky first found an equivalent version of the linearised Einstein equations from the Bianchi identities in terms of the Newman-Penrose or Weyl curvature scalar $\psi_4$ as a variable. This quantity describes the tidal interactions of the small body and the SMBH (Poisson, 2004a). Teukolsky then realised that the linearised Einstein equations could be decoupled into radial and polar ones in the Fourier domain (Teukolsky, 1972, 1973). The solution to the polar equation reduces to spin-weighted spheroidal harmonics, while the solution to the radial equation can be obtained via Green function methods.

For the comparisons in **Sec. 5**, we employ the code described in Hughes, 2000 (modified with the spectral techniques of Fujita and Tagoshi, 2005) to solve for $\psi_4$. This frequency-domain code solves for the multipolarly decomposed $\psi_4$

$$\psi_4 = \frac{1}{r^4} \sum_{\ell m} R_{\ell m \omega}(r) \,_{-2}Y^{\ell m}(\theta, \phi) \, e^{-im\omega t} \tag{5}$$

where $_{-2}Y^{\ell m}(\theta, \phi)$ are spin-weight $-2$ spherical harmonic, $\omega = \sqrt{m_L/r^3}$ is the frequency of circular Schwarzschild orbits, and $(t, r, \theta, \phi)$ are Boyer-Lindquist coordinates. Equation **5** takes on its simple form only for circular orbits in a Schwarzschild background, which we restrict attention to here and in **Sec. 5**. The radial function $R_{\ell m \omega}(r)$ satisfies the radial Teukolsky equation and it can be constructed via a Green function integral over the product of the homogeneous solution and the source to the Teukolsky equation. The homogeneous Teukolsky equation is

$$(r^2 - 2m_L r)^2 \frac{d}{dr}\left(\frac{1}{r^2 - 2m_L r}\frac{dR_{\ell m \omega}}{dr}\right) - V_{\ell m}(r) R_{\ell m \omega} = 0 \,, \tag{6}$$

where we recall that $m$ is a harmonic index (not a mass parameter) and the potential

$$V(r) = -f^{-1}\left[r^2 m^2 \omega^2 + 4im(r - m_L)\omega\right] + 8im\omega r + \lambda_{\ell m} \tag{7}$$

with $f := 1 - 2m_L/r$ the Schwarzschild factor and $\lambda_{\ell m}$ the eigenvalues of the angular sector of the Einstein equations, whose eigenfunctions are $_{-2}Y^{\ell m}(\theta, \phi)$. This equation has two independent solutions, $R^H_{\ell m \omega}$ and $R^\infty_{\ell m \omega}$, which can be used to construct the full solution $R_{\ell m \omega} = R^H_{\ell m \omega} Z^\infty_{\ell m \omega} + R^\infty_{\ell m \omega} Z^H_{\ell m \omega}$, where $Z^\infty_{\ell m \omega}$ are certain Green-function integrals over the product of the homogeneous solutions and the Teukolsky source (see eg. Eq.(4.10) in Hughes, 2000). With these solutions at hand, one can reconstruct $\psi_4$ via Eq. **5**.

Once the Teukolsky equation has been solved, one can relate $\psi_4$ to the metric perturbation. In the radiation zone (several gravitational wavelengths away from the centre of mass of the system), this relation is simply $\psi_4 \to 1/2(\ddot{h}_+ - i\ddot{h}_\times)$ as $r \to \infty$, which in the frequency domain can be trivially inverted to obtain

$$h_+ - ih_\times = \sum_{\ell m} \frac{1}{m\omega} Z^H_{\ell m} {}_{-2}Y^{\ell m}(\theta,\phi) e^{-im\omega t}, \qquad (8)$$

where $h_{+,\times}$ are the plus- and cross-polarisations of the metric perturbation (Misner et al., 1973). Once the metric perturbation has been found, one can construct the GW flux of energy and angular momentum out to infinity and into the SMBH. This parallels the fact that there are two independent solutions to the radial Teukolsky equation: one that describes purely outgoing radiation and one that represents radiation falling into the SMBH. The flux of energy to infinity is given by

$$\left(\frac{dE}{dt}\right)^\infty = \sum_{\ell m} \frac{|Z_{\ell m}|^2}{4\pi m^2 \omega^2}, \qquad (9)$$

while that falling into the SMBH is somewhat more complicated (see eg. Eq.(4.16) in Hughes, 2000). As we shall later be concerned with circular orbits, the angular momentum flux is trivially related to the energy flux via $\dot{L} = \dot{E}/\omega$.

With this information, one can now construct a *radiation-reaction grid* or phase space. That is, for any given geodesic described by a set of orbital elements, one can compute their rates of change. These quantities can be plotted in a phase space diagram, where the orbital elements serve as coordinates and their rates of change become vectors that describe their flow (see eg. Fig. 1 in Hughes, 2001). Once the phase space is constructed, one can "use the arrows to connect the dots," ie. employ interpolation through a sequence of geodesics to obtain the temporal evolution of the orbital elements. Similarly, at each geodesic (or orbital element), one can also store a waveform time-snapshot and then construct its temporal evolution via interpolation.

For the comparisons in **Sec. 5**, we build a radiation-reaction grid from $r = 10,000M$ to the Schwarzschild ISCO at $r_{\text{ISCO}} = 6M$ evenly stepping in $v \equiv \sqrt{M/r}$. This grid then allows us to construct the time-evolution of the waveform, via cubic spline interpolation. A discretisation of 1,000 points over the range $v \in [0.01, 0.408]$ suffices to guarantee a phase accuracy of at least $10^{-2}$ radians after a 2-year evolution. The dominant source of error is the truncation of the $(\ell, m)$ sums in the solution to the Teukolsky equation, and due to discretisation in orbital phase space (apart from issues of numerical precision, which place a lower bound in absolute accuracy of roughly $10^{-14}$). The Teukolsky solution is constructed in such a way as to guarantee that the multipolar sums have converged to a fractional error of no more than $10^{-10}$ (Hughes, 2000, 2001). In practice, we find that the flux is accurate to at least $10^{-13}$ in the low velocity region $v < 0.1$. One should point out that this level of accuracy is at least 6 orders of magnitude greater than previous flux computations (Poisson, 1993a, Cutler et al., 1993, 1994, Poisson and Sasaki, 1995, Berti, 2001 and Yunes and Berti, 2008).

Once the waveforms are computed in this way, one can extract the GW phase and amplitude, which are the quantities we shall later use to compare against the EOB approach. The amplitude is obtained by simply taking the absolute value of the waveforms $A_{GW}^{Teuk} := |h_+ - i\, h_\times|$. Given any function of the form $h_+ - i\, h_\times = A_{GW}^{Teuk} \exp(-i\phi_{GW}^{Teuk})$, one can extract the phase via

$$\phi_{GW}^{Teuk} = -\text{Im}\left[\log\left(\frac{h_+ - i\, h_\times}{A_{GW}^{Teuk}}\right)\right], \qquad (10)$$

where Im stands for the imaginary part operator, after *unwrapping* the phase due to discontinuities at integers of $\pi$. We find this to be more accurate and efficient than solving a differential equation for the GW phase in terms of the GW frequency.

What are the advantages and disadvantages of this approach? Perhaps the greatest advantage is that here one solves the linearised Einstein equations without any approximations beyond the small mass ratio one. As we shall see, all other approximations (apart from the full BH perturbation theory approach described earlier) employ a multipolar expansion to solve for the metric perturbation, which is inherently a weak-field approximation. In spite of this advantage, however, the metric reconstruction from $\psi_4$ is still an open problem in a Kerr background (ie. we can only extract $h_{+,\times}$ near spatial infinity or near the SMBH horizon). Thus, a full self-force calculation is not yet possible from the solution to the Teukolsky equation. In view of this, it is unclear how one could improve on the Teukolsky approach to include conservative corrections to the background. Another disadvantage of this approach is its computational cost. Although the solution to the Teukolsky equation is rather inexpensive for circular orbits in a Schwarzschild SMBH background, its cost increases rapidly for more generic orbits in spinning BH backgrounds, and it is likely the cost becomes at some point prohibitive. Moreover, frequency-domain codes cannot handle trajectories that are highly eccentric or venture inside the Schwarzschild ISCO, as stable circular orbits do not exist there by definition and one therefore cannot infer $\dot{r}$ and $\dot{E}$ (Glampedakis et al., 2002). Time-domain waveforms could handle the plunge, but their accuracy is currently not as good as that of frequency-domain codes for certain orbits (although see Sundararajan et al., 2007, 2008 for recent advances).

Before moving on, let us finally discuss the type of approximation that is employed here. The time-evolution of orbital elements is obtained by balance-law arguments at infinity and at the event horizon (the amount of energy radiated in GWs must be equal and opposite to the amount of energy lost by the orbiting body). Such arguments hold only mon average and not in a point-wise fashion, thus forcing the Teukolsky scheme to be a *radiative approximation*. That is, no conservative effects are accounted for in the Teukolsky approach, as the presence of the small body is ignored when solving the equations of motion (after all, the trajectories employed in the Teukolsky scheme are pure geodesics). These considerations lead to a definite disadvantage: the impossibility to account for higher-order effects. To date, there is no prescription to enhance this scheme via the inclusion of conservative effects or second order dissipative effects. We shall see in **Sec. 5** that such effects might be relevant for parameter estimation given a sufficiently long observation window.

### 3.3 Kludge Approach

The word "kludge" means "a system of poorly matched components" (Merriam-Webster, 2009), which implies from the start that this approach employs a combination of several approximation schemes. Initially proposed by Barack and Cutler, 2004, the *analytic kludge* approach was intended as a fast method to generate waveforms accurate enough for *descoping* exercises. In view of this, it is highly unlikely these waveforms are accurate enough for parameter estimation. To date, there has been only one study of the formal accuracy of analytic kludge waveforms (Babak et al., 2007) that suggests they become inaccurate on a time scale of hours.

In spite of this, let us now review the analytic kludge scheme. First, one approximates the solution to the linearised Einstein equations via a multipolar decomposition, keeping only the leading-order, Newtonian contribution (the so-called Peters and Mathews solutions (Peters and Mathews, 1963)). The waveforms are then given by the quadrupole formula, namely $h_{+,\times} = 2\epsilon^{ij}_{+,\times} \ddot{I}_{ij}/R$, where $R$ is the distance to the source, $\epsilon^{ij}_{+,\times}$ is a polarisation tensor, overhead dots stand for time derivatives and $I^{ij} \propto m_s z^i z^j$ is the reduced, symmetric and trace-free (mass) quadrupole moment, which depends on the small body's trajectory $z^i$.

Analytic kludge waveforms thus require knowledge of the trajectories, which are approximated via a combination of Newtonian and PN theory. The general structure is modelled as purely Keplerian, following again Peters and Mathews, 1963. Such Keplerian ellipses depend on some conserved orbital elements (eg. the semi-major axis, the eccentricity), and thus some additional approximation is required to prescribe their temporal evolution due to radiation-reaction. The analytic kludge approach employs low-order PN expansions to model the evolution of the orbital phase, eccentricity and inclination angle.

With these approximations at hand, the construction of analytic kludge waveforms can be summarised as follows. First, one solves the evolution equations for the orbital elements. Second, one inserts this time-dependent orbital elements in the quadrupole formula to obtain the time-evolution of the waveforms. Third, one constructs the LISA response function, taking special care of Doppler modulation due to the motion of LISA in space.

How do we classify the analytic kludge approach? Clearly, the adiabatic approximation is employed to linearise the field equations in the mass ratio, but moreover a PN scheme is employed to model the trajectories. This PN scheme is used in *average* form; that is, the evolution equations for the trajectories employ the adiabatic approximation to present expressions that have been orbit-averaged (see eg. Arun et al., 2007b, 2007a). As such, the analytic kludge approach employs a *secular approximation*.

The mnumerical kludge approach is an attempt to improve on the analytic kludge scheme (Babak et al., 2007). This approach proposes to model the small body's trajectory via solutions to the geodesic equation supplemented by PN, orbit-averaged, evolution equations for the orbital elements. As in the analytic case, this scheme is also secular, as radiation-reaction is treated in an orbit-averaged fashion. Numerical kludge waveforms are also approximated as a multipolar expansion, but instead of keeping only the leading-order term, one here also keeps the next-to-leading-order term (see e.g., Thorne, 1980).

An important detail that is usually not appreciated when mixing multipole moment solutions with geodesic trajectories is that of coordinates, an issue referred to as the "beads-on-a-wire" problem (Babak et al., 2007). Geodesic evolutions are usually carried out in (spherical) Boyer-Lindquist coordinates, while the metric reconstruction is defined in asymptotically Cartesian, mass centered coordinates (Thorne, 1980). A well-known coordinate transformation needs to be performed before one can employ geodesic trajectories in the quadrupole-octopole formula. A simple flat-space spherical to Cartesian coordinate transformation then forces the small body to move on a curved path like a bead on a wire (Babak et al., 2007), which leads to errors in the waveforms, which could be significant for parameter estimation.

**3.4 Semi-Relativistic Approach**

The semi-relativistic approximation was first introduced by Ruffini and Sasaki, 1981. This scheme approximates wave generation by assuming geodesic motion and flat-space propagation. As such, there is no radiation-reaction in this approach, the small body never plunges into the SMBH and the waveforms are non-chirping. Once the geodesic equation has been solved for the trajectory, this is used in a multipolar solution to the linearised Einstein equations. One could view the semi-relativistic approach as simply the numerical kludge approach without radiation-reaction.

Given these considerations, the semi-relativistic approach can be considered as an extreme adiabatic approximation, where one essentially assumes the mass ratio is zero. Due to the neglect of radiation-reaction, these waveforms are only accurate for very short time periods, perhaps not sufficiently long to be employed in detection strategies. In spite of this, significant work has gone into their development, as one can study completely generic orbits in Kerr backgrounds with relative ease and obtain a qualitative feel for their behaviour.

**3.5 The post-Newtonian Approach**

The theory of PN expansions traces back to the work of Chandrasekhar in the early days of gravitational theory (Chandrasekhar, 1965, 1969 and Chandrasekhar and Nutku, 1969). We shall not attempt to review all of PN theory here, as recent reviews already exist in the literature (Blanchet, 2006). Instead, we here content ourselves with describing why traditional PN techniques fail to model EMRIs.

PN theory has been known to be poorly convergent to extreme-mass ratio systems (Poisson, 1995 and Simone et al., 1995, 1997). When investigating the poorly conservative sector of the theory, Blanchet, 2002 studied the behaviour of the location of the ISCO as a function of PN order to find that indeed the series is purely convergent for $q \ll 1$. Such loss of convergence is related to the rapid increase in magnitude of the coefficients of the PN expansion when $q \ll 1$. Recently, Yunes and Berti, 2008 showed that this poor convergence is associated with the asymptotic nature of the series and is in fact also present in the dissipative sector of the theory (ie. in the energy flux as a function of PN order), even when one enhances the PN approximation with terms up to $O(v^{11})$ (where $v$ is the orbital velocity), which are known in the point-particle limit (Shibata et al., 1995, Tagoshi et al., 1996, Mino et al., 1997 and Tanaka et al., 1996).

Although the poor convergence of the PN series in the EMRI limit does not necessarily imply that PN templates are insufficient for LISA data analysis, recently Mandel and Gair, 2009 argued that this is indeed the case. Their argument rests on comparisons of the accumulated number of cycles as computed with truncated PN expansions at adjacent orders and with or without mass ratio-dependent terms. For example, they find that terms of $O(v^7)$ in the phase can contribute thousands of cycles for EMRIs after a one year evolution. Similarly, they find that mass ratio-dependent terms can contribute up to ten cycles of phase difference after a one year evolution. Of course, the comparison of the GW phase at adjacent PN orders cannot be used to determine the PN series accuracy for data analysis purposes, as we have already established the approximation's asymptotic nature and poor convergence. In spite of this, the speculation of Mandel and Gair, 2009 was recently verified by Yunes et al., 2009 through comparisons of traditional PN templates to Teukolsky-based waveforms. The inaccuracies in the former are likely to be too large for the direct implementation of traditional PN methods for EMRI parameter estimation.

## 4  The Effective-One-Body Approach

So far we have summarised and described some proposed models to approximately construct EMRI GWs; we now concentrate on the new proposed method, based on the effective-one-body (EOB) formalism. This framework is a resummation technique that aims at improving the accuracy of traditional PN expansions. The EOB scheme was first introduced to analytically model GWs produced during inspiral, merger, and ringdown from comparable-mass binaries (Buonanno and Damour, 1999, 2000). Later on, this scheme was extended to higher PN orders (Damour et al., 2000), spinning BHs (Damour, 2001, Buonanno et al., 2006, Damour and Nagar, 2007a, Damour et al., 2008 and Barausse et al., 2009), small mass ratio mergers (Nagar et al., 2007 and Damour and Nagar, 2007a, 2007b), and new factorised resummation techniques (Damour and Nagar, 2007b and Damour et al., 2009). The calibration of some adjustable parameters in the EOB-dynamics and waveforms allow for agreement between EOB and numerical relativity waveforms, at least for the cases so far studied (Damour et al., 2008a, 2008b, Damour and Nagar, 2008, 2009, Buonanno et al., 2007, 2009 and Boyle et al., 2008), possibly providing good templates for data analysis (Damour et al., 2003 and Buonanno et al., 2009)

The first EOB model that studied the late inspiral, plunge and merger of small-mass ratio compact objects were those of Nagar et al., 2007 and Damour and Nagar, 2007a. Nagar et al., 2007 extended the EOB model beyond an adiabatic sequence of circular orbits through insight gained by solving the time-dependent Regge-Wheeler-Zerilli equation for a source moving under the action of an EOB-evaluated radiation reaction force. Later on, Damour and Nagar, 2007a introduced, for the first time, an improved, factorised resummation technique, where certain high-PN order coefficients were calibrated to numerical data, obtained by solving the Regge-Wheeler-Zerilli equations.

Although these pioneering achievements indeed represent the first attempt to model small-mass ratio inspirals within the EOB approach, they did not consider sufficiently long inspirals and sufficiently extreme mass ratios to determine the suitability of the EOB scheme for LISA data analysis applications. In particular, Nagar et al., 2007 and Damour and Nagar, 2007a considered only the last 10 cycles before plunge, including the merger and ringdown, for a particle with mass ratio $m_s/m_L \sim 10^{-2}$. We shall show later that the last 10 cycles of inspiral are negligible for data analysis purposes, since EMRIs are expected to possess millions of cycles in the detector band, with the early inspiral dominating the SNR.

Partially inspired on Nagar et al., 2007 and Damour and Nagar, 2007a, the recent work in Yunes et al., 2009 was specifically aimed at determining whether the EOB scheme could be effective for LISA data analysis purposes. In particular, Yunes et al., 2009 extended the analysis of Nagar et al., 2007 and Damour and Nagar, 2007a through BH perturbation theory insights, employed significantly more accurate numerical BH perturbation theory simulations to calibrate the EOB model and compared in detail the resulting waveforms to Teukolsky ones over a two year period. Such a study concentrated on systems with mass ratio $m_s/m_L \sim 10^{-4}$ and $10^{-5}$ and considered millions of cycles before reaching the ISCO. The plunge, merger and ringdown were here neglected due to their low impact on EMRI data analysis.

In what follows, we present the framework and results of Yunes et al., 2009 in more detail. As a first step, we consider only small compact objects (BHs or NSs) inspiraling *quasi-circularly* into a *non-spinning* SMBH (Amaro-Seoane et al., 2007), although these assumptions can be relaxed. We begin by presenting how the EOB conservative orbital dynamics are treated, and then proceed with a discussion of EOB radiation-reaction. Precisely how the radiation-reaction force is implemented allows us to define a few different EOB models that we compare to Teukolsky waveforms in **Sec. 5**. We then proceed to describe how waveforms are computed in the EOB scheme and how observable quantities are extracted, concluding with a brief discussion of numerical errors. We refer the reader to Damour, 2008 for a more comprehensive review of the EOB scheme.

### 4.1 Orbital Dynamics

Let us first define some notation. We consider a small BH of mass $m_s$ and a SMBH of mass $m_L$, with total mass $M := m_s + m_L$, mass ratio $q := m_s/m_L \ll 1$, reduced mass $\mu := m_L/(1+q)$ and symmetric mass ratio $\nu := q/(1+q)^2 \ll 1$. We further introduce spherical polar (harmonic) coordinates $(t, r, \Theta, \Phi)$ (where $t$ and $r$ are $M$-normalised), and their reduced ($\mu$-normalised) conjugate momenta $(p_t, p_r, p_\Theta, p_\Phi)$. For numerical reasons, we find it more convenient to evolve the reduced conjugate momentum $p_{r^*}$ associated with the tortoise coordinate $r^*$, instead of that associated with the standard radial coordinate. Finally, as we are considering orbits around non-spinning BHs, all motion is constrained to a plane, which we have the freedom to choose due to spherical symmetry, so we set $(\Theta, p_\Theta) = (\pi/2, 0)$ all throughout the evolutions.

With this in mind, the non-spinning EOB Hamiltonian reads (Buonanno and Damour, 1999)

$$H^{\text{real}} = M\sqrt{1 + 2\nu\left[(H^{\text{eff}} - \mu)/\mu\right]} - M, \quad (11)$$

and the *effective* Hamiltonian is (Buonanno and Damour, 1999, Damour et al., 2000 and Damour and Nagar, 2007b)

$$H^{\text{eff}} = \mu\sqrt{p_{r_*}^2 + A(r)\left[1 + \frac{p_\Phi^2}{r^2} + 2(4 - 3\nu)\nu\frac{p_{r_*}^4}{r^2}\right]}, \quad (12)$$

where $A(r)$ and $D(r)$ are the temporal-temporal and radial-radial components of an effective metric, obtained by mapping the two-body PN system to an effective one-body system.

We here employ a Padé-resummed[2] version of these functions (Boyle et al., 2008):

$$A(r) = \frac{\text{Num}(A_3^1)}{\text{Den}(A_3^1)}, \qquad D(r) = \frac{\text{Num}(D_3^0)}{\text{Den}(D_3^0)}, \quad (13)$$

where

$$\text{Num}(A_3^1) := r^2\left[(8\nu - 16) + r(8 - 2\nu)\right],$$
$$\text{Den}(A_3^1) := r^3(8 - 2\nu) + 4\nu r^2 + 8\nu r + 4\nu^2,$$
$$\text{Num}(D_3^0) := r^3,$$
$$\text{Den}(D_3^0) := r^3 + 6\nu r + 2\nu(26 - 3\nu).$$

---

[2] A Padé resummation of $(m, n)$ order, $P_n^m$, is defined as a fraction whose numerator and denominator are polynomials of order $m$ and $n$ respectively (see eg. Appendix A in Damour et al., 1998).

The resummed Hamiltonian is thus accurate to 3 PN order if re-expanded in $m/r \ll 1$ and $v \ll 1$[3]. Also note that we here neglect any 4PN calibration parameter (such as $a_4$ or $d_4$) as these are expected to be proportional to $\nu$. There is no *a priori* formal reason to believe that the same resummation of the Hamiltonian that leads to accurate waveforms in the equal and nearly-equal mass case also succeeds in modelling EMRI waveforms, but they do provide a starting point to begin the analysis.

The EOB Hamilton-Jacobi equations can now be derived in terms of the reduced (i.e., dimensionless) quantity $\widehat{H}^{\text{real}} \equiv H^{\text{real}}/\mu$ (Buonanno and Damour, 2000):

$$\frac{dr}{dt} = \frac{A(r)}{\sqrt{D(r)}} \frac{\partial \widehat{H}^{\text{real}}}{\partial p_{r_*}}, \qquad \frac{d\Phi}{dt} = \frac{\partial \widehat{H}^{\text{real}}}{\partial p_\Phi}, \tag{14}$$

$$\frac{dp_{r_*}}{dt} = -\frac{A(r)}{\sqrt{D(r)}} \frac{\partial \widehat{H}^{\text{real}}}{\partial r}, \qquad \frac{dp_\Phi}{dt} = \widehat{\mathcal{F}}_\Phi. \tag{15}$$

where $\widehat{\mathcal{F}}_\Phi := \mathcal{F}_\Phi/\mu$ is a reduced radiation-reaction force that controls the rate of inspiral. Note that the orbital angular velocity $\Omega := M^{-1}d\Phi/dt = M^{-1}\partial\widehat{H}^{\text{real}}/\partial p_\Phi$.

Unlike traditional PN equations of motion, which are necessarily low-velocity expansions, the above equations reduce exactly to the full geodesic equations in the limit $q = 0$. This is because, in this limit, the Hamiltonian reduces exactly to that of a particle orbiting in a Schwarzschild spacetime. When we compare EOB evolutions to evolutions in the Teukolsky approach (see **Sec. 5**) we initially set $\nu = 0$ and then study the effect of this conservative mass ratio-dependent terms in **Sec. 5.5**.

### 4.2 Radiation-Reaction Force

The radiation-reaction force can be obtained through balance-law arguments. One first realises that the conjugate momentum to the $\Phi$ coordinate, $p_\Phi$, is exactly the PN conserved angular momentum (Damour et al., 2000). Via the Hamilton-Jacobi equations, one then finds that $\widehat{\mathcal{F}}_\Phi = \nu^{-1}\dot{L}$, where $\dot{L}$ is the rate of change of the binary system's orbital angular momentum. Assuming that the binary loses as much energy and angular momentum as that radiated in GWs $\dot{L}_{\text{GW}}$, one can write $\widehat{\mathcal{F}}_\Phi = -\nu^{-1}\dot{L}_{\text{GW}}$. Finally, using the circular orbit relation $\dot{E}_{\text{GW}} = \Omega\dot{L}_{\text{GW}}$ one derives $\widehat{\mathcal{F}}_\Phi = -(\nu\Omega)^{-1}\dot{E}_{\text{GW}}$, where $\dot{E}_{\text{GW}}$ is the GW energy flux. We shall investigate different EOB models that differ only in the expression employed for the GW energy flux when constructing the radiation-reaction force.

#### 4.2.1 The Padé Model

The first model we consider employs a Pade-resummation of the orbit-averaged, Taylor-expanded energy flux $\dot{E}^{\text{P}}_{\text{GW}} = F^p_q(v; v_{\text{pole}})$, where $v = (M\Omega)^{1/3}$ is the Keplerian orbital velocity and $v_{\text{pole}}$ is an adjustable parameter that defines the location of the EOB light-ring, and $p + q$ is twice the approximant's PN order [i.e., $(v/c)^{(p+q)}$] (Damour et al., 1998 and Buonanno et al., 2009):

---

[3] A quantity is said to be of $N$th PN order if it is of $O[(v/c)^{2N}]$.

$$F_q^p(v; v_{\text{pole}}) = \frac{1}{1 - v/v_{\text{pole}}} f_q^p(v), \tag{16}$$

where $v_{\text{pole}}$ formally stands for the location of the light-ring associated with the real Hamiltonian[4], while the Padé-Resummed, log-factorised PN flux is[5]

$$f_q^p(v) = \frac{32}{5} v^{10} v^{10} \left[ 1 + \log(v) \left( \sum_{i \geq 6}^{k} \ell_i v^i \right) \right] \tag{17}$$

$$\times P_q^p \left[ \left( 1 - \frac{v}{v_p} \right) \sum_{i=0}^{k} \mathcal{F}_i^{lf} v^i \right],$$

with $\ell_i$ coefficients proportional to logarithmic terms and $\mathcal{F}_i^{\text{log-fac}}$ coefficients proportional to terms that are log-independent. The log-dependent terms are $\ell_{<6} = 0$ and (Tanaka et al., 1996 and Damour et al., 1998)

$$\ell_6 = -\frac{1712}{105}, \qquad \ell_7 = 0, \qquad \ell_8 = -\frac{34261}{4410},$$
$$\ell_9 = 0, \qquad \ell_{10} = \frac{3278427269}{440082720}, \qquad \ell_{11} = -\frac{298177\pi}{17640},$$
$$\ell_{12} = \ell_{12}^{\text{cal}}, \qquad \ell_{13} = \ell_{13}^{\text{cal}},$$

the $P_n^m[\cdot]$ operator stands for Padé resummation and $\ell_{12,13}^{\text{cal}}$ are calibration parameters. The log-independent terms associated with the GW flux carried out to spatial infinity are (Boyle et al., 2008 and Tanaka et al., 1996)

$$_\infty\mathcal{F}_0^{\text{log-fac}} = 1, \qquad _\infty\mathcal{F}_1^{\text{log-fac}} = 0,$$
$$_\infty\mathcal{F}_2^{\text{log-fac}} = -\frac{1247}{336} - \frac{35}{12}\nu, \qquad _\infty\mathcal{F}_3^{\text{log-fac}} = 4\pi,$$
$$_\infty\mathcal{F}_4^{\text{log-fac}} = -\frac{44711}{9072} + \frac{9271}{504}\nu + \frac{65}{18}\nu^2,$$
$$_\infty\mathcal{F}_5^{\text{log-fac}} = -\left(\frac{8191}{672} + \frac{583}{24}\nu\right)\pi,$$
$$_\infty\mathcal{F}_6^{\text{log-fac}} = \frac{6643739519}{69854400} + \frac{16}{3}\pi^2 - \frac{1712}{105}\gamma_E$$
$$+ \left(-\frac{134543}{7776} + \frac{41}{48}\pi^2\right)\nu - \frac{94403}{3024}\nu^2 - \frac{775}{324}\nu^3,$$
$$_\infty\mathcal{F}_7^{\text{log-fac}} = \left(-\frac{16285}{504} + \frac{214745}{1728}\nu + \frac{193385}{3024}\nu^2\right)\pi,$$
$$_\infty\mathcal{F}_8^{\text{log-fac}} = -\frac{319927174267}{3178375200} + \frac{232597}{4410}\gamma_E - \frac{1369}{126}\pi^2$$
$$+ \frac{39931}{294} \log 2 - \frac{47385}{1568} \log 3,$$

---

[4] When we define phenomenological EOB models, this quantity, $v_{\text{pole}}$, becomes a calibration parameter.
[5] All appearances of $\log(\cdot)$ stand for the natural logarithm.

$$_\infty\mathcal{F}_9^{\text{log-fac}} = \frac{265978667519}{745113600}\pi - \frac{6848}{105}\gamma_E - \frac{13696}{105}\pi\log 2,$$

$$_\infty\mathcal{F}_{10}^{\text{log-fac}} = -\frac{2489533931610883}{2831932303200} + \frac{916628467}{7858620}\gamma_E$$
$$- \frac{424223}{6804}\pi^2 - \frac{83217611}{1122660}\log 2 + \frac{47385}{196}\log 3,$$

$$_\infty\mathcal{F}_{11}^{\text{log-fac}} = \frac{8399309750401}{101708006400}\pi + \frac{177293}{1176}\gamma_E\pi$$
$$+ \frac{8521283}{17640}\pi\log 2 - \frac{142155}{784}\pi\log 3.$$

$$_\infty\mathcal{F}_{12}^{\text{log-fac}} = {}_{\text{cal}}\mathcal{F}_{12}^{\text{log-fac}}, \qquad _\infty\mathcal{F}_{13}^{\text{log-fac}} = {}_{\text{cal}}\mathcal{F}_{13}^{\text{log-fac}}$$

where $_{\text{cal}}\mathcal{F}_{12,13}^{\text{log-fac}}$ are calibration parameters. The terms $(_\infty\mathcal{F}_0,\ldots,{}_\infty\mathcal{F}_7)$ can be obtained using standard PN techniques (see eg. Eqs. (20)-(27) of Boyle et al., 2008), while the terms $(_\infty\mathcal{F}_8,{}_\infty\mathcal{F}_{11})$ can be obtained from a BH perturbation theory calculation ($\nu = 0$) in the PN limit ($v \ll 1$) (Tanaka et al., 1996). Note that unlike $(_\infty\mathcal{F}_0,\ldots,{}_\infty\mathcal{F}_7)$, only the $\nu$-independent part of $(_\infty\mathcal{F}_8,\ldots,{}_\infty\mathcal{F}_{11})$ are known.

One of the main enhancements we make to the EOB scheme is the inclusion of both GW emission to infinity and GW emission into the BH horizon [Eq. (3.1) in Tanaka et al., 1996 plus Eq. (12.31) in Mino et al., 1997]. For non-rotating BHs, these corrections amount to adding to the log-independent terms $_\infty\mathcal{F}_i^{\text{log-fac}}$ the following BH absorption terms $_{\text{abs}}\mathcal{F}_i^{\text{log-fac}}$:

$$_{\text{abs}}\mathcal{F}_8^{\text{log-fac}} = 1, \qquad _{\text{abs}}\mathcal{F}_9^{\text{log-fac}} = 0,$$
$$_{\text{abs}}\mathcal{F}_{10}^{\text{log-fac}} = 4, \qquad _{\text{abs}}\mathcal{F}_{11}^{\text{log-fac}} = 0, \qquad _{\text{abs}}\mathcal{F}_{12}^{\text{log-fac}} = \frac{172}{7}.$$

The energy flux is then given by Eq. **17** with $\mathcal{F}_i^{\text{log-fac}} = {}_\infty\mathcal{F}_i^{\text{log-fac}} + {}_{\text{abs}}\mathcal{F}_i^{\text{log-fac}}$. Notice that for non-rotating BHs, the BH absorption contribution starts at 4PN order, while if we were modelling rotating BHs, it would start at 2.5PN order (see eg. Eq. (12.31) in Mino et al., 1997), suggesting that in such cases their effect is much more dominant. Even for non-rotating BHs, however, we shall see in **Sec. 5** that not including the BH absorption terms in the flux can lead to large disagreements in the GW phase between EOB and Teukolsky waveforms after a two year integration. One could also enhance the energy flux through non-Keplerian corrections that allow for dynamical deviations from Kepler's law $v^2 = M/r$. However, we find that this makes no significant difference in the evolutions, provided one is interested only in GWs emitted outside the Schwarzschild ISCO (for a discussion of the test systems considered see **Sec. 5**).

With these considerations in mind, we are now in a position to define three of the EOB models that we compare in detail in **Sec. 5**. All of these three models employ the Padé-resummed flux of Eq. **16**, but they differ in the Padé matrix element employed and in the values of the calibration parameters $(v_{\text{pole}}, {}_{\text{cal}}\mathcal{F}_{12}^{\text{log-fac}}, {}_{\text{cal}}\mathcal{F}_{13}^{\text{log-fac}}, \ell_{12}^{\text{cal}}, \ell_{13}^{\text{cal}})$[6]:

- **Uncalibrated Padé model at** 5.5 **PN order**: uses the $F_4^7$ Padé element, $v_{\text{pole}} = 1/\sqrt{3}$, $({}_{\text{cal}}\mathcal{F}_{12}^{\text{log-fac}}, {}_{\text{cal}}\mathcal{F}_{13}^{\text{log-fac}}) = (0,0)$, and $(\ell_{12}^{\text{cal}}, \ell_{13}^{\text{cal}}) = (0,0)$.

- **Calibrated Padé model at** 6 **PN order**: uses the $F_6^6$ Padé element, $v_{\text{pole}} = 0.548247$, $({}_{\text{cal}}\mathcal{F}_{12}^{\text{log-fac}}, {}_{\text{cal}}\mathcal{F}_{13}^{\text{log-fac}}) = (2155.93, 0)$ and $(\ell_{12}^{\text{cal}}, \ell_{13}^{\text{cal}}) = (-202.055, 0)$.

- **Calibrated Padé model at** 6.5 **PN order**: uses the $F_6^7$ Padé element, and $v_{\text{pole}} = 0.548337$, $({}_{\text{cal}}\mathcal{F}_{12}^{\text{log-fac}}, {}_{\text{cal}}\mathcal{F}_{13}^{\text{log-fac}}) = (2311.17, -9905.66)$, and $(\ell_{12}^{\text{cal}}, \ell_{13}^{\text{cal}}) = (-155.542, -693.876)$.

As the name suggests, the uncalibrated model is fully determined without fitting to any numerical calculation ($v_{\text{pole}}$ is set to the value of the orbital velocity at the Schwarzschild light-ring), while the calibrated ones are actually *phenomenological* models, with calibration parameters $[(v_{\text{pole}}, {}_{\text{cal}}\mathcal{F}_{12}^{\text{log-fac}}, \ell_{12}^{\text{cal}})$ at 6PN order and $(v_{\text{pole}}, {}_{\text{cal}}\mathcal{F}_{12}^{\text{log-fac}}, {}_{\text{cal}}\mathcal{F}_{13}^{\text{log-fac}}, \ell_{12}^{\text{cal}}, \ell_{13}^{\text{cal}})$ at 6.5PN order] found by fitting the $\nu = 0$ limit of the Padé-resumed flux to a numerical flux computed in the Teukolsky approach (see **Sec. 5** for more details about the fitting). One can show that after Taylor expanding this Padé resummations, the 6.5PN calibrated model adds the following terms to the Taylor expansion of the flux: if $\dot{E}_{\text{GW}} = \sum_\ell \dot{E}_\ell^{\text{GW}} v^\ell$, then $\dot{E}_{12}^{\text{GW}} = 5152.53 - 1015.94\log(16v^2)$ and $\dot{E}_{13}^{\text{GW}} = -12053.6 + 774.703\log(16v^2)$. The matrix element chosen is such that no spurious poles appear in the flux function and so that the agreement with the Teukolsky flux is best[7].

Before proceeding, one should note that the expressions employed for the energy flux have been orbit-averaged so as to be expressible via resummed Taylor expansions of the velocity parameter. In general, the energy flux is a function of certain combination of time derivatives of multipole moments. For the case of circular orbits, one can orbit-average these expressions to obtain the analytic, closed-form quantities presented above, a procedure that is not possible in closed-form for eccentric orbits (Arun et al., 2007b). We see that this averaging does not seem to induce disagreement with Teukolsky waveforms for the quasi-circular inspirals considered in **Sec. 5**, although this need not necessarily hold for more generic orbits. Finally, as discussed in **Sec. 4.1**, we remind the reader that unless otherwise specified, we set relative $\nu$ terms initially to zero when comparing to Teukolsky waveforms, as the latter cannot account for $\nu$ corrections in their evolutions. We study the effect of these dissipative $\nu$-dependent terms in **Sec. 5.5**.

---

[6] In Yunes et al., 2009, only the first and third models were analysed. We here also consider an intermediate model (the second one) that adds only two calibration parameters, instead of four as in the third model.

[7] We could have used the Padé element $F_6^5$ instead of $F_4^7$ in the uncalibrated model, but the latter presents slightly better agreement in the high velocity region relative to the Teukolsky flux.

### 4.2.2 The $\rho_{\ell m}$ Model

Another class of EOB models that we consider consist of employing the recently proposed, factorised resummation for the energy flux (Damour and Nagar, 2007b and Damour et al., 2009), namely $\dot{E}^\rho_{\text{GW}} = F^\rho$ with

$$F^\rho = \frac{2}{16\pi} \sum_{\ell=2}^{\ell=8} \sum_{m=1}^{m=\ell} (m\Omega)^2 |R\, h_{\ell m}|^2, \qquad (18)$$

where $R$ is the distance to the observer and $h_{\ell m}$ is the harmonically decomposed GW metric perturbation.

Such an expression for the flux can be thought of as deriving from Isaacsson's GW effective stress-energy tensor in the short-wavelength approximation,

$$T^{\text{GW}}_{\mu\nu} = \frac{1}{32\pi} \left\langle \left(\partial_\mu h^{ij}_{\text{TT}}\right) \left(\partial_\nu h^{\text{TT}}_{ij}\right) \right\rangle, \qquad (19)$$

where the angle-brackets stand for averaging over several GW wavelengths and *TT* stands for the transverse-traceless projection (Isaacson, 1968a, 1968b and Misner et al., 1973). The energy flux carried out to infinity $\dot{E}^{\text{GW}}_\infty$ can then be obtained from the integral of the ($t\,i$)-components of this effective stress-energy tensor:

$$\dot{E}^{\text{GW}}_\infty = -\lim_{R\to\infty} R^2 \int_{S^2_\infty} dS\, T^{\text{GW}}_{ti} n^i, \qquad (20)$$

where $S_\infty$ is a 2-surface at spatial infinity and $n^i$ is the space-like normal to $S_\infty$. Using the harmonic gauge condition, one can rewrite this equation entirely in terms of time derivatives of the metric perturbation

$$\dot{E}^{\text{GW}}_\infty = \frac{1}{32\pi} \lim_{R\to\infty} R^2 \int_{S^2_\infty} dS\, \left\langle \dot{h}^{ij}_{\text{TT}} \dot{h}^{\text{TT}}_{ij} \right\rangle n^i, \qquad (21)$$

One then assumes that the waveform has a dependence of the form $\exp(-i\,m\,\Omega\,t)$, with $\Omega$ time-independent, to obtain Eq. **18**, after using symmetry arguments to remove the sum over negative $m$ harmonic number.

We see then that the energy flux at infinity in the $\rho_{\ell m}$-model is controlled exclusively by the waveform, which we define in **Sec. 4.3**. As such, this scheme is perhaps more self-consistent that the Padé one, as no orbit averaging is explicitly assumed. However, as the expressions for the GW luminosity are computed formally at spatial infinity, it is unclear how one would account for BH absorption. Although it is clear that such contributions are subdominant for some orbits (for circular orbits in Schwarzschild, BH absorption enters at $O(v^8)$), for generic orbits BH absorption could be much more important (for circular orbits in Kerr, BH absorption enters at $O(v^5)$).

Without specifying the functional form of $h^{\ell m}$ first, we cannot define the three additional models of $\rho_{\ell m}$ class that we investigate. We thus discuss the waveforms next.

### 4.3 Waveform

We employ a product-decomposed, factorised-resummation of the waveform (Damour and Nagar, 2007b and Damour et al., 2009). Let us then define the harmonically-decomposed waveform via

$$h^{\ell m}_{(k)} = h^{\ell m}_N \, _{pd}h^{\ell m}_{(k)}. \tag{22}$$

The quantity $h^{\ell m}_N$ is the Newtonian (leading-order) expression for the waveform (Damour et al., 2009)

$$h^{\ell m}_N := \frac{M\nu}{R} n^{(\epsilon)}_{\ell m} c_{\ell+\epsilon}(\nu) (M\Omega)^{(\ell+\epsilon)/3} Y^{\ell-\epsilon,-m}\left(\frac{\pi}{2}, \Phi\right), \tag{23}$$

where $\epsilon$ is the parity of the multipolar waveforms ($\epsilon = 0$ when $\ell + m$ is even or $\epsilon = 1$ for $\ell + m$ os odd), $Y^{\ell,m}$ are scalar spherical harmonics, while $n_{\ell m}$ and $c_{\ell+\epsilon}$ are $\nu$-independent and $\nu$-dependent coefficients respectively (see eg. Eqs. (5)-(7) of Damour et al., 2009). The quantity $_{pd}h^{\ell m}_{(k)}$ is a product-decomposed resummation of the PN corrections (Damour et al., 2009)

$$_{pd}h^{\ell m}_{(k)} = S_{\text{eff}} \, T_{\ell m} \, e^{i\delta_{\ell m}} \left(\rho^{(k)}_{\ell m}\right)^\ell. \tag{24}$$

where $k$ is the order of the PN approximation, $S_{\text{eff}} = H_{\text{eff}}/\mu$ if $\ell + m$ is even or $S_{\text{eff}} = \Omega^{1/3} p_\Phi$ if $\ell + m$ is odd, $T_{\ell m}$ is a resummed tail term (see eg. Eq. (19) in Damour et al., 2009), $\delta_{\ell m}$ is a phase correction (see eg. Eqs. (20)-(29) in Damour et al., 2009, where we here choose $\bar{y} = \Omega^{2/3}$), and $\rho_{\ell m} = \sum_k c^{\rho_{\ell m}}_k (M\Omega)^{2k/3}$ is a Taylor-expanded amplitude correction (see eg. Appendix C of Damour et al., 2009).

We can now define three models of $\rho_{\ell m}$ class[8]:

- ***Uncalibrated*** $\rho_{\ell m}$ ***model at*** 5 ***PN order***: uses $k = 5$.

- ***Calibrated*** $\rho_{\ell m}$ ***model at*** 6 ***PN order only***: uses $k = 6$ with the calibrated 6 PN term $c^{\rho_{22}}_6 = 152.7 - 35.61 \, \text{eulerlog}_2(v^2)$ in $\rho_{22}$[9].

- ***Calibrated*** $\rho_{\ell m}$ ***model at*** 6 ***and*** 5 ***PN order***: uses $k = 6$ with the calibrated 6 PN term $c^{\rho_{22}}_6 = 640.6 - 453.7 \, \text{eulerlog}_2(v^2)$ in $\rho_{22}$ and calibrated 5 PN term $c^{\rho_{33}}_5 = -918.3 + 679.7 \, \text{eulerlog}_2(v^2)$ in $\rho_{33}$.

---

[8] In Yunes et al., 2009, only the first and third models were analysed, where the latter was referred to as *calibrated $\rho_{\ell m}$-model at 6PN order*. We here refer to that model as a *calibrated one at 6 and 5 PN order*. We additionally consider an intermediate one that adds only 2 calibration parameters and we refer to it as *calibrated $\rho_{\ell m}$-model at 6PN order only*.

[9] The eulerlog is a function of the Euler number, defined eg. in Eq. (36) of Damour et al., 2009.

As before, the uncalibrated model is fully determined without fitting it to any numerical result, while the calibrated model is actually a phenomenological one fully specified once one fits the EOB $\rho_{\ell m}$ flux in the $\nu = 0$ limit to a numerically-computed Teukolsky flux to find the calibration parameters $c_6^{\rho_{22}}$ and $c_5^{\rho_{33}}$ (see **Sec. 5** for more details about the fitting). In the calibrated model, all $c_6^{\rho>2,m} = 0$ and $c_5^{\rho>3,m} = 0$. In both models, the sum in Eq. **18** is taken up to $\ell = 8$, since this is the highest multipole order known in the PN approximation (including the point-particle, BH perturbation theory terms, but not the BH absorption ones).

The different Padé and $\rho_{\ell m}$ models differ only in the way the radiation-reaction is computed. As in the case of the Padé models, we set all relative $\nu$ terms initially to zero when comparing to Teukolsky waveforms in **Sec. 5**. Later on in **Sec. 5.5** we study the effect of these $\nu$-dependent terms.

Finally, when we compare EOB and Teukolsky amplitudes and phases, we need to extract an EOB phase and an amplitude from some EOB waveform. We here always employ the uncalibrated $\rho_{\ell m}$ model to compute the waveform, from which the EOB phase and amplitude are extracted.

### 4.4 Initial Data

The evolution of the Hamilton-Jacobi equations (Eqs. **14**- **15**) requires initial data. The initial phase can always be initialised to zero, $\Phi_0 = 0$, while the initial radius $r_0$ is chosen according to the systems that are to be evolved (see **Sec. 5.1**). In the post-circular initial data setup (Buonanno and Damour, 2000 and Damour et al., 2003), the conjugate momenta are chosen via (Damour et al., 2003)

$$p_0^\Phi = \left.\sqrt{-\frac{A'}{u(2A + uA')}}\right|_{u=u_0, \Phi=\Phi_0}, \quad (25)$$

$$p_0^{r^*} = \left[\frac{u(2A + uA')A'D}{A[2uA'^2 + AA' - uAA'']}\frac{\dot{E}_{\text{GW}}}{\omega_0}\right]_{u=u_0, \Phi=\Phi_0}, \quad (26)$$

where $u = r^{-1}$, the primes stands for differentiation with respect to $u$ and $\Omega_0$ is the orbital frequency associated with the Hamilton-Jacobi equations:

$$\Omega_0 = u^{3/2} \left.\sqrt{\frac{-A'/2}{1 + 2\nu\left(\frac{A}{\sqrt{A+uA'/2}} - 1\right)}}\right|_{u=u_0, \Phi=\Phi_0}. \quad (27)$$

Technically, Eq. **26** is actually appropriate only for $p_0^r$ (not $p_0^{r^*}$), but these two momenta are asymptotic to each other when $u \ll 1$. Note that $\dot{E}_{\text{GW}}$ depends on the effective Hamiltonian in the $\rho_{\ell m}$ models, which in turn depends on $(p_{r^*}, p_\Phi)$. When initialising the $\rho_{\ell m}$ models, we set $(p_{r^*}, p_\Phi)$ in the right-hand sides of Eqs. **25**-**26** to their lower-order values: $p_{r^*}^{(c)} = 0$ and $p_\Phi^{(c)}$ given in Eq. **25**.

One could initialise evolutions with these conditions, but we want to avoid introducing error due to improper choices of initial data. Recall that the above initial conditions are approximately valid only when $u \ll 1$, while some of the evolutions we consider in **Sec. 5.1** start at $u_0 \sim 0.1$, which we have found could introduce error in the waveforms. One way to fix this problem is to use more accurate initial data, as that prescribed in the post-post-circular scheme (Damour and Nagar, 2008 and Damour et al., 2008a), but since these data is also approximate it might also lead to some error in the waveforms for small initial separations. Instead, we here chose to perform a *mock* evolution, ie. an evolution whose sole purpose is to provide accurate initial data for the true evolution. We perform a fake evolution with initial separation $r_0 = 100$ and initial momenta described above, and then read-off from this evolution initial data at some new $r_0$ appropriate for the systems defined in **Sec. 5**. The mock evolution provides highly accurate initial conditions beyond the post-post-circular approximation, limited only by the numerical accuracy to which the Hamilton-Jacobi equations are solved.

**4.5 High-Order EOB Corrections**

The EOB framework described above employs quantities known up to 3 PN order ($\nu$-dependent) and up to 5.5 PN order ($\nu$-independent), which are later resummed in different ways. This resummation naturally introduces an infinite number of analytically-unknown, higher-order PN corrections. If the EOB expressions were re-expanded and one knew what this higher-order terms were, one could verify whether the higher-order terms introduced via resummation agree or disagree with the expected, analytical value.

In the absence of such analytic results, one can instead compare the EOB model with numerical simulations, which have been performed for nearly equal-mass binaries (Boyle et al., 2008, Buonanno et al., 2009 and Damour and Nagar, 2009). These comparisons revealed a certain, small disagreement in the phase and amplitude of the EOB GWs relative to the numerical results. Presumably, this disagreement is due to the uncontrolled, higher-order terms introduced via resummation.

Such uncontrolled, higher-order terms, however, can be corrected by analytically introducing calibration parameters at higher than 3PN terms, which are found by fitting the waveforms to numerical simulations. One such correction involves modifying the Hamiltonian, where one adds a certain $\nu$-dependent, 4 PN correction ($a_5$) to the $A$ metric function. Another such correction involves the radiation-reaction force, where one adds a phenomenological, *radial* radiation-reaction force component and also non-Keplerian deviations to Kepler's third law. We do not include any of such corrections here, as we find they have no visible effect in the extreme-mass ratio orbits we considered, particularly as all evolutions are stopped outside the Schwarzschild ISCO.

### 4.6 Extraction of Observables, Systematic Errors and Properties of the EOB model

The Hamilton-Jacobi equations (Eq. **15**) are solved with the initial data of **Sec. 4.4** and via the `NDSolve` command of `Mathematica`, which employs an adaptive scheme that guarantees an accuracy determined by two flags: `AccuracyGoal` and `PrecisionGoal`. The higher these flags are set to, the longer it takes to solve the system of differential equations, but the more accurate the result. We have found that setting these flags to 11 guarantees an accuracy of better than $\delta\Phi_{22} \lesssim 0.03$ radians in the dominant mode of the waveform's phase and $\delta h_{22}/h_{22} \lesssim 10^{-7}$ in the dominant mode of the waveform's normalised amplitude after a two year evolution. Other sources of error are due to initial data (accuracy of the numerical mock evolution, initial mock radius, interpolation to obtain data at a particular radius), but are always smaller than the dominant numerical error in the `NDSolve` routine. We expect that if the EOB scheme were to be exported to the `C` programming language, some of these error could be minimised and the speed of its execution could be increased.

Once the Hamilton-Jacobi equations have been solved, one obtains interpolating functions for the radius, orbital phase and conjugate momenta as a function of time, which can be used to evaluate the uncalibrated, product-decomposed waveform as a function of time. This waveform is then used to extract the GW phase and amplitude. The latter is simply the absolute magnitude of the waveform, $A_{GW,\ell m}^{EOB} = |h_{\ell m}|$, while the former is obtained by solving

$$\dot{\phi}_{GW,\ell m}^{EOB} = -\frac{1}{m} \text{Im}\left(\frac{\dot{h}_{\ell m}}{h_{\ell m}}\right), \tag{28}$$

where Im stands for the imaginary part operator. The solution to this differential equation is also obtained via the `NDSolve` command of `Mathematica`, this time with `AccuracyGoal` and `PrecisionGoal` flags set to 13, which guarantees numerical error below that of the numerical solution of the Hamilton-Jacobi equations.

Let us now discuss the type of approximation the EOB scheme belongs to in the classification discussed in **Sec. 2.2**. In the $\nu = 0$ limit, both the Padé and the $\rho_{\ell m}$ models are dissipative approximations, that is, approximations that neglect conservative corrections ($\nu$-dependent terms) in the background (or Hamiltonian). Moreover, the Padé scheme employs orbit-averaged expressions to model radiation-reaction, and as such, it is a secular approximation. The $\rho_{\ell m}$ model, on the other hand, employs a mixture of orbit-averaged terms (those in $\rho_{\ell m}$) and non orbit-averaged terms (those in $S_{\ell m}$ and $T_{\ell m}$ due to the presence of the real Hamiltonian). As such, the $\rho_{\ell m}$ model is somewhere in between a secular and radiative approximation.

One of the advantages of the EOB scheme, however, is that one can easily turn on the $\nu$ terms in the Hamiltonian to model (in a PN sense) conservative corrections to the equations of motion. When these terms are turned on, the EOB scheme goes beyond a secular or radiative approximation, for the first time accounting consistently for conservative deformations of the background geometry due to the small body. We denote these terms *PN conservative self-force* ones, and we investigate their effect in the Padé and $\rho_{\ell m}$ models in **Sec. 5.5**.

In addition to $\nu$ terms in the Hamiltonian, one can also turn on relative $\nu$ terms in the radiation-reaction force (ie. in the energy flux), thus accounting for second-order and third-order corrections to the dissipative dynamics, albeit in a PN sense. We denote such terms *PN dissipative self-force* terms. Of course, in the $\rho_{\ell m}$ model there is not a clean separation of the conservative and dissipative self-force terms, as the Hamiltonian enters directly in the radiation-reaction force, through the source terms $S_{\ell m}$ in the waveform.

One could study the effect of $\nu^n$ terms in the PN conservative or dissipative self-force contributions to the waveform as a function of $n$, but we do not attempt to do so here. In fact, one should notice that due to the resummation of the Hamiltonian, when $\nu$ terms are turned on, one is actually including an infinite number of $\nu$ powers, although terms higher than 3PN order are not formally valid.

## 5 Comparison of Approaches

In this section we compare waveforms computed in the EOB scheme described in **Sec. 4** versus waveforms computed in the Teukolsky approach described in **Sec. 3.2**. We begin by describing the EMRI systems we consider and how the comparisons are performed. We then proceed to compare the fluxes and the waveforms.

Before beginning with the comparisons, it is worthwhile to briefly think about what quantities to compare. One can always construct and compare coordinate and gauge invariant quantities, but just because these are invariants does not necessarily mean they are observable, in a *practical* sense. For this reason, we here concentrate on observables that affect the GW response function and thus can be measured by planned GW detectors in the near future.

An example of such an observable is the ISCO location and its GW imprint. The Teukolsky and EOB (in the $\nu = 0$ limit) approaches, naturally agree on the ISCO location by construction, as they both evolve geodesics on a fixed Schwarzschild background. Another example is to compare directly the GW phase and amplitude, as these are the ready-to-use quantities of most interest to GW astronomy. We concentrate on these in the remainder of this section.

### 5.1 Systems for Comparison

We consider two fiducial EMRI systems that sample different regions of the LISA sensitivity curve:

- **system-I**: $(m_1, m_2) = (10^5, 10) M_\odot$.

- **system-II**: $(m_1, m_2) = (10^6, 10) M_\odot$.

Clearly, the reduced mass for these systems is $q = (10^{-4}, 10^{-5})$ respectively, the symmetric mass ratio is $\nu = 9.9998 \times (10^{-5}, 10^{-6})$ respectively, and the chirp mass is $\mathcal{M} := \mu^{3/5} M^{2/5} \approx (400, 1000) M_\odot$ respectively. Lower or higher total mass systems are not considered because they would either reach the ISCO outside the LISA optimal sensitivity band (LISA's noise raises sharply above $\sim 10^{-2}$ Hz), or fall inside the white-dwarf confusion limit (below $\sim 0.003$ Hz (Farmer and Phinney, 2003)).

As for the initial conditions, we choose system-II ($q = 10^{-5}$) to begin at an initial separation of $r_{\text{in}} \simeq 10.6 M$ and terminate at the ISCO, sweeping GW frequencies in $f_{\text{GW}} \in [1.8 \times 10^{-3}, 4.4 \times 10^{-3}]$ Hz. Beginning the evolution at this initial separation guarantees that system-II is in band for a period of 2 years. As for system-I ($q = 10^{-4}$), we choose an initial separation of $r_{\text{in}} \simeq 29.34 M$ and we terminate the evolutions at $r_{\text{fin}} \simeq 16.1 M$. These initial and final separations are chosen such that system-II sweeps frequencies in $f_{\text{GW}} \in [4 \times 10^{-3}, 10^{-2}]$ Hz in a period of two years. With these choices, we can study EMRI dynamics and GWs, while guaranteeing the GW signal remains in the most sensitive region of the LISA band.

We terminate the evolution of system-II ($q = 10^{-5}$) at the ISCO for two reasons. First, the Teukolsky waveforms we compare against are computed with a frequency domain code, which fails inside of the ISCO as there are no stable circular orbits there (computing Teukolsky waveforms with a time-domain code could remedy this). Second, there are very few cycles between the Schwarzschild ISCO and the Schwarzschild light-ring for systems-II after a two year evolution. More precisely, for system-II there are only $\simeq 16$ GW cycles between the ISCO and the light-ring, out of $\simeq 1.5 \times 10^5$ GW cycles between the initial separation and the ISCO. System-I reaches the ISCO outside the most sensitive region of the LISA frequency band, which is why evolutions for this system are stopped at $r_{\text{fin}} \simeq 16.1 M$.

### 5.2 Flux Comparisons

Figure **1** compares the absolute value of the difference between the Teukolsky and EOB (*uncalibrated* and *calibrated*) fluxes, normalised by the Newtonian, leading order flux ($F_{\text{Newt}} = 32 \nu^2 v^{10} / 5$), as a function of orbital velocity. The solid blue line corresponds to using the Taylor-expanded flux (ie. Eqs. (19)-(27) in Boyle et al., 2008), which leads to a residual at least two orders of magnitude larger than when one uses the EOB scheme. Such a large residual leads to large accumulated disagreement in the waveform, which is consistent with the results of Poisson, 1995, Simone et al., 1995, 1997, Blanchet, 2002, Yunes and Berti, 2008 and Mandel and Gair, 2009.

Figure **1** also allows for comparisons between EOB models. Observe that the uncalibrated EOB models (solid lines) do equally well in the high-velocity region, although the Padé one performs better at low velocities. Also observe that the calibrated models (dashed and dotted lines) perform better than the uncalibrated ones,

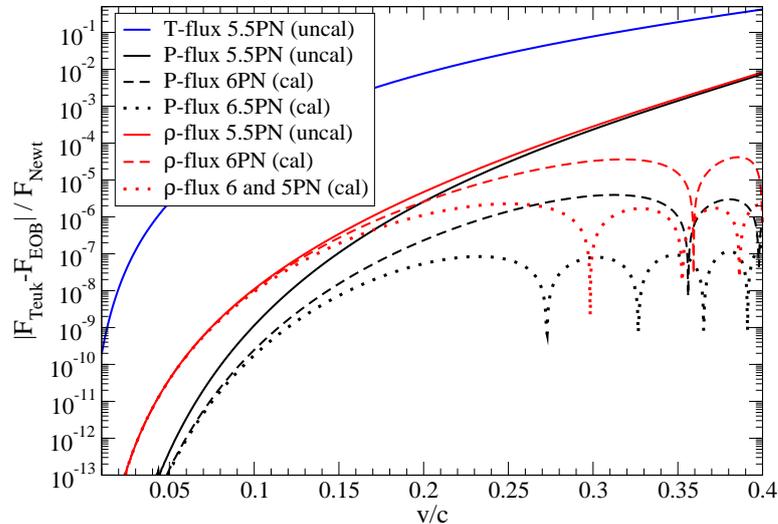

**Fig. 1** Absolute value of the difference in the Newtonian normalised Teukolsky and EOB fluxes as a function of orbital velocity. Calibrating the Padé or $\rho_{\ell m}$-flux improves the agreement by orders of magnitude at the orbital speeds of BHs close to the ISCO.

a result to be expected by construction. Interestingly, the calibrated Padé-models perform consistently better than the calibrated $\rho_{\ell m}$-models, presumably because the former include BH absorption corrections that the latter neglect. For velocities $v \in [0.01, 0.1]$ (not shown in the figure) the agreement is generically better than $10^{-8}$ with a best agreement of $10^{-13}$ near $v = 0.01$ for all models.

Before proceeding, we should note that flux comparisons have already been performed in the literature in the extreme mass-ratio limit. Notably, Yunes and Berti, 2008 compared a Teukolsky flux to a Taylor expanded flux, while Damour and Nagar, 2008, 2007b compared the former to an EOB resummed flux. Both of these studies, however, employed a Teukolsky flux only accurate to one part in $10^6$. The flux employed here is accurate to at least one part in $10^{10}$ in the entire velocity range, with accuracies on the order of one part in $10^{13}$ for $v < 0.1$. Moreover, the EOB fluxes employed here include the effect of BH GW absorption in the Padé models, an effect not accounted for in the past.

**5.3 Strategy for Dynamical Comparisons**

Before any formal comparison can be made, one must make sure the comparison is fair. Whenever comparing waveforms, there is always a time offset and phase offset that can be tuned to maximise the overlap. Physically, one can think of these offsets as the time of coalescence and phase at time of coalescence, two extrinsic parameters that are maximised over when performing matched filtering. In the comparisons we study below, we present results after such a maximisation has been carried out.

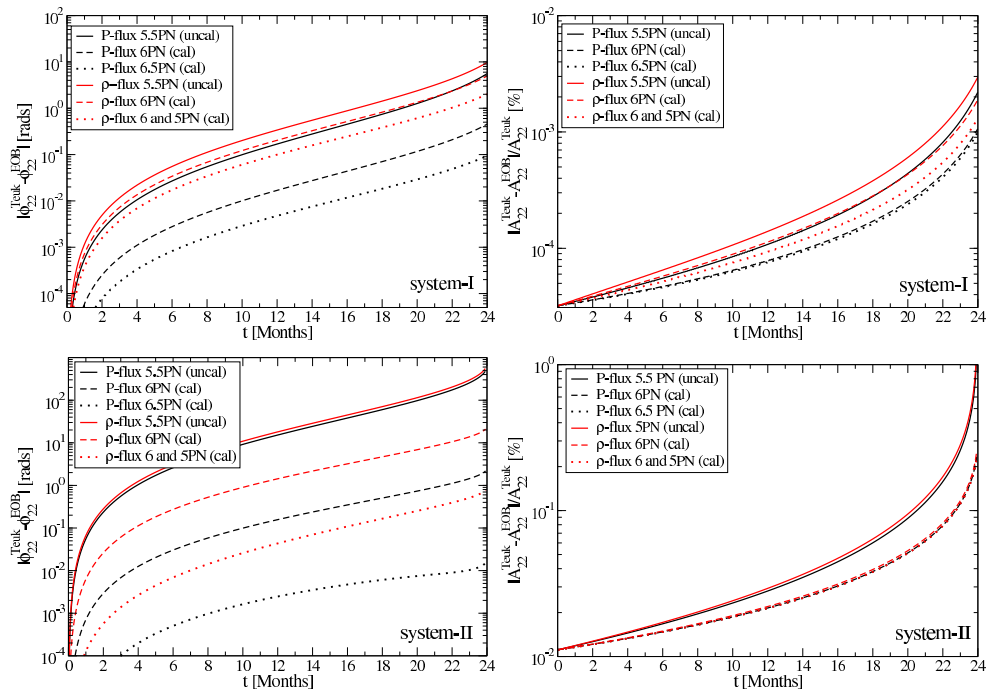

**Fig. 2** Absolute value of the dephasing (left) and fractional amplitude difference (right) of the dominant GW $(2,2)$ mode as a function of time in months. The solid curves use the uncalibrated EOB models, the dashed curves use the calibrated models at 6PN order only, and the dotted curves use the calibrated models at either and 6.5PN order or at 6 and 5 PN order. Red curves use the Padé models and black curves the $\rho_{\ell m}$ ones. As before, the introduction of calibrated higher-order terms, reduces the differences even over a full two year coherent integration. Top panels refer to system-I and bottom panels to system-II.

Maximising over the time and phase offsets is equivalent to minimising the following integral

$$\Theta(\delta t, \delta \phi) = \int_{t_1}^{t_2} \left[\phi_{Teuk}(t) - \phi_{EOB}(t - \delta t) - \delta\phi\right]^2 dt. \qquad (29)$$

over $\delta t$ and $\delta \phi$ simultaneously, which is equivalent to maximising the fitting factor over time and phase of coalescence in a matched filtering calculation with white noise (Buonanno et al., 2009). Because of the particular behaviour of the phase function (a monotonically increasing function of time), one can implement this minimisation in two steps. First, one minimises

$$\Theta_1(\delta t) = \int_{t_1}^{t_2} \left[\dot{\phi}_{Teuk}(t) - \dot{\phi}_{EOB}(t - \delta t)\right]^2 dt. \tag{30}$$

over $\delta t$ to find $\delta t_{best}$, which corresponds to a vertical alignment of the waveforms. Second, one minimises

$$\Theta_2(\delta \phi) = \int_{t_1}^{t_2} \left[\dot{\phi}_{Teuk}(t) - \dot{\phi}_{EOB}(t - \delta t_{best}) - \delta \phi\right]^2 dt \tag{31}$$

over $\delta\phi$ to find $\delta\phi_{best}$, which corresponds to a horizontal alignment of the waveforms. Such alignments are performed until the waveforms agree within the alignment window $\Delta t := t_2 - t_1$ to a level of $10^{-10}$ [$10^{-6}$] radians in the phase for systems-I [system-II].

This minimisation routine depends sensitively and by construction on the alignment window. Of course, one obtains different results if one performs matched filtering over a small window in the time domain versus over an entire two-year window. Here, we are interested in studying the accumulated difference between EOB and Teukolsky waveforms, and therefore, we must perform the alignment in the early stages of the evolution. We have studied how the phase difference behaves as a function of this window, with the parametrisation $\Delta t = 2^n \lambda_{GW}$, where $\lambda_{GW} \sim \pi/\dot{\Phi}$ and $\dot{\Phi}$ is the orbital frequency[10]. We find that when $n < 9$, the final phase and normalised amplitude difference between EOB and Teukolsky waveforms is less than $10^{-3}$ radians and $10^{-6}$ respectively. When $\Delta t \simeq 2$ years, we find the disagreement between waveforms becomes slightly smaller but is now spread out over the entire observation window (as opposed to presenting a monotonically increasing behaviour).

Henceforth, we set $n = 6$, corresponding to $\Delta t =\simeq 0.006$ [$\Delta t \simeq 0.013$] months for system-I [system-II]. All comparisons performed below assume the alignment has already been carried out.

**5.4 Waveform Comparisons**

We now compare the phase and amplitude of EOB and Teukolsky-based waveforms. The left panels of Fig. 2 plot the absolute value of the (dominant) GW phase difference (or dephasing) in radians as a function of time in months. For reference (not shown in the figure) the dephasing when employing a Taylor expanded flux can be up to $\sim 40$ (3000) radians for system-I (system-II) after two-years of observation, a result in qualitative agreement with Mandel and Gair, 2009. On the other hand, with the uncalibrated Padé or $\rho_{\ell m}$-model, the dephasing goes down by a factor of six, roughly to $\sim$ 5-10 (530) radians for system-I (system-II). With the calibrated Padé models, the dephasing can be reduced to ∼ 0.5 (2) radians with 6PN calibration and ∼ 0.1 (0.02) radians with 6.5PN calibration for system-I (system-II) after two-years. With the calibrated $\rho_{\ell m}$ models, the dephasing is not reduced quite as much; it goes down to ∼ 5 (20) radians if we calibrate the 6PN term only and to ∼ 2 (0.8) radians if we calibrate both the 6 and 5 PN terms for system-I (system-II).

---

[10] Formally, one should really use the GW frequency, instead of the orbital frequency to define the GW wavelength, but we find these quantities are almost identical, and the orbital definition suffices for this parametrisation.

Could a GW detector differentiate between Teukolsky and EOB waveforms? A measure for this was estimated in Lindblom et al., 2008: less than 0.07 ($10^{-4}$) radians of dephasing is necessary for detection (parameters estimation) assuming an SNR of order a thousand. After a 1 year observation, the Padé model calibrated at 6.5PN order deviates from the Teukolsky one at a level of ≤ 0.004 radians for both systems, which suggests they are sufficiently similar that LISA would only distinguish them via a parameter estimation study given a sufficiently high SNR. On the other hand, LISA should be able to easily discern between the $\rho_{\ell m}$-models and the Teukolsky ones, (even the most calibrated ones we study), because the dephasing in this case is at least ten times larger even after one year (∼ 0.05 radians). The reason for the Padé model agreeing better is probably a symptom of the importance of the inclusion of BH GW absorption terms in the radiation-reaction force. Of course, if one wishes to employ any of these templates for parameter estimation, then a much shorter observation time could be employed. For example, the Padé model at 6.5PN order deviates from Teukolsky waveforms by less then $10^{-4}$ within the first two (three) months of observation for system-I (system-II).

The right panels of Fig. 2 show the absolute value of the normalised difference in the (dominant) GW amplitude. As before, the calibrated models do better than the uncalibrated ones, but this time the improvement is not as dramatic. The uncalibrated models lead to amplitude differences of approximately $3 \times 10^{-5}$ ($10^{-3}$) for system-I (system-II) after two years, while the calibrated ones improve this agreement to ∼ $10^{-5}$ (∼ $2 \times 10^{-4}$). The agreement presented here is remarkable when one realises that in a 2-year observation window, the system evolves for over ∼ $2 \times 10^6$ (∼ $9 \times 10^5$) radians.

One might be tempted to compare a fixed model across systems, but we must warn the reader that such a comparison is highly misleading. This is because systems-I and II truly sample different regimes of the EMRI phase space. For example, system-I always remains in a more weakly-gravitating region that system-II, as the former's evolution is terminated at ∼ 16$M$, while the latter's ends at the ISCO. But although system-I is less relativistic in this sense (its orbital velocity is smaller than system-II's), the mass ratio of system-I is more extreme, and thus radiation-reaction is weaker. In turn, this implies that system-I evolves over a longer time period than system-II, thus yielding more GW cycles.

We have also investigated subdominant modes of the GW function. Higher-harmonics present essentially the same behaviour, with dephasing of ∼ 0.14 (0.07) and ∼ 0.18 (0.09) radians, and normalised amplitude differences of ∼ $6 \times 10^{-5}$ $(4 \times 10^{-3})$ and $\sim 3 \times 10^{-4}$ $(9 \times 10^{-3})$, for system-I (system-II) and harmonics $(\ell, m) =$ $(3, 3)$ and $(\ell, m) = (4, 4)$ respectively. The similarities between these differences and those found for the dominant mode are rooted in the similarities between the GW and orbital phases. In fact, we find that these two quantities agree to $\sim 1$ radian over a 2-year integration for both systems-I and II. This implies that comparisons of the GW phase are nothing but comparisons of $m$ times the orbital phase, and thus, all dephasings are really controlled by the trajectories of the small body.

One might wonder whether it is necessary to sum over all modes before comparing the waveforms. The answer to this question depends on the SNR contained in higher harmonics for the systems considered. We calculated a measure of the SNR, namely

$$\rho^2 = 4 \int_{f_l}^{f_h} \frac{|\tilde{h}_+|^2 + |\tilde{h}_\times|^2}{S_n(f)} df, \qquad (32)$$

where $\tilde{h}_{+,\times}$ stands for the Fourier transform of the plus or cross GW polarisation and $S_n(f)$ is LISA's spectral noise density curve, including white-dwarf confusion noise (Barack and Cutler, 2004). For simplicity, we have not included here the beam-pattern functions, so the $\rho^2$ really overestimates the true SNR, as it assumes the detectors sees 100% of the power in both polarisations. Nonetheless, as we are interested in ratios of $\rho^2$, this simplification does not matter. The limits of integration $f_l$ and $f_h$ are set to the initial and final frequencies of the EMRI system under consideration (see **Sec. 5.1**).

With this measure in mind, we find that including up to the $\ell = 5$ ($\ell = 7$) harmonic for system-I (system-II) guarantees a recovery of 97% of the total SNR, with the $\ell = m$ modes the most dominant. Table **1** shows the fraction of the power contained in individual $(\ell, m)$ modes for system-A, $F_{sA}^{\ell m}$, with the sum of the squares equal to unity.

The fractional SNR recovered after summing over all $m$ harmonics is then 97.42% ($\ell = 2$), 21.91% ($\ell = 3$), 5.219% ($\ell = 4$), 1.271% ($\ell = 4$) for system-I, and 91.11% ($\ell = 2$), 36.90% ($\ell = 3$), 16.34% ($\ell = 4$), 7.420% ($\ell = 4$) for system-II. For more generic, for example eccentric orbits (see eg. (Yunes et al., 2009)), it is known that one must include many more harmonics in the waveform, than those considered here, in which case it might be more appropriate to study the GW phase and amplitude constructed from the full, harmonically recomposed waveform.

**5.5 Inclusion of PN Self-Force Terms**

We now consider the effect of the inclusion of $\nu$-dependent terms in the GW phase and amplitude. Of course, the comparison to Teukolsky waveforms is then slightly unfair as the latter does not include such effects. Therefore, the inclusion of these terms gives us a sense of the modification such self-force contributions could have on the waveform and LISA data analysis.

| $\ell$ | $m$ | $F_{sI}^{\ell m}$ | $F_{sII}^{\ell m}$ | $\ell$ | $m$ | $F_{sI}^{\ell m}$ | $F_{sII}^{\ell m}$ |
|---|---|---|---|---|---|---|---|
| 2 | 1 | 6.063% | 5.467% | 5 | 1 | 0.0002% | 0.0002% |
| 2 | 2 | 97.42% | 9.095% | 5 | 2 | 0.005% | 0.011% |
| 3 | 1 | 0.335% | 0.283% | 5 | 3 | 0.079% | 0.238% |
| 3 | 2 | 2.026% | 3.411% | 5 | 4 | 0.133% | 0.844% |
| 3 | 3 | 21.81% | 36.74% | 5 | 5 | 1.261% | 7.368% |
| 4 | 1 | 0.005% | 0.005% | 6 | 1 | 0.000003% | 0.000003 % |
| 4 | 2 | 2.191% | 0.340% | 6 | 2 | 0.0003% | 0.00008% |
| 4 | 3 | 0.530% | 1.744% | 6 | 3 | 0.0022% | 0.010% |
| 4 | 4 | 5.187% | 16.24% | 6 | 4 | 0.0024% | 0.141% |
| | | | | 6 | 5 | 0.0034% | 0.399% |
| | | | | 6 | 6 | 0.3137% | 3.336% |

**Table 1** Fractional SNR contained in individual $(\ell, m)$ modes for system-A, $F_{sA}^{\ell m}$. The sum of the squares of these fractions equals unity. Notice that the $\ell = m$ terms are dominant and that including up to the $\ell = 5$ ($\ell = 7$) harmonic for system-I (system-II) guarantees a recovery of 97% of the total SNR.

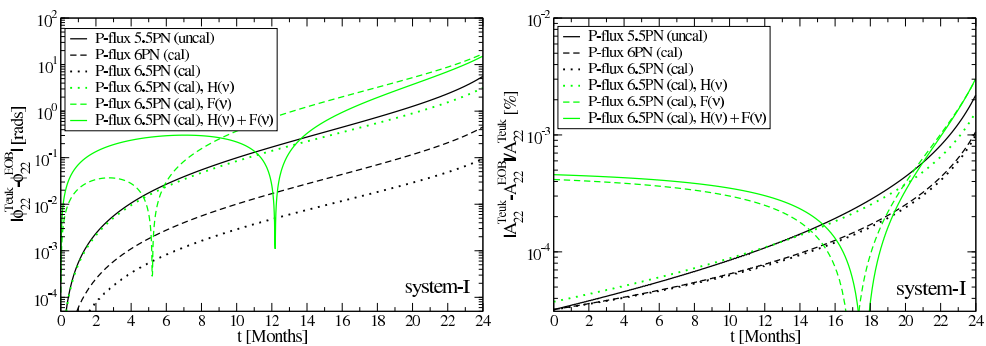

**Fig. 3** Same as Fig. 2 but with the inclusion of self-force, $\nu$-dependent terms in either the conservative Hamiltonian only (dotted green), the radiation-reaction force and energy flux only (dashed green), or in both (solid green). Observe that the dissipative PN self-force terms lead to the largest changes in the GW phase.

Figure 3 shows the dephasing and normalised amplitude difference as a function of time for the Padé models with and without PN self-force terms for system-I (similar results are obtained for system-II). We choose here to include PN self-force terms in the Padé model, instead of the $\rho_{\ell m}$ as done in (Yunes et al., 2009), since the former allows for a clean split between PN conservative (Hamiltonian-dependent) and dissipative (radiation-reaction force-dependent) effects. Observe that the PN dissipative self-force terms dominate for the type of orbits we considered (possibly this will be different for eccentric orbits), a result consistent with Pound and Poisson, 2008 for circular orbits. Also observe that self-force contributions can lead to large errors, almost $10^3$ times larger than those obtained when not including them. Finally, observe that the PN dissipative self-force correction builds up very fast, affecting waveforms substantially only after one month of evolution.

Of course, $\nu$-dependent corrections to the PN conservative and dissipative dynamics are expected to lead to modifications in the waveform relative to Teukolsky based ones. The question then becomes whether these $\nu$-corrections truly correspond to the deformations produced by the small object. The answer to this question can only be ascertained by comparing the EOB approach or PN theory to self-force calculations, comparisons which have already began to take place (Blanchet et al., 2009 and Damour, 2009). In the slow-velocity (far-separation), small-mass regime, PN/self-force comparisons have recently shown that essentially the conservative sector of the metric tensor agrees to 3PN order (Blanchet et al., 2009).

In the fast-velocity (small-separation), small-mass regime, however, PN and self-force calculation agree only qualitatively and not quantitatively. Barack and Sago, 2009 have shown that the PN ISCO (computed either in the Taylor formalism or within the EOB approach) does not agree quantitatively with the ISCO shift predicted by self-force calculations (an observation recently confirmed in (Damour, 2009)). Further investigation is required to determine the impact of such a disagreement on EMRI waveforms over large integration times.

## 6 Data Analysis Implications

The above sections have demonstrated that the EOB approach, when augmented by BH GW absorption corrections, can model GWs from EMRIs in a circular orbit around Schwarzschild extremely accurately and efficiently. Moreover, this scheme allows for the trivial inclusion of PN-inspired, conservative and dissipative self-force contributions. Before one can argue that the EOB approach is definitely applicable to LISA data analysis, one must first verify that this flexibility is preserved in more complex and generic orbits.

Nevertheless, one can ask the question of how many templates would be required to cover the EMRI parameter space with EOB templates. Let us begin by defining the Fourier transform of a time-signal in both continuous and discrete form:

$$\tilde{h} = \int_{-\infty}^{\infty} h(t) e^{-2\pi i f t} dt, \qquad \tilde{h}_k = \sum_{j=0}^{N-1} h_j e^{-2\pi i j k / N}, \tag{33}$$

where $h_j = h(t_j)$ and we follow the conventions of Schutz, 1997. Such a discrete Fourier transform can only be computed once the time-domain waveform has been discretised into a time-series, but care must be taken during discretisation, since classic data-analysis pitfall could drastically affect the frequency-domain results. One of the classic problems is that of *aliasing*, which is caused by an under-sampling of the time-series, leading to power in unphysical (alias) frequencies. Aliasing can be prevented by making sure that the time-domain discretisation is sufficiently fine. In GW data analysis, this translates to the following condition on the time-discretisation stencil $\Delta$:

$$\Delta_j < \frac{1}{2F_{GW}^{max}} \sim \frac{\pi}{m\dot{\Phi}_{max}}, \tag{34}$$

where $F_{GW}^{max}$ and $\dot{\Phi}_{max}$ stand for the maximum GW frequency and the maximum orbital frequency in the evolution of the $m$-harmonic GW mode. We here choose a time discretisation a few orders of magnitude under this Nyquist minimum.

Given any real-valued, time-domain series, the discrete Fourier transform produces a complex result. The EOB waveforms defined above, however, are naturally complex, so we take the real part of these functions. With this in mind, one can define the inner product

$$\langle A, B \rangle := \mathfrak{R} \int_0^\infty \frac{\tilde{A}\tilde{B}}{S_n(f)} df, \tag{35}$$

for any time-series $A$ or $B$ and spectral noise density $S_n(f)$. For example, for the waveforms under consideration, this statistic becomes

$$\langle h_1, h_2 \rangle = \int_{f_l}^{f_h} \frac{\tilde{h}_{1,R}\tilde{h}_{2,R} + \tilde{h}_{1,I}\tilde{h}_{2,I}}{S_n(f)} df, \tag{36}$$

because the time-domain waveform is complex, and thus, the real and imaginary parts have independent Fourier transforms. For realistic data analysis, one must fold into these waveforms the detector's beam-pattern functions, but we neglect this here for simplicity.

The spectral noise density curve is modelled via the following fitting function (see eg. Yunes et al., 2009)

$$S_n(f) = 4.2 \times 10^{-47} f^{-7/3} + \text{Min}\left[S_{n,1}, S_{n,2}\right] \tag{37}$$

where

$$\begin{aligned}
S_{n,1}(f) &= \left(9.18 \times 10^{-52} f^{-4} + 1.59 \times 10^{-41} \right. \\
&\quad \left. + 9.18 \times 10^{-38} f^2\right) \exp\left(9 \times 10^{-3} f^{-11/3} T^{-1}\right), \\
S_{n,2}(f) &= 9.18 \times 10^{-52} f^{-4} + 1.59 \times 10^{-41} \\
&\quad + 9.18 \times 10^{-38} f^2 + 2.1 \times 10^{-45} f^{-7/3},
\end{aligned}$$

and where $T$ is the length of integration, which we take to be 2 weeks, as this is the nominal value that is currently expected to be used in LISA data analysis.

Before one can compute the overlaps, however, one must ensure that the best value of the overlap is obtained, by working with orthonormalised waveforms and maximising over time and phase offsets. These overlaps allow us to compute the template submanifold (Owen, 1996). We shall here follow the prescription laid out in Appendix B of Damour et al., 1998, which for completeness we summarise below. The orthonormalisation of the waveforms can be achieved via the standard Gram-Schimdt procedure, where here we apply this directly to the Fourier transformed waveforms

$$\tilde{e}_{A,R} = \frac{\tilde{h}_{A,R}}{\|h_{A,R}\|},$$
$$\tilde{e}_{A,I} = \frac{1}{\|h_{A,R}\|} \left[ \|h_{A,R}\|^2 \|h_{A,I}\|^2 - \langle h_{A,R}, h_{A,I} \rangle^2 \right]^{-1/2}$$
$$\times \left[ \|h_{A,R}\|^2 \tilde{h}_{A,I} - \langle h_{A,R}, h_{A,I} \rangle \tilde{h}_{A,R} \right],$$

These quantities can be thought of as vectors, since they are time-series in the time-shift parameter. Finally, with these shorthands, one constructs the statistic

$$O(t) = \sqrt{\frac{A(t) + B(t)}{2} - \alpha \sqrt{\frac{(A(t) + B(t))^2}{4} + C(t)^2}}. \tag{38}$$

When $\alpha = 1$, the maximum value of $O$ returns the mini-max overlap, while when $\alpha = -1$, the maximum value of $O$ returns the max-max overlap.

In Yunes et al., 2009, we employed the above algorithm to compute the number of templates required to cover a parameter space in the total mass range $(10^5–10^6)M_\odot$ and mass ratio range $(10^{-4}–10^{-5})$ for a minimal overlap of 0.97. We find that one requires less than $10^7$ EOB templates to cover the template bank in this way. The `Mathematica` routine we have constructed can construct one single template in approximately a few CPU seconds, but this greatly under-estimates the efficiency of the method. One can expect an efficiency of approximately at least a hundred times better if one were to implement the EOB scheme in C code.

## 7  Discussion and Future Directions

The construction of accurate and efficient waveform models for LISA parameter estimation of EMRI signals remains an open problem. We have here reviewed and contrasted some of the proposed models. We have then concentrated on a new model, recently proposed in Yunes et al., 2009, which combines EOB theory with insights from BH perturbation theory.

We have demonstrated that the EOB approach is sufficiently flexible to accurately estimate the waveform from quasi-circular EMRIs in a non-spinning SMBH background. Errors in the phase and amplitude are well-under those required for detection and possibly also for parameter estimation, when employing the calibrated models that include BH GW absorption. This calibration is performed through the inclusion of higher-order PN terms in the energy flux, which are then determined via fitting to a Teukolsky flux, accurate to at least ten orders of magnitude. After such calibration, the evolution of the system leads to waveforms with incredible high accuracy, relative to Teukolsky ones. Moreover, the EOB scheme allows for an efficient inclusion of PN conservative and dissipative self-force corrections.

One then wonders whether such waveforms are also accurate relative to more realistic BH perturbation theory waveforms, such as those constructed with knowledge of the self-force. Unfortunately, the latter are not available in the literature and comparisons are thus not yet possible. However, preliminary comparisons of the ISCO shift in BH perturbation theory and the EOB approach suggest that the latter is flexible enough to account for the former's conservative corrections via high-order PN calibration parameters in the conservative EOB Hamiltonian. A more complete and direct comparison of the waveforms themselves is required to verify these claims in more detail, a fruitful direction for future research.

For the first time in the EMRI literature, the EOB scheme has also allowed us to compute the template submanifold in practice and estimate the number of EOB templates required to cover the template space with a given mismatch. Estimates, of course, exist in the literature that suggest one requires on the order of $10^{40}$ templates to cover the parameter space. Although we here find that $10^7$ templates are sufficient, one must recall that the EMRIs we studied are overly-simplified: quasi-circular orbits around a non-spinning BH. If one were to include spin in the SMBH and eccentricity plus inclination in the small compact object's orbit, then the number of templates required to cover the space would probably approach the less favourable estimate mentioned above.

A final possible avenue for future research could involve the coupling of information from numerical simulations of equal, or nearly-equal mass systems in the EOB calibration. Perhaps calibrating the linear-in-$\nu$, high-order PN parameters might lead to important GW contributions that need not be negligible for parameter estimation. With the advent of the LISA detector, new PN techniques and practical results from the BH perturbation theory community, the future of EMRI theory looks promising.

## Acknowledgements

This paper is based on Yunes et al., 2009, which was published in collaboration with Alessandra Buonanno, Scott Hughes, Cole Miller, and Yi Pan. We are thus grateful to these authors for granting authorisation to reproduce and extend some of the results from that collaboration to be presented in this article. We are grateful to Luis Lehner, Eric Poisson and Carlos Sopuerta for useful suggestions and comments when proof-reading this article. Finally, we thank Pau Amaro-Seoane for aiding in the reformatting of this article to abide with GW Notes standards. We acknowledge the hospitality of the Perimeter Institute, where some of this work was completed, as well as support from NSF grant PHY-0745779.

# References in the highlight article

*Selected abstracts*

*July to September 2009*

## Measuring the Spin of GRS 1915+105 with Relativistic Disk Reflection

**Authors:** Blum, J. L.; Miller, J. M.; Fabian, A. C.; Miller, M. C.; Homan, J.; van der Klis, M.; Cackett, E. M.; Reis, R. C.

**Eprint:** http://arxiv.org/abs/0909.5383

**Keywords:** astro-ph.HE; astrophysics; spin; supermassive black holes

**Abstract:** GRS 1915+105 harbors one of the most massive known stellar black holes in the Galaxy. In May 2007, we observed GRS 1915+105 for 117 ksec in the low/hard state using Suzaku. We collected and analyzed the data with the HXD/PIN and XIS cameras spanning the energy range from 2.3-55 keV. Fits to the spectra with simple models reveal strong disk reflection through an Fe K emission line and a Compton back-scattering hump. We report constraints on the spin parameter of the black hole in GRS 1915+105 using relativistic disk reflection models. The model for the soft X-ray spectrum (i.e. < 10 keV) suggests $a/M = 0.56(2)$ and excludes zero spin at the 4 sigma level of confidence. The model for the full broadband spectrum suggests that the spin may be higher, $a/M = 0.98(1)$ (1 sigma confidence), and again excludes zero spin at the 2 sigma level of confidence. We discuss these results in the context of other spin constraints and inner disk studies in GRS 1915+105.

## Self-gravitating warped discs around supermassive black holes

**Authors:** Ulubay-Siddiki, A.; Gerhard, O.; Arnaboldi, M.

**Eprint:** http://arxiv.org/abs/0909.5333

**Keywords:** accretion discs; astro-ph.CO; astro-ph.HE; astrophysics

**Abstract:** We consider warped equilibrium configurations for stellar and gaseous disks in the Keplerian force-field of a supermassive black hole, assuming that the self-gravity of the disk provides the only acting torques. Modeling the disk as a collection of concentric circular rings, and computing the torques in the non-linear regime, we show that stable, strongly warped precessing equilibria are possible. These solutions exist for a wide range of disk-to-black hole mass ratios $M_d/M_{bh}$, can span large warp angles of up to $\pm \sim 120$ deg, have inner and outer boundaries, and extend over a radial range of a factor of typically two to four. These equilibrium configurations obey a scaling relation such that in good approximation $\dot{\phi}/\Omega \propto M_d/M_{bh}$ where $\dot{\phi}$ is the (retrograde) precession frequency and $\Omega$ is a characteristic orbital frequency in the disk. Stability was determined using linear perturbation theory and, in a few cases, confirmed by numerical integration of the equations of motion. Most of the precessing equilibria are found to be stable, but some are unstable. The main result of this study is that highly warped disks near black holes can persist for long times without any persistent forcing other than by their self-gravity. The possible relevance of this to galactic nuclei is briefly discussed.

## High-velocity runaway stars from three-body encounters

**Authors:** Gvaramadze, V. V.; Gualandris, A.; Zwart, S. Portegies

**Eprint:** <http://arxiv.org/abs/0909.4929>

**Keywords:** astro-ph.SR; astrophysics; EMRI; stellar dynamics

**Abstract:** We performed numerical simulations of dynamical encounters between hard massive binaries and a very massive star (VMS; formed through runaway mergers of ordinary stars in the dense core of a young massive star cluster), in order to explore the hypothesis that this dynamical process could be responsible for the origin of high-velocity (≥200-400 km/s) early or late B-type stars. We estimated the typical velocities produced in encounters between very tight massive binaries and VMSs (of mass of ≥200 Msun) and found that about 3-4 per cent of all encounters produce velocities of ≥400 km/s, while in about 2 per cent of encounters the escapers attain velocities exceeding the Milky Ways's escape velocity. We therefore argue that the origin of high-velocity (≥200-400 km/s) runaway stars and at least some so-called hypervelocity stars could be associated with dynamical encounters between the tightest massive binaries and VMSs formed in the cores of star clusters. We also simulated dynamical encounters between tight massive binaries and single ordinary 50-100 Msun stars. We found that from 1 to ≃4 per cent of these encounters can produce runaway stars with velocities of ≥300-400 km/s (typical of the bound population of high-velocity halo B-type stars) and occasionally (in less than 1 per cent of encounters) produce hypervelocity (≥700 km/s) late B-type escapers.

## Modeling Extreme Mass Ratio Inspirals within the Effective-One-Body Approach

**Authors:** Yunes, Nicolas; Buonanno, Alessandra; Hughes, Scott A.; Miller, M. Coleman; Pan, Yi

**Eprint:** http://arxiv.org/abs/0909.4263

**Keywords:** astro-ph.CO; astro-ph.GA; astro-ph.HE; Effective one body; EMRI; gr-qc; waveforms

**Abstract:** We present the first models of extreme-mass-ratio inspirals within the effective-one-body (EOB) formalism, focusing on quasi-circular orbits into non-rotating black holes. We show that the phase difference and (Newtonian normalized) amplitude difference between EOB and Teukolsky-based gravitational waveforms can be reduced to < 0.1 rads and < 0.002, respectively, after a 2-year evolution. The inclusion of post-Newtonian self-force terms in the EOB approach leads to a phase disagreement of ∼ 6-27 rads after a 2-year evolution. Such inclusion could also allow for the EOB modeling of waveforms from intermediate-mass ratio, quasi-circular inspirals.

## An exploration of CUDA and CBEA for a gravitational wave source-modelling application

**Authors:** Khanna, Gaurav; McKennon, Justin

**Eprint:** http://arxiv.org/abs/0909.4039

**Keywords:** EMRI; general relativity; GPU; gr-qc; numerical relativity; supermassive black holes; waveforms

**Abstract:** In this paper, we accelerate a gravitational physics numerical modelling application using hardware accelerators – Cell processor and Tesla CUDA GPU. We describe these new technologies and our approach in detail, and then present our final performance results. We obtain well over an order-of-magnitude performance gain in our application by making use of these many-core architectures.

## Simulations of Binary Black Hole Mergers Using Spectral Methods

**Authors:** Szilágyi, Béla; Lindblom, Lee; Scheel, Mark A.

**Abstract:** Several improvements in numerical methods and gauge choice are presented that make it possible now to perform simulations of the merger and ringdown phases of "generic" binary black-hole evolutions using the pseudo-spectral evolution code SpEC. These improvements include the use of a new damped-wave gauge condition, a new grid structure with appropriate filtering that improves stability, and better adaptivity in conforming the grid structures to the shapes and sizes of the black holes. Simulations illustrating the success of these new methods are presented for a variety of binary black-hole systems. These include fairly "generic" systems with unequal masses (up to 2:1 mass ratios), and spins (with magnitudes up to 0.4 $M^2$) pointing in various directions.

## Fundamental Theoretical Bias in Gravitational Wave Astrophysics and the Parameterized Post-Einsteinian Framework

**Authors:** Yunes, Nicolas; Pretorius, Frans

**Abstract:** We consider the concept of fundamental bias in gravitational wave astrophysics as the assumption that general relativity is the correct theory of gravity during the entire wave-generation and propagation regime. Such an assumption is valid in the weak-field, as verified by precision experiments and observations, but it need not hold in the dynamical strong-field regime where tests are lacking. Fundamental bias can cause systematic errors in the detection and parameter estimation of signals, which can lead to a mischaracterization of the universe through incorrect inferences about source event rates and populations. We propose a remedy through the introduction of the parameterized post-Einsteinian framework, which consists of the enhancement of waveform templates via the inclusion of post-Einsteinian parameters. These parameters would ostensibly be designed to interpolate between templates constructed in general relativity and well-motivated alternative theories of gravity, and also include extrapolations that follow sound theoretical principles, such as consistency with conservation laws and symmetries. As an example, we construct parameterized post-Einsteinian templates for the binary coalescence of equal-mass, non-spinning compact objects in a quasi-circular inspiral. The parametrized post-Einsteinian framework should allow matched filtered data to select a specific set of post-Einsteinian parameters without a priori assuming the validity of the former, thus either verifying general relativity or pointing to possible dynamical strong-field deviations.

## Constructing the self-force

**Authors:** Poisson, Eric

**Eprint:** <http://arxiv.org/abs/0909.2994>

**Keywords:** gr-qc; self force

**Abstract:** I present an overview of the methods involved in the computation of the scalar, electromagnetic, and gravitational self-forces acting on a point particle moving in a curved spacetime. For simplicity, the focus here will be on the scalar self-force. The lecture follows closely my review article on this subject published in Living Reviews in Relativity. I begin with a review of geometrical elements (Synge's world function, the parallel propagator). Next I introduce useful coordinate systems (Fermi normal coordinates and retarded light-cone coordinates) in a neighborhood of the particle's world line. I then present the wave equation for a scalar field in curved spacetime and the equations of motion for a particle endowed with a scalar charge. The wave equation is solved by means of a Green's function, and the self-force is constructed from the field gradient. Because the retarded field is singular on the world line, the self-force must involve a regularized version of the field gradient, and I describe how the regular piece of the self-field can be identified. In the penultimate section of the lecture I put the construction of the self-force on a sophisticated axiomatic basis, and in the concluding section I explain how one can do better by abandoning the dangerous fiction of a point particle

## "Complete" gravitational waveforms for black-hole binaries with non-precessing spins

**Authors:** Ajith, P.; Hannam, M.; Husa, S.; Chen, Y.; Bruegmann, B.; Dorband, N.; Mueller, D.; Ohme, F.; Pollney, D.; Reisswig, C.; Santamaria, L.; Seiler, J.

**Eprint:** <http://arxiv.org/abs/0909.2867>

**Keywords:** astro-ph.CO; gr-qc; numerical relativity; post-Newtonian theory; spin; waveforms

**Abstract:** We present the first analytical inspiral-merger-ringdown gravitational waveforms from black-hole (BH) binaries with non-precessing spins. By matching a post-Newtonian description of the inspiral to a set of numerical calculations performed in full general relativity, we obtain a waveform family with a conveniently small number of physical parameters. The physical content of these waveforms includes the "orbital hang-up" effect, when BHs are spinning rapidly along the direction of the orbital angular momentum. These waveforms will allow us to detect a larger parameter space of BH binary coalescence, to explore various scientific questions related to GW astronomy, and could dramatically improve the expected detection rates of GW detectors.

### The influence of short term variations in AM CVn systems on LISA measurements

**Authors:** Stroeer, Alexander; Nelemans, Gijs

**Eprint:** http://arxiv.org/abs/0909.1796

**Keywords:** astro-ph.SR; back/foreground; gr-qc

**Abstract:** We study the effect of short term variations of the evolution of AM CVn systems on their gravitational wave emissions and in particular LISA observations. We model the systems according to their equilibrium mass-transfer evolution as driven by gravitational wave emission and tidal interaction, and determine their reaction to a sudden perturbation of the system. This is inspired by the suggestion to explain the orbital period evolution of the ultra-compact binary systems V407 Vul and RX-J0806+1527 by non-equilibrium mass transfer. The characteristics of the emitted gravitational wave signal are deduced from a Taylor expansion of a Newtonian quadrupolar emission model, and the changes in signal structure as visible to the LISA mission are determined. We show that short term variations can significantly change the higher order terms in the expansion, and thus lead to spurious (non) detection of frequency derivatives. This may hamper the estimation of the parameters of the system, in particular their masses and distances. However, we find that overall detection is still secured as signals still can be described by general templates. We conclude that a better modelling of the effects of short term variations is needed to prepare the community for astrophysical evaluations of real gravitational wave data of AM CVn systems.

### High accuracy simulations of black hole binaries:spins anti-aligned with the orbital angular momentum

**Authors:** Chu, Tony; Pfeiffer, Harald P.; Scheel, Mark A.

**Eprint:** http://arxiv.org/abs/0909.1313

**Keywords:** gr-qc; massive binaries of black holes; numerical relativity; spin

**Abstract:** High-accuracy binary black hole simulations are presented for black holes with spins anti-aligned with the orbital angular momentum. The particular case studied represents an equal-mass binary with spins of equal magnitude $S/m^2 = 0.43757 \pm 0.00001$. The system has initial orbital eccentricity $\sim 4 \cdot 10^{-5}$, and is evolved through 10.6 orbits plus merger and ringdown. The remnant mass and spin are $M_f = (0.961109 \pm 0.000003)M$ and $S_f/M_f^2 = 0.54781 \pm 0.00001$, respectively, where M is the mass during early inspiral. The gravitational waveforms have accumulated numerical phase errors of $<\sim 0.1$ radians without any time or phase shifts, and $<\sim 0.01$ radians when the waveforms are aligned with suitable time and phase shifts. The waveform is extrapolated to infinity using a procedure accurate to $<\sim 0.01$ radians in phase, and the extrapolated waveform differs by up to 0.13 radians in phase and about one percent in amplitude from the waveform extracted at finite radius r=350M. The simulations employ different choices for the constraint damping parameters in the wave zone; this greatly reduces the effects of junk radiation, allowing the extraction of a clean gravitational wave signal even very early in the simulation.

## Gravitomagnetic corrections on gravitational waves

**Authors:** Capozziello, S.; De Laurentis, M.; Forte, L.; Garufi, F.; Milano, L.

**Eprint:** http://arxiv.org/abs/0909.0895

**Keywords:** astro-ph.CO; general relativity; gr-qc; hep-th; waveforms

**Abstract:** Gravitational waveforms and production could be considerably affected by gravitomagnetic corrections considered in relativistic theory of orbits. Beside the standard periastron effect of General Relativity, new nutation effects come out when $c^{-3}$ corrections are taken into account. Such corrections emerge as soon as matter-current densities and vector gravitational potentials cannot be discarded into dynamics. We study the gravitational waves emitted through the capture, in the gravitational field of massive binary systems (e.g. a very massive black hole on which a stellar object is inspiralling) via the quadrupole approximation, considering precession and nutation effects. We present a numerical study to obtain the gravitational wave luminosity, the total energy output and the gravitational radiation amplitude. From a crude estimate of the expected number of events towards peculiar targets (e.g. globular clusters) and in particular, the rate of events per year for dense stellar clusters at the Galactic Center (SgrA*), we conclude that this type of capture could give signatures to be revealed by interferometric GW antennas, in particular by the forthcoming laser interferometer space antenna LISA.

### The Future of Direct Supermassive Black Hole Mass Estimates

**Authors:** Batcheldor, D.; Koekemoer, A. M.

**Eprint:** http://arxiv.org/abs/0909.4245

**Keywords:** astro-ph.CO; astrophysics; massive binaries of black holes

**Abstract:** (Abridged) The repeated discovery of supermassive black holes (SMBHs) at the centers of galactic bulges, and the discovery of relations between the SMBH mass (M) and the properties of these bulges, has been fundamental in directing our understanding of both galaxy and SMBH formation and evolution. However, there are still many questions surrounding the SMBH - galaxy relations. For example, are the scaling relations linear and constant throughout cosmic history, and do all SMBHs lie on the scaling relations? These questions can only be answered by further high quality direct M estimates from a wide range in redshift. In this paper we determine the observational requirements necessary to directly determine SMBH masses, across cosmological distances, using current M modeling techniques. We also discuss the SMBH detection abilities of future facilities. We find that if different M modeling techniques, using different spectral features, can be shown to be consistent, then both 30 m ground- and 16 m space-based telescopes will be able to sample M $10^9 \, M_\odot$ across $\sim$ 95% of cosmic history. However, we find that the abilities of ground-based telescopes critically depend on future advancements in adaptive optics systems; more limited AO systems will result in limited effective spatial resolutions, and forces observations towards the near-infrared where spectral features are weaker and more susceptible to sky features. Ground-based AO systems will always be constrained by relatively bright sky backgrounds and atmospheric transmission. The latter forces the use of multiple spectral features and dramatically impacts the SMBH detection efficiency. The most efficient way to advance our database of direct SMBH masses is therefore through the use of a large (16 m) space-based UVOIR telescope.

### X-Ray Variability of AGN and Relationship to Galactic Black Hole Binary Systems

**Authors:** McHardy, I. M.

**Eprint:** http://arxiv.org/abs/0909.2579

**Keywords:** astro-ph.HE; astrophysics; massive binaries of black holes

**Abstract:** Over the last 12 years, AGN monitoring by RXTE, has revolutionised our understanding of the X-ray variability of AGN, of the relationship between AGN and Galactic black hole X-ray binaries (BHBs) and hence of the accretion process itself, which fuels the emission in AGN and BHBs and is the major source of power in the universe. In this paper I review our current understanding of these topics. I begin by considering whether AGN and BHBs show the same X-ray spectral-timing 'states' (e.g. low-flux, hard-spectrum or 'hard' and high-flux, soft-spectrum or 'soft'). Observational selection effects mean that most of the AGN which we have monitored will probably be 'soft state' objects, but AGN are found in the other BHB states, although possibly with different critical transition accretion rates. I examine timescale scaling relationships between AGN and BHBs. I show that characteristic power spectral 'bend' timescales, $T_B$, scale approximately with black hole mass, $M_BH$, but inversely with accretion rate, $\dot{m}_E$, (in units of the Eddington accretion rate) probably signifying that $T_B$, arises at the inner edge of the accretion disc. The relationship $T_B$ proportional to $M_{BH}/\dot{m}_E$ is a good fit, implying that no other potential variable, e.g. black hole spin, varies significantly. Lags between hard and soft X-ray bands as a function of Fourier timescale follow similar patterns in AGN and BHBs. [Abridged]

### Jets in Active Galactic Nuclei

**Authors:** Marscher, A. P.

**Eprint:** <http://arxiv.org/abs/0909.2576>

**Keywords:** astro-ph.HE; astrophysics; massive binaries of black holes

**Abstract:** The jets of active galactic nuclei can carry a large fraction of the accreted power of the black-hole system into interstellar and even extragalactic space. They radiate profusely from radio to X-ray and gamma-ray frequencies. In the most extreme cases, the outward flow speeds correspond to high Lorentz factors that can reach 40 or more. This chapter displays images at various wavebands as well as light curves and continuum spectra that illustrate the variability with location, time, and frequency of the emission from compact, parsec- and subparsec-scale jets. It presents a physical framework for investigating many aspects of the structure and dynamical processes from such data.

### From Multiwavelength to Mass Scaling: Accretion and Ejection in Microquasars and AGN

**Authors:** Markoff, S.

**Eprint:** `http://arxiv.org/abs/0909.2574`

**Keywords:** astro-ph.HE; astrophysics; massive binaries of black holes

**Abstract:** A solid theoretical understanding of how inflowing, accreting plasma around black holes and other compact objects gives rise to outflowing winds and jets is still lacking, despite decades of observations. The fact that similar processes and morphologies are observed in both X-ray binaries as well as active galactic nuclei has led to suggestions that the underlying physics could scale with black hole mass, which could provide a new handle on the problem. In the last decade, simultaneous broadband campaigns of the fast-varying X-ray binaries particularly in their microquasar state have driven the development of, and in some cases altered, our ideas about the inflow/outflow connection in accreting black holes. Specifically the discovery of correlations between the radio, infrared and X-ray bands has revealed a remarkable connectivity between the various emission regions, and argued for a more holistic approach to tackling questions about accretion. This article reviews the recent major observational and theoretical advances that focus specifically on the relation between the two "sides" of the accretion process in black holes, with an emphasis on how new tools can be derived for comparisons across the mass scale.

### 'Disc-jet' coupling in black hole X-ray binaries and active galactic nuclei

**Authors:** Fender, R. P.

**Eprint:** `http://arxiv.org/abs/0909.2572`

**Keywords:** astro-ph.HE; astrophysics; massive binaries of black holes

**Abstract:** In this chapter I will review the status of our phenomenological understanding of the relation between accretion and outflows in accreting black hole systems. This understanding arises primarily from observing the relation between X-ray and longer wavelength (infrared, radio) emission. The view is necessarily a biased one, beginning with observations of X-ray binary systems, and attempting to see if they match with the general observational properties of active galactic nuclei.

### X-ray emission from black-hole binaries

**Authors:** Gilfanov, M.

**Eprint:** `http://arxiv.org/abs/0909.2567`

**Abstract:** The properties of X-ray emission from accreting black holes are reviewed. The contemporary observational picture and current status of theoretical understanding of accretion and formation of X-ray radiation in the vicinity of the compact object are equally in the focus of this chapter. The emphasis is made primarily on common properties and trends rather than on peculiarities of individual objects and details of particular theoretical models. The chapter starts with discussion of the geometry of the accretion flow, spectral components in X-ray emission and black hole spectral states. The prospects and diagnostic potential of X-ray polarimetry are emphasized. Significant attention is paid to the discussion of variability of X-ray emission in general and of different spectral components – emission of the accretion disk, Comptonized radiation and reflected component. Correlations between spectral and timing characteristics of X-ray emission are reviewed and discussed in the context of theoretical models. Finally, a comparison with accreting neutron stars is made.

## States and transitions in black-hole binaries

**Authors:** Belloni, T. M.

**Eprint:** http://arxiv.org/abs/0909.2474

**Keywords:** astro-ph.HE; astrophysics; massive binaries of black holes

**Abstract:** With the availability of the large database of black-hole transients from the Rossi X-Ray Timing Explorer, the observed phenomenology has become very complex. The original classification of the properties of these systems in a series of static states sorted by mass accretion rate proved not to be able to encompass the new picture. I outline here a summary of the current situation and show that a coherent picture emerges when simple properties such as X-ray spectral hardness and fractional variability are considered. In particular, fast transition in the properties of the fast time variability appear to be crucial to describe the evolution of black-hole transients. Based on this picture, I present a state-classification which takes into account the observed transitions. I show that, in addition to transients systems, other black-hole binaries and Active Galactic Nuclei can be interpreted within this framework. The association between these states and the physics of the accretion flow around black holes will be possible only through modeling of the full time evolution of galactic transient systems.

### Tidal break-up of binary stars at the Galactic center and its consequences

**Authors:** Antonini, Fabio; Faber, Joshua; Gualandris, Alessia; Merritt, David

**Eprint:** http://arxiv.org/abs/0909.1959

**Keywords:** astro-ph.CO; astro-ph.GA; astrophysics; EMRI; stellar dynamics

**Abstract:** The tidal breakup of binary star systems by the supermassive black hole (SMBH) in the center of the galaxy has been suggested as the source of both the observed sample of hypervelocity stars (HVSs) in the halo of the Galaxy and the S-stars that remain in tight orbits around Sgr A*. Here, we use a post-Newtonian N-body code to study the dynamics of main-sequence binaries on highly elliptical bound orbits whose periapses lie close to the SMBH, determining the properties of ejected and bound stars as well as collision products. Unlike previous studies, we follow binaries that remain bound for several revolutions around the SMBH, finding that in the case of relatively large periapses and highly inclined binaries the Kozai resonance can lead to large periodic oscillations in the internal binary eccentricity and inclination. Collisions and mergers of the binary elements are found to increase significantly for multiple orbits around the SMBH, while HVSs are primarily produced during a binary's first passage. This process can lead to stellar coalescence and eventually serve as an important source of young stars at the galactic center.

### The Distribution of Stars and Stellar Remnants at the Galactic Center

**Authors:** Merritt, David

**Eprint:** http://arxiv.org/abs/0909.1318

**Keywords:** astro-ph.CO; astro-ph.GA; astrophysics; EMRI; Sagittarius A*; stellar dynamics

**Abstract:** Motivated by recent observations that suggest a low density of old stars around the Milky Way supermassive black hole, evolutionary models for the nuclear star cluster are considered that postulate a parsec-scale core as initial conditions. Gravitational encounters cause the core to shrink; a core of initial radius 1-1.5 pc evolves to a size of 0.5 pc after 10 Gyr, roughly the size of the observed core. The absence of a Bahcall-Wolf cusp is naturally explained. In these models, the time for a 10-Solar-mass black hole to spiral in to the Galactic center from an initial distance of 5 pc can be much greater than 10 Gyr. Assuming that the stellar black holes had the same phase-space distribution initially as the stars, their density after 5-10 Gyr is predicted to rise very steeply going into the stellar core, but to remain substantially below the densities inferred from steady-state models that include a steep density cusp in the stars. The implications of these models are discussed for the rates of gravitational wave inspiral events and of other physical processes that depend on a high density of stars or stellar mass black holes near Sagittarius A*.

## Supercomputing and stellar dynamics

**Authors:** Capuzzo-Dolcetta, R.

**Eprint:** http://arxiv.org/abs/0909.0879

**Keywords:** astro-ph.GA; astro-ph.IM; astrophysics; GPU; stellar dynamics

**Abstract:** In this paper I will outline some of the aspects and problems of modern celestial mechanics and stellar dynamics, in the context of the quickly growing computing facilities. I will point the attention on the great advantages in using, for astrophysical simulations, the modern, fast and cheap Graphic Processing Units (GPUs) acting as true supercomputers. Finally, I present and discuss some characteristics and performances of a new double-parallel code exploiting the joint power of multicore CPUs and GPUs.

## Oct-tree Method on GPU

**Authors:** Nakasato, N.

**Eprint:** http://arxiv.org/abs/0909.0541

**Keywords:** astro-ph.GA; astro-ph.IM; astrophysics; GPU; stellar dynamics

**Abstract:** The kd-tree is a fundamental tool in computer science. Among others, an application of the kd-tree search (oct-tree method) to fast evaluation of particle interactions and neighbor search is highly important since computational complexity of these problems are reduced from $O(N^2)$ with a brute force method to $O(NlogN)$ with the tree method where N is a number of particles. In this paper, we present a parallel implementation of the tree method running on a graphic processor unit (GPU). We successfully run a simulation of structure formation in the universe very efficiently. On our system, which costs roughly 900, the run with $N \sim 2.87x10^6$ particles took 5.79 hours and executed $1.2x10^{13}$ force evaluations in total. We obtained the sustained computing speed of 21.8 Gflops and the cost per Gflops of 41.6/Gflops that is two and half times better than the previous record in 2006.

### An overview of the Laser Interferometer Space Antenna

**Authors:** Shaddock, Daniel A.

**Eprint:** http://arxiv.org/abs/0909.0650

**Keywords:** gr-qc

**Abstract:** The Laser Interferometer Space Antenna will detect gravitational waves with frequencies from 0.1 mHz to 1 Hz. This article provides a brief overview of LISA's science goals followed by a tutorial of the LISA measurement concept.

### Hot high-mass accretion disk candidates

**Authors:** Beuther, H.; Walsh, A. J.; Longmore, S. N.

**Eprint:** http://arxiv.org/abs/0909.0691

**Keywords:** accretion discs; astro-ph.SR; astrophysics; observations

**Abstract:** To better understand the physical properties of accretion disks in high-mass star formation, we present a study of a 12 high-mass accretion disk candidates observed at high spatial resolution with the Australia Telescope Compact Array (ATCA) in the NH3 (4,4) and (5,5) lines. Almost all sources were detected in NH3, directly associated with CH3OH Class II maser emission. From the remaining eleven sources, six show clear signatures of rotation and/or infall motions. These signatures vary from velocity gradients perpendicular to the outflows, to infall signatures in absorption against ultracompact HII regions, to more spherical infall signatures in emission. Although our spatial resolution is ~ 1000AU, we do not find clear Keplerian signatures in any of the sources. Furthermore, we also do not find flattened structures. In contrast to this, in several of the sources with rotational signatures, the spatial structure is approximately spherical with sizes exceeding $10^4$ AU, showing considerable clumpy sub-structure at even smaller scales. This implies that on average typical Keplerian accretion disks – if they exist as expected – should be confined to regions usually smaller than 1000AU. It is likely that these disks are fed by the larger-scale rotating envelope structure we observe here. Furthermore, we do detect 1.25cm continuum emission in most fields of view.

## The stochastic gravitational wave background from turbulence and magnetic fields generated by a first-order phase transition

**Authors:** Caprini, Chiara; Durrer, Ruth; Servant, Geraldine

**Eprint:** http://arxiv.org/abs/0909.0622

**Keywords:** astro-ph.CO; back/foreground; cosmology

**Abstract:** We analytically derive the spectrum of gravitational waves due to magneto-hydrodynamical turbulence generated by bubble collisions in a first-order phase transition. In contrast to previous studies, we take into account the fact that turbulence and magnetic fields act as sources of gravitational waves for many Hubble times after the phase transition is completed. This modifies the gravitational wave spectrum at large scales. We also model the initial stirring phase preceding the Kolmogorov cascade, while earlier works assume that the Kolmogorov spectrum is set in instantaneously. The continuity in time of the source is relevant for a correct determination of the peak position of the gravitational wave spectrum. We discuss how the results depend on assumptions about the unequal-time correlation of the source and motivate a realistic choice for it. Our treatment gives a similar peak frequency to previous analyses but the amplitude of the signal is reduced due to the use of a more realistic power spectrum for the MHD turbulence. For a strongly first-order electroweak phase transition, the signal is observable by LISA.

## Emission Lines as a Tool in Search for Supermassive Black Hole Binaries and Recoiling Black Holes

**Authors:** Bogdanovic, Tamara; Eracleous, Michael; Sigurdsson, Steinn

**Eprint:** http://arxiv.org/abs/0909.0516

**Keywords:** astro-ph.CO; astro-ph.HE; astrophysics; EM counterparts; massive binaries of black holes

**Abstract:** Detection of electromagnetic (EM) counterparts of pre-coalescence binaries has very important implications for our understanding of the evolution of these systems as well as the associated accretion physics. In addition, a combination of EM and gravitational wave signatures observed from coalescing supermassive black hole binaries (SBHBs) would provide independent measurements of redshift and luminosity distance, thus allowing for high precision cosmological measurements. However, a statistically significant sample of these objects is yet to be attained and finding them observationally has proven to be a difficult task. Here we discuss existing observational evidence and how further advancements in the theoretical understanding of observational signatures of SBHBs before and after the coalescence can help in future searches.

### A Relationship between Supermassive Black Hole Mass and the Total Gravitational Mass of the Host Galaxy

**Authors:** Bandara, Kaushala; Crampton, David; Simard, Luc

**Eprint:** http://arxiv.org/abs/0909.0269

**Keywords:** astro-ph.GA; astrophysics; cosmology; observations; supermassive black holes

**Abstract:** We investigate the correlation between the mass of a central supermassive black hole and the total gravitational mass of the host galaxy ($M_{\text{tot}}$). The results are based on 43 galaxy-scale strong gravitational lenses from the Sloan Lens ACS (SLACS) Survey whose black hole masses were estimated through two scaling relations: the relation between black hole mass and Sersic index ($M_{bh} - n$) and the relation between black hole mass and stellar velocity dispersion ($M_{bh} - \sigma$). We use the enclosed mass within $R_{200}$, the radius within which the density profile of the early type galaxy exceeds the critical density of the Universe by a factor of 200, determined by gravitational lens models fitted to HST imaging data, as a tracer of the total gravitational mass. The best fit correlation, where $M_{bh}$ is determined from $M_{bh} - \sigma$ relation, is $\log(M_{bh}) = (8.18 +/- 0.11) + (1.55 +/- 0.31)(\log(M_{\text{tot}}) - 13.0)$ over 2 orders of magnitude in $M_{bh}$. From a variety of tests, we find that we cannot reliably infer a connection between $M_{bh}$ and $M_{\text{tot}}$ from the $M_{bh} - n$ relation. The $M_{bh} - -M_{\text{tot}}$ relation provides some of the first, direct observational evidence to test the prediction that supermassive black hole properties are determined by the halo properties of the host galaxy.

### An Electromagnetic Signature of Galactic Black Hole Binaries That Enter Their Gravitational-Wave Induced Inspiral

**Authors:** Loeb, Abraham

**Eprint:** http://arxiv.org/abs/0909.0261

**Keywords:** astro-ph.CO; astro-ph.GA; astrophysics; EM counterparts; massive binaries of black holes; supermassive black holes

**Abstract:** Mergers of gas-rich galaxies lead to black hole binaries that coalesce as a result of dynamical friction on the ambient gas. Once the binary tightens to < $10^3$ Schwarzschild radii, its merger is driven by the emission of gravitational waves (GWs). We show that this transition occurs generically at orbital periods of ~ 1-10 years and an orbital velocity V of a few thousand km/s, with a very weak dependence on the supply rate of gas (V proportional to $\dot{M}^{1/8}$). Therefore, as binaries enter their GW-dominated inspiral, they inevitably induce large periodic shifts in the broad emission lines of any associated quasar(s). The probability of finding a binary in tighter configurations scales as $V^{-8}$ owing to their much shorter lifetimes. Systematic monitoring of the broad emission lines of quasars on timescales of months to decades can set a lower limit on the expected rate of GW sources for LISA.

### The Quasar SDSS J153636.22+044127.0: A Double-Peaked Emitter in a Candidate Binary Black-Hole System

**Authors:** Tang, Sumin; Grindlay, Jonathan

**Eprint:** http://arxiv.org/abs/0909.0258

**Keywords:** astro-ph.CO; astrophysics; massive binaries of black holes; observations

**Abstract:** Double-peaked emission lines are believed to be originated from accretion disks around supermassive black holes (SMBHs), and about 3% of z<0.33 AGNs are found to be double-peaked emitters. The quasar SDSS J153636.22+044127.0 has recently been identified with peculiar broad-line emission systems exhibiting multiple redshifts. We decompose the H$\alpha$ and H$\beta$ profiles into a circular Keplerian disk line component and other Gaussian components. We propose that the system is both a double-peaked emitter and a binary SMBH system, where the extra-flux in the blue peaks of the broad lines comes from the region around the secondary black hole. We suggest that such black hole binary systems might also exist in many known double-peaked emitters, where the tidal torques from the secondary black hole clear the outer region of the disk around the primary black hole, similar to the gap in a protostellar disk due to the process of planetary migration, and might also stimulate the formation of a vertical extended source in the inner region around the primary which illuminates the disk. However, most secondary SMBHs in such systems might be too small to maintain a detectable broad line region (BLR), so that the disk line from the primary dominates.

## Renormalized spin coefficients in the accumulated orbital phase for unequal mass black hole binaries

**Authors:** Gergely, László Á.; Biermann, Peter L.; Mikóczi, Balázs; Keresztes, Zoltán

**Eprint:** http://arxiv.org/abs/0909.0487

**Keywords:** astro-ph.HE; gr-qc; massive binaries of black holes; post-Newtonian theory; spin

**Abstract:** We analyze galactic black hole mergers and their emitted gravitational waves. Such mergers have typically unequal masses with mass ratio of the order 1/10. The emitted gravitational waves carry the inprint of spins and mass quadrupoles of the binary components. Among these contributions, we consider here the quasi-precessional evolution of the spins. A method of taking into account these third post-Newtonian (3PN) effects by renormalizing (redefining) the 1.5 PN and 2PN accurate spin contributions to the accumulated orbital phase is developed.

## Theory and modeling of the magnetic field measurement in LISA PathFinder

**Authors:** Diaz-Aguilo, M; Garcia-Berro, E; Lobo, A

**Eprint:** http://arxiv.org/abs/0908.4564

**Keywords:** gr-qc; instruments; interferometers

**Abstract:** The magnetic diagnostics subsystem of the LISA Technology Package (LTP) on board the LISA PathFinder (LPF) spacecraft includes a set of four tri-axial fluxgate magnetometers, intended to measure with high precision the magnetic field at their respective positions. However, their readouts do not provide a direct measurement of the magnetic field at the positions of the test masses, and hence an interpolation method must be designed and implemented to obtain the values of the magnetic field at these positions. However, such interpolation process faces serious difficulties. Indeed, the size of the interpolation region is excessive for a linear interpolation to be reliable while, on the other hand, the number of magnetometer channels does not provide sufficient data to go beyond the linear approximation. We describe an alternative method to address this issue, by means of neural network algorithms. The key point in this approach is the ability of neural networks to learn from suitable training data representing the behavior of the magnetic field. Despite the relatively large distance between the test masses and the magnetometers, and the insufficient number of data channels, we find that our artificial neural network algorithm is able to reduce the estimation errors of the field and gradient down to levels below 10%, a quite satisfactory result. Learning efficiency can be best improved by making use of data obtained in on-ground measurements prior to mission launch in all relevant satellite locations and in real operation conditions. Reliable information on that appears to be essential for a meaningful assessment of magnetic noise in the LTP.

## Tidal interaction of a small black hole in the field of a large Kerr black hole

**Authors:** Comeau, Simon; Poisson, Eric

**Eprint:** http://arxiv.org/abs/0908.4518

**Keywords:** EMRI; general relativity; gr-qc; spin

**Abstract:** The rates at which the mass and angular momentum of a small black hole change as a result of a tidal interaction with a much larger black hole are calculated to leading order in the small mass ratio. The small black hole is either rotating or nonrotating, and it moves on a circular orbit in the equatorial plane of the large Kerr black hole. The orbits are fully relativistic, and the rates are computed to all orders in the orbital velocity $V < V_{isco}$, which is limited only by the size of the innermost stable circular orbit. We show that as $V \to V_{isco}$, the rates take on a limiting value that depends only on $V_{isco}$ and not on the spin parameter of the large black hole.

## Elementary development of the gravitational self-force

**Authors:** Detweiler, Steven

**Eprint:** http://arxiv.org/abs/0908.4363

**Keywords:** general relativity; gr-qc; self force

**Abstract:** The gravitational field of a particle of small mass $\mu$ moving through curved spacetime, with metric $g_{ab}$, is naturally and easily decomposed into two parts each of which satisfies the perturbed Einstein equations through $O(\mu)$. One part is an inhomogeneous field $h^S_{ab}$ which, near the particle, looks like the Coulomb $\mu/r$ field with tidal distortion from the local Riemann tensor. This singular field is defined in a neighborhood of the small particle and does not depend upon boundary conditions or upon the behavior of the source in either the past or the future. The other part is a homogeneous field $h^R_{ab}$. In a perturbative analysis, the motion of the particle is then best described as being a geodesic in the metric $g_{ab} + h^R_{ab}$. This geodesic motion includes all of the effects which might be called radiation reaction and conservative effects as well.

## Radial Distribution of X-ray Point Sources near the Galactic Center

**Authors:** Hong, Jaesub; Berg, Maureen van den; Grindlay, Jonathan E.; Laycock, Silas

**Eprint:** http://arxiv.org/abs/0908.4306

**Keywords:** astro-ph.HE; astrophysics; observations; Sagittarius A*; supermassive black holes

**Abstract:** (Abridged) We present the LogN-LogS and spatial distributions of X-ray point sources in seven Galactic Bulge (GB) fields within 4 deg from the Galactic Center (GC). We compare the properties of 1159 X-ray point sources discovered in our deep (100 ks) Chandra observations of three low extinction Window fields near the GC with the X-ray sources in the other GB fields centered around Sgr B2, Sgr C, the Arches Cluster and Sgr A* using Chandra archival data. To reduce the systematic errors induced by the uncertain X-ray spectra of the sources coupled with field-and-distance dependent extinction, we classify the X-ray sources using quantile analysis and estimate their fluxes accordingly. The result indicates the GB X-ray population is highly concentrated at the center, more heavily than the stellar distribution models. We also compare the total X-ray and infrared surface brightness using the Chandra and Spitzer observations of the regions. The radial distribution of the total infrared surface brightness from the 3.6 band $\mu$m images appears to resemble the radial distribution of the X-ray point sources better than predicted by the stellar distribution models. Assuming a simple power law model for the X-ray spectra, the closer to the GC the intrinsically harder the X-ray spectra appear, but adding an iron emission line at 6.7 keV in the model allows the spectra of the GB X-ray sources to be largely consistent across the region. This implies that the majority of these GB X-ray sources can be of the same or similar type. Their X-ray luminosity and spectral properties support the idea that the most likely candidate is magnetic cataclysmic variables (CVs), primarily intermediate polars (IPs). Their observed number density is also consistent with the majority being IPs.

## Two phase galaxy formation: The Gas Content of Normal Galaxies

**Authors:** Cook, M.; Evoli, C.; Barausse, E.; Granato, G. L.; Lapi, A.

**Abstract:** We investigate the atomic ($HI$) and molecular ($H_2$) Hydrogen content of normal galaxies by combining observational studies linking galaxy stellar and gas budgets to their host dark matter (DM) properties, with a physically grounded galaxy formation model. This enables us to analyse empirical relationships between the virial, stellar, and gaseous masses of galaxies and explore their physical origins. Utilising a semi-analytic model (SAM) to study the evolution of baryonic material within evolving DM halos, we study the effects of baryonic infall, star formation and various feedback mechanisms on the properties of formed galaxies using the most up-to-date physical recipes. We find that in order to significantly improve agreement with observations of low-mass galaxies we must suppress the infall of baryonic material and exploit a two-phase interstellar medium (ISM), where the ratio of $HI$ to $H_2$ is determined by the galactic disk structure. Modifying the standard Schmidt-Kennicutt star formation law by correlating the star formation activity with the $H_2$ gas mass allows us to simultaneously reproduce stellar, $HI$ and $H_2$ mass functions for normal galaxies.

## Integrating Post-Newtonian Equations on Graphics Processing Units

**Authors:** Herrmann, Frank; Silberholz, John; Bellone, Matias; Guerberoff, Gustavo; Tiglio, Manuel

**Abstract:** We report on early results of a numerical and statistical study of binary black hole inspirals. The two black holes are evolved using post-Newtonian approximations starting with initially randomly distributed spin vectors. We characterize certain aspects of the distribution shortly before merger. In particular we note the uniform distribution of black hole spin vector dot products shortly before merger and a high correlation between the initial and final black hole spin vector dot products in the equal-mass, maximally spinning case. These simulations were performed on Graphics Processing Units, and we demonstrate a speed-up of a factor 50 over a more conventional CPU implementation.

### Third post-Newtonian angular momentum flux and the secular evolution of orbital elements for inspiralling compact binaries in quasi-elliptical orbits

**Authors:** Arun, K. G.; Blanchet, Luc; Iyer, Bala R.; Sinha, Siddhartha

**Eprint:** http://arxiv.org/abs/0908.3854

**Keywords:** general relativity; gr-qc; massive binaries of black holes; post-Newtonian theory

**Abstract:** The angular momentum flux from an inspiralling binary system of compact objects moving in quasi-elliptical orbits is computed at the third post-Newtonian (3PN) order using the multipolar post-Minkowskian wave generation formalism. The 3PN angular momentum flux involves the instantaneous, tail, and tail-of-tails contributions as for the 3PN energy flux, and in addition a contribution due to non-linear memory. We average the angular momentum flux over the binary's orbit using the 3PN quasi-Keplerian representation of elliptical orbits. The averaged angular momentum flux provides the final input needed for gravitational wave phasing of binaries moving in quasi-elliptical orbits. We obtain the evolution of orbital elements under 3PN gravitational radiation reaction in the quasi-elliptic case. For small eccentricities, we give simpler limiting expressions relevant for phasing up to order $e^2$. This work is important for the construction of templates for quasi-eccentric binaries, and for the comparison of post-Newtonian results with the numerical relativity simulations of the plunge and merger of eccentric binaries.

### Searching for Galactic White Dwarf Binaries in the Second Mock LISA Data Challenge using an F-Statistic Template Bank

**Authors:** Whelan, John T.; Prix, Reinhard; Khurana, Deepak

**Eprint:** http://arxiv.org/abs/0908.3766

**Keywords:** back/foreground; data analysis; gr-qc; MLDC; parameter estimation; waveforms

**Abstract:** We describe the application of an F-statistic search for continuous gravitational waves to the search for galactic white-dwarf binaries in the Second Mock LISA Data Challenge. The search method employs a hierarchical template-grid based exploration of the parameter space, using a coincidence step to distinguish between primary ("true") and secondary maxima, followed by a final (multi-TDI)

"zoom" stage to provide an accurate parameter estimation of the final candidates. Suitably tuned, the pipeline is able to extract 1989 true signals with only 5 false alarms. The use of the rigid adiabatic approximation allows recovery of signal parameters comparable to statistical expectations, although there is still some systematic excess above expected statistical errors due to Gaussian noise. An experimental iterative pipeline with seven rounds of subtraction and re-analysis allows us to increase the number of signals recovered, up to a total of 3419 with 29 false alarms.

## Gravitational Waves and Light Cosmic Strings

**Authors:** Depies, Matthew R

**Eprint:** <http://arxiv.org/abs/0908.3680>

**Keywords:** astro-ph.CO; cosmology; general relativity; gr-qc

**Abstract:** Gravitational wave signatures from cosmic strings are analyzed numerically. Cosmic string networks form during phase transistions in the early universe and these networks of long cosmic strings break into loops that radiate energy in the form of gravitational waves until they decay. The gravitational waves come in the form of harmonic modes from individual string loops, a "confusion noise" from galactic loops, and a stochastic background of gravitational waves from a network of loops. In this study string loops of larger size $\alpha$ and lower string tensions $G\mu$ (where $\mu$ is the mass per unit length of the string) are investigated than in previous studies. Several detectors are currently searching for gravitational waves and a space based satellite, the Laser Interferometer Space Antenna (LISA), is in the final stages of pre-flight. The results for large loop sizes ($\alpha = 0.1$) put an upper limit of about $G\mu < 10^{-9}$ and indicate that gravitational waves from string loops down to $G\mu \approx 10^{-20}$ could be detectabe by LISA. The string tension is related to the energy scale of the phase transition and the Planck mass via $G\mu = \Lambda_s^2/m_{pl}^2$, so the limits on $G\mu$ set the energy scale of any phase transition to $\Lambda_s < 10^{-4.5} m_{pl}$. Our results indicate that loops may form a significant gravitational wave signal, even for string tensions too low to have larger cosmological effects.

## Three Dimensional Simulations of Vertical Magnetic Flux in the Immediate Vicinity of Black Holes

**Authors:** Punsly, Brian; Igumenshchev, Igor V.; Hirose, Shigenobu

**Eprint:** <http://arxiv.org/abs/0908.3697>

**Keywords:** accretion discs; astro-ph.CO; astro-ph.HE; astrophysics; supermassive black holes

**Abstract:** This article reports on three-dimensional (3-D) MHD simulations of non-rotating and rapidly rotating black holes and the adjacent black hole accretion disk magnetospheres. A particular emphasis is placed on the vertical magnetic flux that is advected inward from large radii and threads the equatorial plane near the event horizon. In both cases of non-rotating and rotating black holes, the existence of a significant vertical magnetic field in this region is like a switch that creates powerful jets. There are many similarities in the vertical flux dynamics in these two cases in spite of the tremendous enhancement of azimuthal twisting of the field lines and enhancement of the jet power because of an "ergospheric disk" in the Kerr metric. A 3-D approach is essential because two-dimensional axisymmetric flows are incapable of revealing the nature of vertical flux near a black hole. Poloidal field lines from the ergospheric accretion region have been visualized in 3-D and much of the article is devoted to a formal classification of the different manifestations of vertical flux in the Kerr case.

## Black Holes in Active Galactic Nuclei

**Authors:** Valtonen, M. J.; Mikkola, S.; Merritt, D.; Gopakumar, A.; Lehto, H. J.; Hyvönen, T.; Rampadarath, H.; Saunders, R.; Basta, M.; Hudec, R.

**Eprint:** http://arxiv.org/abs/0908.2706

**Keywords:** astro-ph.GA; astrophysics; general relativity; gr-qc; massive binaries of black holes; observations

**Abstract:**

Supermassive black holes are common in centers of galaxies. Among the active galaxies, quasars are the most extreme, and their black hole masses range as high as to $6 \cdot 10^{10} M_\odot$. Binary black holes are of special interest but so far OJ287 is the only confirmed case with known orbital elements. In OJ287, the binary nature is confirmed by periodic radiation pulses. The period is twelve years with two pulses per period. The last four pulses have been correctly predicted with the accuracy of few weeks, the latest in 2007 with the accuracy of one day. This accuracy is high enough that one may test the higher order terms in the Post Newtonian approximation to General Relativity. The precession rate per period is $39°.1 \pm 0°.1$, by far the largest rate in any known binary, and the $(1.83 \pm 0.01) \cdot 10^{10} M_\odot$ primary is among the dozen biggest black holes known. We will discuss the various Post Newtonian terms and their effect on the orbit solution.

The over 100 year data base of optical variations in OJ287 puts limits on these terms and thus tests the ability of Einstein's General Relativity to describe, for the first time, dynamic binary black hole spacetime in the strong field regime. The quadrupole-moment contributions to the equations of motion allows us to constrain the 'no-hair' parameter to be 1.0±0.3 which supports the black hole no-hair theorem within the achievable precision.

## The mass function of nearby black hole candidates

**Authors:** Caramete, Laurentiu I.; Biermann, Peter L.

**Eprint:** http://arxiv.org/abs/0908.2764

**Keywords:** astro-ph.CO; astro-ph.HE; astrophysics; observations; supermassive black holes

**Abstract:** The mass function of super-massive black holes in our cosmic neighborhood is required to understand the statistics of activity, specifically the production of ultra high energy particles. We determine a mass function of black hole candidates from the entire sky outside the Galactic plane. Using the 2MASS catalogue as a starting point, and the well established correlation between black hole mass and the old bulge population of stars, we derive a list of nearby black hole candidates within the redshift range z $10^7 M_\odot$ has 5,634 entries. Here we use this catalogue to derive the mass function. We correct for volume, so that this mass function is a volume limited distribution to redshift 0.025. The mass function of nearby black hole candidates is a straight simple power-law, extending down into the mass range, where nuclear star clusters may replace the super-massive black holes. The slope of this mass function can be explained in a simple merger picture. Integrating this mass function over the redshift range, from which it has been derived, gives a total number of black holes z $10^7 M_\odot$ of about $2.8 \times 10^5$, or on the sky 7 for every square degree, if we just average uniformly. In some models many of these, if not all, are candidates to produce ultra high energy particles. So whatever the possibly very small fraction of super-massive black holes is that gives rise to ultra high energy cosmic rays, their directional arrival statistics may reflect the highly inhomogeneous distribution of galaxies and their super-massive black holes, barring complete scattering mixing of the particles by intergalactic magnetic fields.

## The Black Hole Mass and Magnetic Field Correlation in AGN: Testing by Optical Polarimetry

**Authors:** Silant'ev, N. A.; Piotrovich, M. Yu.; Gnedin, Yu. N.; Natsvlishvili, T. M.

**Eprint:** http://arxiv.org/abs/0908.2725

**Keywords:** accretion discs; astro-ph.GA; astrophysics; supermassive black holes

**Abstract:** We consider the integral light polarization from optically thick accretion disks. Basic mechanism is the multiple light scattering on free electrons (Milne's problem) in magnetized atmosphere. The Faraday rotation of the polarization plane changes both the value of integral polarization degree $p$ and the position angle $\chi$. Besides, the characteristic spectra of these values appear. We are testing the known relation between magnetic field of black hole at the horizon $B_{BH}$ and its mass $M_{BH}$, and the usual power-law distribution inside the accretion disk. The formulae for $p(\lambda)$ and $\chi(\lambda)$ depend on a number of parameters describing the particular dependence of magnetic field in accretion disk (the index of power-law distribution, the spin of the black hole, etc.). Comparison of our theoretical values of $p$ and $\chi$ with observed polarization can help us to choice more realistic values of parameters if the accretion disk mechanism gives the main contribution to the observed integral polarization. The main content is connected with estimation of validity of the relation between $B_{BH}$ and $M_{BH}$. We found for the AGN NGC 4258 that such procedure does not confirm the mentioned correlation between magnetic field and mass of black hole.

## Type 2 AGNs with Double-Peaked [O III] Lines: Narrow Line Region Kinematics or Merging Supermassive Black Hole Pairs?

**Authors:** Liu, Xin; Shen, Yue; Strauss, Michael A.; Greene, Jenny E.

**Eprint:** http://arxiv.org/abs/0908.2426

**Keywords:** astro-ph.CO; astrophysics; massive binaries of black holes; observations

**Abstract:** We present a sample of 167 type 2 AGNs with double-peaked [O III]4959,5007 narrow emission lines, selected from the seventh data release of the Sloan Digital Sky Survey. The double-peaked profiles can be well modeled by two velocity components, blueshifted and redshifted from the systemic velocity. Half of these objects have a more prominent redshifted component. In cases where the H-beta emission line is strong, it also shows two velocity components whose line-of-sight (LOS) velocity offsets are consistent with those of [O III]. The relative LOS velocity offset between the two components is typically a few hundred km/s, larger by a factor of ~ 1.5 than the line full width at half maximum of each component. The offset correlates with the host stellar velocity dispersion sigma. The host galaxies of this sample show systematically larger sigmas, stellar masses, and concentrations, and older luminosity-weighted mean stellar ages than a regular type 2 AGN sample matched in redshift, [O III]5007 equivalent width and luminosity; they show no significant difference in radio properties. These double-peaked features could be due to narrow-line region kinematics, or binary black holes. We suggest that a considerable fraction of the sample could be binaries, but spatially resolved optical imaging, spectroscopy, radio or X-ray follow up are needed to draw firm conclusions.

## Self-force with (3+1) codes: a primer for numerical relativists

**Authors:** Vega, Ian; Diener, Peter; Tichy, Wolfgang; Detweiler, Steven

**Eprint:** http://arxiv.org/abs/0908.2138

**Keywords:** EMRI; gr-qc; numerical relativity; self force

**Abstract:** Prescriptions for numerical self-force calculations have traditionally been designed for frequency-domain or (1+1) time-domain codes which employ a mode decomposition to facilitate in carrying out a delicate regularization scheme. This has prevented self-force analyses from benefiting from the powerful suite of tools developed and used by numerical relativists for simulations of the evolution of comparable-mass black hole binaries. In this work, we revisit a previously-introduced (3+1) method for self-force calculations, and demonstrate its viability by applying it to the test case of a scalar charge moving in a circular orbit around a Schwarzschild black hole. Two (3+1) codes originally developed for numerical relativity applications were independently employed, and in each we were able to compute the two independent components of the self-force and the energy flux correctly to within < 1%. We also demonstrate consistency between *t*-component of the self-force and the scalar energy flux. Our results constitute the first successful calculation of a self-force in a (3+1) framework, and thus open opportunities for the numerical relativity community in self-force analyses and the perturbative modeling of extreme-mass-ratio inspirals.

## Role of emission angular directionality in spin determination of accreting black holes with broad iron line

**Authors:** Svoboda, J.; Dovciak, M.; Goosmann, R. W.; Karas, V.

**Eprint:** http://arxiv.org/abs/0908.2387

**Keywords:** accretion discs; astro-ph.GA; astro-ph.HE; astrophysics; spin; supermassive black holes

**Abstract:** Spin of an accreting black hole can be determined by spectroscopy of the emission and absorption features produced in the inner regions of an accretion disc. We discuss the method employing the relativistic line profiles of iron in the X-ray domain, where the emergent spectrum is blurred by general relativistic effects. Precision of spectra fitting procedure could be compromised by inappropriate account of the angular distribution of the disc emission. Often a unique profile is assumed, invariable over the entire range of radii in the disc and energy in the spectral band. We study how sensitive the spin determination is to the assumptions about the intrinsic angular distribution of the emitted photons. We find that the uncertainty of the directional emission distribution translates to 20% uncertainty in determination of the marginally stable orbit. By assuming a rotating black hole in the centre of an accretion disc, we perform radiation transfer computations of an X-ray irradiated disc atmosphere to determine the directionality of outgoing X-rays in the 2-10 keV energy band. We implemented the simulation results as a new extension to the KY software package for X-ray spectra fitting of relativistic accretion disc models. Although the parameter space is rather complex, leading to a rich variety of possible outcomes, we find that on average the isotropic directionality reproduces our model data to the best precision. Our results also suggest that an improper usage of limb darkening can partly mimic a steeper profile of radial emissivity. We demonstrate these results on the case of XMM-Newton observation of the Seyfert galaxy MCG-6-30-15, for which we construct confidence levels of chi squared statistics, and on the simulated data for the future X-ray IXO mission.

### A stochastic template placement algorithm for gravitational wave data analysis

**Authors:** Harry, Ian; Allen, Bruce; Sathyaprakash, B. S.

**Eprint:** http://arxiv.org/abs/0908.2090

**Keywords:** data analysis; gr-qc; search algorithms; waveforms

**Abstract:** This paper presents an algorithm for constructing matched-filter template banks in an arbitrary parameter space. The method places templates at random, then removes those which are "too close" together. The properties and optimality of stochastic template banks generated in this manner are investigated for some simple models. The effectiveness of these template banks for gravitational wave searches for binary inspiral waveforms is also examined. The properties of a stochastic template bank are then compared to the deterministically placed template banks that are currently used in gravitational wave data analysis.

## Shapiro delay of asteroids on LISA

**Authors:** Chauvineau, Bertrand; Pireaux, Sophie; Regimbau, Tania

**Eprint:** http://arxiv.org/abs/0908.2043

**Keywords:** gr-qc; instruments; physics.space-ph

**Abstract:** In this paper, we examine the Shapiro delay caused by the close approach of an asteroid to the LISA constellation. We find that the probability that such an event occurs at a detectable level during the time interval of the mission is smaller than 1 %.

## The black hole mass-bulge mass correlation: bulges versus pseudo-bulges

**Authors:** Hu, Jian

**Eprint:** http://arxiv.org/abs/0908.2028

**Keywords:** astro-ph.GA; astro-ph.HE; astrophysics; observations; supermassive black holes

**Abstract:** We investigate the scaling relations between the supermassive black holes mass ($M_{bh}$) and the host bulge mass in elliptical galaxies, classical bulges, and pseudo-bulges. We use two-dimensional image analysis software BUDDA to obtain the structural parameters of 57 galaxies with dynamical $M_{bh}$ measurement, and determine the bulge K-band luminosities ($L_{bul,K}$), stellar masses ($M_s$), and dynamical masses ($M_d$). The updated $M_{bh} - L_{bul,K}$, $M_{bh} - M_s$, and $M_{bh} - M_d$ correlations for elliptical galaxies and classical bulges give $M_{bh} \sim 0.006 M_s$ or $0.003 M_d$. The most tight relationship is $\log(M_{bh}/M_\odot) = a + b \log(M_d/10^1 1 M_\odot)$, with a=8.46+/-0.05, b=0.90+/-0.06, and intrinsic scatter 0.27 dex. The pseudo-bulges follow their own relations, they harbor an order of magnitude smaller black holes than those in the same massive classical bulges, i.e. $M_{bh} \sim 0.0003 M_s$ or $0.0002 M_d$. Besides the $M_{bh} - \sigma$ (bulge stellar velocity dispersion) relation, these bulge type dependent $M_{bh} - M_{bul}$ scaling relations provide information for the growth and coevolution histories of SMBHs and their host bulges. We also find the core elliptical galaxies obey the same $M_{bh} - M_d$ relation with other normal elliptical galaxies, that is expected in the dissipationless merger scenario.

### The Millennium Galaxy Catalogue: The $M_{bh}$–$L_{spheroid}$ derived supermassive black hole mass function

**Authors:** Vika, Marina; Driver, Simon P.; Graham, Alister W.; Liske, Jochen

**Eprint:** http://arxiv.org/abs/0908.2102

**Keywords:** astro-ph.CO; astro-ph.GA; astrophysics; observations; supermassive black holes

**Abstract:** Supermassive black hole mass estimates are derived for 1743 galaxies from the Millennium Galaxy Catalogue using the recently revised empirical relation between supermassive black hole mass and the luminosity of the host spheroid. The MGC spheroid luminosities are based on $R^{1/n}$-bulge plus exponential-disc decompositions. The majority of black hole masses reside between $10^6 M_\odot$ and an upper limit of $2 \times 10^9 M_\odot$. Using previously determined space density weights, we derive the SMBH mass function which we fit with a Schechter-like function. Integrating the black hole mass function over $10^6 < M_{bh}/M_\odot < 10^{10}$ gives a supermassive black hole mass density of $(3.8 \pm 0.6) \times 10^5 h_{70}^3 M_\odot$ Mpc$^{-3}$ for early-type galaxies and $(0.96 \pm 0.2) \times 10^5 h_{70}^3 M_\odot$ Mpc$^{-3}$ for late-type galaxies. The errors are estimated from Monte Carlo simulations which include the uncertainties in the $M_{bh}$–$L$ relation, the luminosity of the host spheroid and the intrinsic scatter of the $M_{bh}$–$L$ relation. Assuming supermassive black holes form via baryonic accretion we find that $(0.008 \pm 0.002)h_{70}^3$ per cent of the Universe's baryons are currently locked up in supermassive black holes. This result is consistent with our previous estimate based on the $M_{bh}$–$n$ (Sérsic index) relation.

### Gravitational self force in extreme mass-ratio inspirals

**Authors:** Barack, Leor

**Eprint:** http://arxiv.org/abs/0908.1664

**Keywords:** astro-ph.HE; EMRI; general relativity; geodesic motion; gr-qc; self force

**Abstract:** This review is concerned with the gravitational self-force acting on a mass particle in orbit around a large black hole. Renewed interest in this old problem is driven by the prospects of detecting gravitational waves from strongly gravitating binaries with extreme mass ratios. We begin here with a summary of recent advances in the theory of gravitational self-interaction in curved spacetime, and proceed to survey some of the ideas and computational strategies devised for implementing this theory in the case of a particle orbiting a Kerr black hole. We review in detail two of these methods: (i) the standard mode-sum method, in which the metric perturbation is regularized mode-by-mode in a multipole decomposition, and (ii) $m$-mode regularization, whereby individual azimuthal modes of the metric perturbation are regularized in 2+1 dimensions. We discuss several practical issues that arise, including the choice of gauge, the numerical representation of the particle singularity, and how high-frequency contributions near the particle are dealt with in frequency-domain calculations. As an example of a full end-to-end implementation of the mode-sum method, we discuss the computation of the gravitational self-force for eccentric geodesic orbits in Schwarzschild, using a direct integration of the Lorenz-gauge perturbation equations in the time domain. With the computational framework now in place, researchers have recently turned to explore the physical consequences of the gravitational self force; we will describe some preliminary results in this area. An appendix to this review presents, for the first time, a detailed derivation of the regularization parameters necessary for implementing the mode-sum method in Kerr spacetime.

## Nonlinear Dynamical Friction in a Gaseous Medium

**Authors:** Kim, Hyosun; Kim, Woong-Tae

**Eprint:** <http://arxiv.org/abs/0908.1391>

**Keywords:** astro-ph.GA; astrophysics; stellar dynamics; supermassive black holes

**Abstract:** Using high-resolution, two-dimensional hydrodynamic simulations, we investigate nonlinear gravitational responses of gas to, and the resulting drag force on, a very massive perturber $M_p$ moving at velocity $V_p$ through a uniform gaseous medium of adiabatic sound speed $a_0$. We model the perturber as a Plummer potential with softening radius $r_s$, and run various models with differing $A = GM_p/(a_0^2 r_s)$ and $M = V_p/a_0$ by imposing cylindrical symmetry with respect to the line of perturber motion. For supersonic cases, a massive perturber quickly develops nonlinear flows that produce a detached bow shock and a vortex ring, which is unlike in the linear cases where Mach cones are bounded by low-amplitude Mach waves. The flows behind the shock are initially non-steady, displaying quasi-periodic, over-stable oscillations of the vortex ring and the shock. The vortex ring is eventually shed downstream and the flows evolve toward a quasi-steady state where the density wake near the perturber is in near hydrostatic equilibrium. We find that the detached shock distance $\delta$ and the nonlinear drag force F depend solely on $\eta = A/(M^2 - 1)$ such that $\delta/r_s = \eta$ and $F/F_{lin} = (\eta/2)^{-0.45}$ for $\eta > 2$, where $F_{lin}$ is the linear drag force of Ostriker (1999). The reduction of F compared with $F_{lin}$ is caused by front-back symmetry in the nonlinear density wakes. In subsonic cases, the flows without involving a shock do not readily reach a steady state. Nevertheless, the subsonic density wake near a perturber is close to being hydrostatic, resulting in the drag force similar to the linear case. Our results suggest that dynamical friction of a very massive object as in a merger of black holes near a galaxy center will take considerably longer than the linear prediction.

## Bowen-York trumpet data and black-hole simulations

**Authors:** Hannam, Mark; Husa, Sascha; Murchadha, Niall Ó

**Eprint:** http://arxiv.org/abs/0908.1063

**Keywords:** gr-qc; massive binaries of black holes; numerical relativity

**Abstract:** The most popular method to construct initial data for black-hole-binary simulations is the puncture method, in which compactified wormholes are given linear and angular momentum via the Bowen-York extrinsic curvature. When these data are evolved, they quickly approach a "trumpet" topology, suggesting that it would be preferable to use data that are in trumpet form from the outset. To achieve this, we extend the puncture method to allow the construction of Bowen-York trumpets, including an existence and uniqueness proof of the solutions. We construct boosted, spinning and binary Bowen-York puncture trumpets using a single-domain pseudospectral elliptic solver, and evolve the binary data and compare with standard wormhole-data results. We also show that the black-hole mass can be prescribed a priori, without recourse to the iterative procedure that is necessary for wormhole data.

## Coalescing binaries as possible standard candles

**Authors:** Capozziello, S.; De Laurentis, M.; Formisano, M.

**Eprint:** http://arxiv.org/abs/0908.0961

**Keywords:** astro-ph.CO; cosmology; general relativity; massive binaries of black holes

**Abstract:** Gravitational waves detected from well-localized inspiraling binaries would allow to determine, directly and independently, both binary luminosity and redshift. In this case, such systems could behave as "standard candles" providing an excellent probe of cosmic distances up to $z < 0.1$ and thus complementing other indicators of cosmological distance ladder.

## Nuclear Disk Formation by Direct Collisions of Gas Clouds with the Central Black Hole

**Authors:** Alig, Christian; Burkert, Andreas; Johansson, Peter H.; Schartmann, Marc

**Eprint:** http://arxiv.org/abs/0908.1100

**Keywords:** accretion discs; astro-ph.GA; astrophysics; EM counterparts; Sagittarius A*; supermassive black holes

**Abstract:** We simulate clouds in the Galactic Centre (GC) crossing over the black hole in parts and present this as a possible formation mechanism for the observed stellar disks in the GC through the redistribution of angular momentum by colliding material with opposite angular momentum. A parameter study using six high resolution simulations of an isothermal cloud of constant density falling onto the black hole and crossing over it in parts demonstrates that this mechanism is able to reproduce the observed disk properties in the GC. The evolution of the ensuing accretion disks is highly non-linear with the redistribution of the angular momentum through dissipative processes being a dominant effect. We analyse the resulting Toomre unstable, eccentric gaseous disk and show that this already yields a good comparison with the observed stellar disk size and eccentricity in the GC. The best simulation results in an outer radius of 1 pc, a mass of $10^4$ $M_\odot$ and an eccentricity of 0.24 for the Toomre unstable disk, which compares well with the observations.

## Empirical Constraints on the Evolution of the Relationship between Black Hole and Galaxy Mass: Scatter Matters

**Authors:** Somerville, Rachel S.

**Eprint:** http://arxiv.org/abs/0908.0927

**Keywords:** astro-ph.CO; astrophysics; cosmology; observations; supermassive black holes

**Abstract:** I investigate whether useful constraints on the evolution of the relationship between galaxy mass and black hole (BH) mass can be obtained from recent measurements of galaxy stellar mass functions and QSO bolometric luminosity functions at high redshift. I assume a simple power-law relationship between galaxy mass and BH mass, as implied by BH mass measurements at low redshift, and consider only evolution in the zero-point of the relation. I argue that one can obtain a lower limit on the zero-point evolution by assuming that every galaxy hosts a BH, shining at its Eddington rate. One can obtain an upper limit by requiring that the number of massive BH at high redshift does not exceed that observed locally. I find that, under these assumptions, and neglecting scatter in the BH-galaxy mass relation, BH must have been a factor of about 2 times larger at z=1 and 5 to 6 times more massive relative to their host galaxies at z=2. However, accounting for intrinsic scatter in the BH-galaxy mass relationship considerably relaxes these constraints. With a logarithmic scatter of 0.3 to 0.5 dex in black hole mass at fixed galaxy mass, similar to estimates of the intrinsic scatter in the observed relation today, there are enough massive BH to produce the observed population of luminous QSOs at z=2 even in the absence of any zero-point evolution. Adopting more realistic estimates for the fraction of galaxies that host active BH and the Eddington ratios of the associated quasars, I find that the zero-point of the BH-galaxy mass relation at z=2 cannot be much more than a factor of two times larger than the present-day value, as the number of luminous quasars predicted would exceed the observed population.

## Binaries of massive black holes in rotating clusters: Dynamics, gravitational waves, detection and the role of eccentricity

**Authors:** Amaro-Seoane, Pau; Eichhorn, Christoph; Porter, Ed; Spurzem, Rainer

**Eprint:** http://arxiv.org/abs/0908.0755

**Keywords:** astro-ph.GA; astrophysics; globular clusters; intermediate-mass black holes; massive binaries of black holes; parameter estimation; stellar dynamics

**Abstract:** The dynamical evolution of binaries of intermediate-massive black holes (IMBHs, massive black holes with a mass ranging between $10^2$ and $10^4 M_\odot$) in stellar clusters has recently received an increasing amount of attention. This is at least partially due to the fact that if the binary is hard enough to evolve to the phase at which it will start emitting gravitational waves (GWs) efficiently, there is a good probability that it will be detectable by future space-borne detectors like LISA. We study this evolution in the presence of rotation in the cluster. The eccentricity is strongly connected to the initial IMBHs velocities, and values of ∼ 0.7 up to 0.9 are reached for low initial velocities, while almost circular orbits result if the initial velocities are increased. A Monte Carlo study indicates that these sources will be detectable by a detector such as LISA with median signal to noise ratios of between 10 and 20 over a three year period, although some events had signal to noise ratios of 300 or greater. Furthermore, one should also be able to estimate the chirp-mass with median fractional errors of $10^{-4}$, reduced mass on the order of $10^{-3}$ and luminosity distance on the order of $10^{-1}$. Finally, these sources will have a median angular resolution in the LISA detector of about 3 square degrees, putting events firmly in the field of view of future electromagnetic detectors such as LSST.

## Non-Gaussianity analysis of GW background made by short-duration burst signals

**Authors:** Seto, Naoki

**Eprint:** http://arxiv.org/abs/0908.0228

**Keywords:** astro-ph.CO; back/foreground; bursts; gr-qc

**Abstract:** We study an observational method to analyze non-Gaussianity of a gravitational wave (GW) background made by superposition of weak burst signals. The proposed method is based on fourth-order correlations of data from four detectors, and might be useful to discriminate the origin of a GW background. With a formulation newly developed to discuss geometrical aspects of the correlations, it is found that the method provides us with linear combinations of two interesting parameters, $I_2$ and $V_2$ defined by the Stokes parameters of individual GW burst signals. We also evaluate sensitivities of specific detector networks to these parameters.

## The performance of arm locking in LISA

**Authors:** McKenzie, Kirk; Spero, Robert E.; Shaddock, Daniel A.

**Eprint:** http://arxiv.org/abs/0908.0290

**Keywords:** gr-qc; instruments; interferometers; noise: instrumental

**Abstract:** For the laser interferometer space antenna (LISA) to reach it's design sensitivity, the coupling of the free running laser frequency noise to the signal readout must be reduced by more than 14 orders of magnitude. One technique employed to reduce the laser frequency noise will be arm locking, where the laser frequency is locked to the LISA arm length. This paper details an implementation of arm locking, studies orbital effects, the impact of errors in the Doppler knowledge, and noise limits. The noise performance of arm locking is calculated with the inclusion of the dominant expected noise sources: ultra stable oscillator (clock) noise, spacecraft motion, and shot noise. Studying these issues reveals that although dual arm locking [A. Sutton & D. A Shaddock, Phys. Rev. D 78, 082001 (2008).] has advantages over single (or common) arm locking in terms of allowing high gain, it has disadvantages in both laser frequency pulling and noise performance. We address this by proposing a hybrid sensor, retaining the benefits of common and dual arm locking sensors. We present a detailed design of an arm locking controller and perform an analysis of the expected performance when used with and without laser pre-stabilization.

We observe that the sensor phase changes beneficially near unity-gain frequencies of the arm-locking controller, allowing a factor of 10 more gain than previously believed, without degrading stability. We show that the LISA frequency noise goal can be realized with arm locking and Time-Delay Interferometry only, without any form of pre-stabilization.

## The Lick AGN Monitoring Project: Broad-Line Region Radii and Black Hole Masses from Reverberation Mapping of Hbeta

**Authors:** Bentz, Misty C.; Walsh, Jonelle L.; Barth, Aaron J.; Baliber, Nairn; Bennert, Nicola; Canalizo, Gabriela; Filippenko, Alexei V.; Ganeshalingam, Mohan; Gates, Elinor L.; Greene, Jenny E.; Hidas, Marton G.; Hiner, Kyle D.; Lee, Nicholas; Li, Weidong; Malkan, Matthew A.; Minezaki, Takeo; Sakata, Yu; Serduke, Frank J. D.; Silverman, Jeffrey M.; Steele, Thea N.; Stern, Daniel; Street, Rachel A.; Thornton, Carol E.; Treu, Tommaso; Wang, Xiaofeng; Woo, Jong-Hak; Yoshii, Yuzuru

**Eprint:** http://arxiv.org/abs/0908.0003

**Keywords:** astro-ph.CO; astrophysics; observations; supermassive black holes

**Abstract:** We have recently completed a 64-night spectroscopic monitoring campaign at the Lick Observatory 3-m Shane telescope with the aim of measuring the masses of the black holes in 12 nearby (z < 0.05) Seyfert 1 galaxies with expected masses in the range $\sim 10^6 - 10^7 M_\odot$ and also the well-studied nearby active galactic nucleus (AGN) NGC 5548. Nine of the objects in the sample (including NGC 5548) showed optical variability of sufficient strength during the monitoring campaign to allow for a time lag to be measured between the continuum fluctuations and the response to these fluctuations in the broad Hbeta emission. We present here the light curves for the objects in this sample and the subsequent Hbeta time lags for the nine objects where these measurements were possible. The Hbeta lag time is directly related to the size of the broad-line region, and by combining the lag time with the measured width of the Hbeta emission line in the variable part of the spectrum, we determine the virial mass of the central supermassive black hole in these nine AGNs. The absolute calibration of the black hole masses is based on the normalization derived by Onken et al. We also examine the time lag response as a function of velocity across the Hbeta line profile for six of the AGNs. The analysis of four leads to ambiguous results with relatively flat time lags as a function of velocity. However, SBS 1116+583A exhibits a symmetric time lag response around the line center reminiscent of simple models for circularly orbiting broad-line region (BLR) clouds, and Arp 151 shows an asymmetric profile that is most easily explained by a simple gravitational infall model. Further investigation will be necessary to fully understand the constraints placed on physical models of the BLR by the velocity-resolved response in these objects.

## Constraining the Spin of the Black Hole in Fairall 9 with Suzaku

**Authors:** Schmoll, S.; Miller, J. M.; Volonteri, M.; Cackett, E.; Reynolds, C. S.; Fabian, A. C.; Brenneman, L. W.; Miniutti, G.; Gallo, L. C.

**Eprint:** http://arxiv.org/abs/0908.0013

**Keywords:** astro-ph.GA; astro-ph.HE; astrophysics; observations; spin; supermassive black holes

**Abstract:** We report on the results of spectral fits made to data obtained from a 168 ksec Suzaku observation of the Seyfert-1 galaxy Fairall 9. The source is clearly detected out to 30 keV. The observed spectrum is fairly simple; it is well-described by a power-law with a soft excess and disk reflection. A broad iron line is detected, and easily separated from distinct narrow components owing to the resolution of the CCDs in the X-ray Imaging Spectrometer (XIS). The broad line is revealed to be asymmetric, consistent with a disk origin. We fit the XIS and Hard X-ray Detector (HXD) spectra with relativistically-blurred disk reflection models. With the assumption that the inner disk extends to the innermost stable circular orbit, the best-fit model implies a black hole spin parameter of a = 0.60(7) and excludes extremal values at a high level of confidence. We discuss this result in the context of Seyfert observations and models of the cosmic distribution of black hole spin.

## Modeling Flows Around Merging Black Hole Binaries

**Authors:** van Meter, James R.; Wise, John H.; Miller, M. Coleman; Reynolds, Christopher S.; Centrella, Joan M.; Baker, John G.; Boggs, William D.; Kelly, Bernard J.; McWilliams, Sean T.

**Eprint:** http://arxiv.org/abs/0908.0023

**Keywords:** astro-ph.HE; EM counterparts; massive binaries of black holes; numerical relativity

**Abstract:** Coalescing massive black hole binaries are produced by the mergers of galaxies. The final stages of the black hole coalescence produce strong gravitational radiation that can be detected by the space-borne LISA. In cases where the black hole merger takes place in the presence of gas and magnetic fields, various types of electromagnetic signals may also be produced. Modeling such electromagnetic counterparts of the final merger requires evolving the behavior of both gas and fields in the strong-field regions around the black holes. We have taken a step towards solving this problem by mapping the flow of pressureless matter in the dynamic, 3-D general relativistic spacetime around the merging black holes. We find qualitative differences in collision and outflow speeds, including a signature of the merger when the net angular momentum of the matter is low, between the results from single and binary black holes, and between nonrotating and rotating holes in binaries. If future magnetohydrodynamic results confirm these differences, it may allow assessment of the properties of the binaries as well as yielding an identifiable electromagnetic counterpart to the attendant gravitational wave signal.

### Tidal effects in the vicinity of a black hole

**Authors:** Cadez, A.; Kostic, U.; Calvani, M.

**Eprint:** http://arxiv.org/abs/0908.0117

**Keywords:** astro-ph.GA; astro-ph.HE; astrophysics; observations; Sagittarius A*; stellar dynamics; supermassive black holes

**Abstract:** The discovery that the Galactic centre emits flares at various wavelengths represents a puzzle concerning their origin, but at the same time it is a relevant opportunity to investigate the environment of the nearest super-massive black hole. In this paper we shall review some of our recent results concerning the tidal evolution of the orbits of low mass satellites around black holes, and the tidal effect during their in-fall. We show that tidal interaction can offer an explanation for transient phenomena like near infra-red and X-ray flares from Sgr A*.

### High angular resolution integral-field spectroscopy of the Galaxy's nuclear cluster: a missing stellar cusp?

**Authors:** Do, Tuan; Ghez, Andrea M.; Morris, Mark R.; Lu, Jessica R.; Matthews, Keith; Yelda, Sylvana; Larkin, James

**Eprint:** http://arxiv.org/abs/0908.0311

**Keywords:** astro-ph.GA; astrophysics; observations; stellar dynamics; supermassive black holes

**Abstract:** We report on the structure of the nuclear star cluster in the innermost 0.16 pc of the Galaxy as measured by the number density profile of late-type giants. Using laser guide star adaptive optics in conjunction with the integral field spectrograph, OSIRIS, at the Keck II telescope, we are able to differentiate between the older, late-type (∼ 1 Gyr) stars, which are presumed to be dynamically relaxed, and the unrelaxed young (∼ 6 Myr) population. This distinction is crucial for testing models of stellar cusp formation in the vicinity of a black hole, as the models assume that the cusp stars are in dynamical equilibrium in the black hole potential. Based on the late-type stars alone, the surface stellar number density profile, $\Sigma(R) \propto R^{-\Gamma}$, is flat, with $\Gamma = -0.27 \pm 0.19$. Monte Carlo simulations of the possible de-projected volume density profile, $n(r) \propto r^{-\gamma}$, show that $\gamma$ is less than 1.0 at the 99.73 % confidence level. These results are consistent with the nuclear star cluster having no cusp, with a core profile that is significantly flatter than predicted by most cusp formation theories, and even allows for the presence of a central hole in the stellar distribution. Of the possible dynamical interactions that can lead to the depletion of the red giants observable in this survey – stellar collisions, mass segregation from stellar remnants, or a recent merger event – mass segregation is the only one that can be ruled out as the dominant depletion mechanism. The lack of a stellar cusp around a supermassive black hole would have important implications for black hole growth models and inferences on the presence of a black hole based upon stellar distributions.

## Probing the nature of the massive black hole binary candidate SDSS J1536+0441

**Authors:** Decarli, R.; Dotti, M.; Falomo, R.; Treves, A.; Colpi, M.; Kotilainen, J. K.; Montuori, C.; Uslenghi, M.

**Eprint:** http://arxiv.org/abs/0907.5414

**Keywords:** astro-ph.CO; astrophysics; massive binaries of black holes; observations

**Abstract:** We present an imaging study of the black hole binary candidate SDSS J1536+0441 (z=0.3893), based on deep, high resolution VzK images collected at the ESO/VLT. The images clearly show an asymmetric elongation, indicating the presence of a companion source at ∼ 1" (∼ 5 kpc projected distance) East from the quasar. The host galaxy of the quasar is marginally resolved. We find that the companion source is a luminous galaxy, the light profile of which suggests the presence of an unresolved, faint nucleus (either an obscured AGN or a compact stellar bulge). The study of the environment around the quasar indicates the occurrence of a significant over-density of galaxies with a redshift compatible with z∼ 0.4. This suggests that it resides in a moderately rich cluster of galaxies.

### Polarized Emission of Sagittarius A*

**Authors:** Huang, Lei; Liu, Siming; Shen, Zhi-Qiang; Yuan, Ye-Fei; Cai, Mike J.; Li, Hui; Fryer, Christopher L.

**Eprint:** http://arxiv.org/abs/0907.5463

**Keywords:** astro-ph.GA; astrophysics; observations; Sagittarius A*; supermassive black holes

**Abstract:** We explore the parameter space of the two temperature pseudo-Newtonian Keplerian accretion flow model for the millimeter and shorter wavelength emission from Sagittarius A*. A general relativistic ray-tracing code is used to treat the radiative transfer of polarized synchrotron emission from the flow. The synchrotron self-Comptonization and bremsstrahlung emission components are also included. It is shown that the model can readily account for the millimeter to sub-millimeter emission characteristics with an accretion rate of $\sim 6 \times 10^{17} g\,s^{-1}$ and an inclination angle of ~ 40 deg. However, the corresponding model predicted near-infrared and X-ray fluxes are more than one order of magnitude lower than the observed 'quiescent' state values. While the extended quiescent-state X-ray emission has been attributed to thermal emission from the large-scale accretion flow, the NIR emission and flares are likely dominated by emission regions either within the last stable orbit of a Schwarzschild black hole or associated with outflows. With the viscous parameter derived from numerical simulations, there is still a degeneracy between the electron heating rate and the magnetic parameter. A fully general relativistic treatment with the black hole spin incorporated will resolve these issues.

### Near-infrared Polarimetry of flares from Sgr A* with Subaru/CIAO

**Authors:** Nishiyama, Shogo; Tamura, Motohide; Hatano, Hirofumi; Nagata, Tetsuya; Kudo, Tomoyuki; Ishii, Miki; Schödel, Rainer; Eckart, Andreas

**Eprint:** http://arxiv.org/abs/0907.5466

**Keywords:** astro-ph.GA; astrophysics; observations; Sagittarius A*; supermassive black holes

**Abstract:** We have performed near-infrared monitoring observations of Sgr A*, the Galactic center radio source associated with a super-massive black hole, with the near-infrared camera CIAO and the 36-element AO system on the Subaru telescope.

We observed three flares in the Ks band (2.15micron) during 220 min monitoring on 2008 May 28, and confirmed the flare emission is highly polarized, supporting the synchrotron radiation nature of the near-infrared emission. Clear variations in the degree and position angle of polarization were also detected: an increase of the degree of polarization of about 20 %, and a swing of the position angle of about 60 - 70 degrees in the declining phase of the flares. The correlation between the flux and the degree of polarization can be well explained by the flare emission coming from hotspot(s) orbiting Sgr A*. Comparison with calculations in the literature gives a constraint to the inclination angle i of the orbit of the hotspot around Sgr A*, as 45 < i < 90 degrees (close to edge-on).

## Galaxies in COSMOS: Evolution of Black hole vs. bulge mass but not vs. total stellar mass over the last 9 Gyrs?

**Authors:** Jahnke, Knud; Bongiorno, Angela; Brusa, Marcella; Capak, Peter; Cappelluti, Nico; Cisternas, Mauricio; Civano, Francesca; Colbert, James; Comastri, Andrea; Elvis, Martin; Hasinger, Günther; Impey, Chris; Inskip, Katherine; Koekemoer, Anton M.; Lilly, Simon; Maier, Christian; Merloni, Andrea; Riechers, Dominik; Salvato, Mara; Schinnerer, Eva; Scoville, Nick Z.; Silverman, John; Taniguchi, Yoshi; Trump, Jonathan R.; Yan, Lin

**Eprint:** http://arxiv.org/abs/0907.5199

**Keywords:** astro-ph.CO; astrophysics; cosmology; observations; supermassive black holes

**Abstract:**

We constrain the ratio of black hole (BH) mass to total stellar mass of type-1 AGN in the COSMOS survey at 1<z<2. For 10 AGN at mean redshift $z \sim 1.4$ with both HST/ACS and HST/NICMOS imaging data we are able to compute total stellar mass $M_{*,total}$, based on restframe UV-to-optical host galaxy colors which constrain mass-to-light ratios. All objects have virial BH mass-estimates available from the COSMOS Magellan/IMACS and zCOSMOS surveys. We find zero difference between the $M_{BH} - M_{*,total}$-relation at $z \sim 1.4$ and the $M_{BH} - M_{*,bulge}$-relation in the local Universe.

Our interpretation is: (a) If our objects were purely bulge-dominated, the $M_{BH} - M_{*,bulge}$-relation has not evolved since $z \sim 1.4$. However, (b) since we have evidence for substantial disk components, the bulges of massive galaxies (log $M_{*,total}$=11.1+-0.25 or log $M_{BH} \sim$ 8.3+-0.2) must have grown over the last 9 Gyrs predominantly by redistribution of disk- into bulge-mass. Since all necessary stellar mass exists in the galaxy at z=1.4, no star-formation or addition of external stellar material is required, only a redistribution e.g. induced by minor and major merging or through disk instabilities. Merging, in addition to redistributing mass in the galaxy, will add both BH and stellar/bulge mass, but does not change the overall final $M_{BH}/M_{*,bulge}$ ratio.

Since the overall cosmic stellar and BH mass buildup trace each other tightly over time, our scenario of bulge-formation in massive galaxies is independent of any strong BH-feedback and means that the mechanism coupling BH and bulge mass until the present is very indirect.

### The Demography of Super-Massive Black Holes: Growing Monsters at the Heart of Galaxies

**Authors:** Shankar, Francesco

**Eprint:** http://arxiv.org/abs/0907.5213

**Keywords:** astro-ph.CO; astrophysics; cosmology; supermassive black holes

**Abstract:** Supermassive black holes (BHs) appear to be ubiquitous at the center of all galaxies which have been observed at high enough sensitivities and resolution with the Hubble Space Telescope. Their masses are found to be tightly linked with the masses and velocity dispersions of their host galaxies. On the other hand, BHs are widely held to constitute the central engines of quasars and active galactic nuclei (AGN) in general. It is however still unclear how BHs have grown, and whether they have co-evolved with their hosts. In this Review I discuss how, in ways independent of specific models, constraints on the growth history of BHs and their host galaxies have been set by matching the statistics of local BHs to the emissivity, number density, and clustering properties of AGNs at different cosmological epochs. I also present some new results obtained through a novel numerical code which evolves the BH mass function and clustering adopting broad distributions of Eddington ratios. I finally review BH evolution in a wider cosmological context, connecting BH growth to galaxy evolution.

### Quantifying the Coexistence of Massive Black Holes and Dense Nuclear Star Clusters

**Authors:** Graham, Alister W.; Spitler, Lee

**Eprint:** http://arxiv.org/abs/0907.5250

**Keywords:** astro-ph.CO; astro-ph.GA; astrophysics; intermediate-mass black holes; observations; supermassive black holes

**Abstract:** In large spheroidal stellar systems, such as elliptical galaxies, one invariably finds a $10^6 - 10^9 M_\odot$ supermassive black hole at their centre. In contrast, within dwarf elliptical galaxies one predominantly observes a $10^5 - 10^7 M_\odot$ nuclear star cluster. To date, few galaxies have been found with both type of nuclei coexisting and even less have had the masses determined for both central components. Here we identify one dozen galaxies housing nuclear star clusters and supermassive black holes whose masses have been measured. This doubles the known number of such hermaphrodite nuclei - which are expected to be fruitful sources of gravitational radiation. Over the host spheroid (stellar) mass range from $10^8$ to $10^{11} M_\odot$, we find that a galaxy's nucleus-to-spheroid (baryon) mass ratio is not a constant value but decreases from a few percent to ~ 0.3 percent such that $\log[(M_{BH} + M_{NC})/M_{sph}] = -(0.39 +/- 0.07) \log[M_{sph}/10^{10} M_\odot] - (2.18 +/- 0.07)$. Once dry merging has commenced by $M_{sph} \sim 10^{11} M_\odot$ and the nuclear star clusters have disappeared, this ratio is expected to become a constant value.

As a byproduct of our investigation, we have found that the projected flux from resolved nuclear star clusters can be well approximated with Sersic functions having a range of indices from ~ 0.5 to ~ 3, the latter index describing the Milky Way's nuclear star cluster.

## Hamiltonian of a spinning test-particle in curved spacetime

**Authors:** Barausse, E.; Racine, E.; Buonanno, A.

**Eprint:** http://arxiv.org/abs/0907.4745

**Keywords:** EMRI; general relativity; geodesic motion; post-Newtonian theory

**Abstract:** Using a Legendre transformation, we compute the unconstrained Hamiltonian of a spinning test-particle in a curved spacetime at linear order in the particle spin. The equations of motion of this unconstrained Hamiltonian coincide with the Mathisson-Papapetrou-Pirani equations. We then use the formalism of Dirac brackets to derive the constrained Hamiltonian and the corresponding phase-space algebra in the Newton-Wigner spin supplementary condition (SSC), suitably generalized to curved spacetime, and find that the phase-space algebra (q,p,S) is canonical at linear order in the particle spin. We provide explicit expressions for this Hamiltonian in a spherically symmetric spacetime, both in isotropic and spherical coordinates, and in the Kerr spacetime in Boyer-Lindquist coordinates. Furthermore, we find that our Hamiltonian, when expanded in Post-Newtonian (PN) orders, agrees with the Arnowitt-Deser-Misner (ADM) canonical Hamiltonian computed in PN theory in the test-particle limit. Notably, we recover the known spin-orbit couplings through 2.5PN order and the spin-spin couplings of type $S_{Kerr} S$ (and $S^2_{Kerr}$) through 3PN order, $S_{Kerr}$ being the spin of the Kerr spacetime. Our method allows one to compute the PN Hamiltonian at any order, in the test-particle limit and at linear order in the particle spin. As an application we compute it at 3.5PN order.

## Precessing supermassive black hole binaries and dark energy measurements with LISA

**Authors:** Stavridis, Adamantios; Arun, K. G.; Will, Clifford M.

**Eprint:** http://arxiv.org/abs/0907.4686

**Keywords:** cosmology; massive binaries of black holes

**Abstract:** Spin induced precessional modulations of gravitational wave signals from supermassive black hole binaries can improve the estimation of luminosity distance to the source by space based gravitational wave missions like the Laser Interferometer Space Antenna (LISA). We study how this impacts the ablity of LISA to do cosmology, specifically, to measure the dark energy equation of state (EOS) parameter $w$. Using the $\Lambda$CDM model of cosmology, we show that observations of precessing binaries by LISA, combined with a redshift measurement, can improve the determination of $w$ up to an order of magnitude with respect to the non precessing case depending on the masses, mass ratio and the redshift.

## Energy Level Diagrams for Black Hole Orbits

**Authors:** Levin, Janna

**Eprint:** http://arxiv.org/abs/0907.5195

**Keywords:** astro-ph.CO; EMRI; general relativity; geodesic motion; gr-qc; hep-th

**Abstract:** A spinning black hole with a much smaller black hole companion forms a fundamental gravitational system, like a colossal classical analog to an atom. In an appealing if imperfect analogy to atomic physics, this gravitational atom can be understood through a discrete spectrum of periodic orbits. Through a correspondence between the set of periodic orbits and the set of rational numbers, we are able to construct periodic tables of orbits and energy level diagrams of the accessible states around black holes. We also present a closed form expression for the rational q, thereby quantifying zoom-whirl behavior in terms of spin, energy, and angular momentum. The black hole atom is not just a theoretical construct, but corresponds to extant astrophysical systems detectable by future gravitational wave observatories.

### Gravitational wave forms for a three-body system in Lagrange's orbit: parameter determinations and a binary source test

**Authors:** Asada, Hideki

**Eprint:** http://arxiv.org/abs/0907.5091

**Keywords:** general relativity; gr-qc; parameter estimation; waveforms

**Abstract:** Continuing work initiated in an earlier publication [Torigoe et al. Phys. Rev. Lett. **102**, 251101 (2009)], gravitational wave forms for a three-body system in Lagrange's orbit are considered especially in an analytic method. First, we derive an expression of the three-body wave forms at the mass quadrupole, octupole and current quadrupole orders. By using the expressions, we solve a gravitational-wave *inverse* problem of determining the source parameters to this particular configuration (three masses, a distance of the source to an observer, and the orbital inclination angle to the line of sight) through observations of the gravitational wave forms alone. For this purpose, the chirp mass to a three-body system in the particular configuration is expressed in terms of only the mass ratios by deleting initial angle positions. We discuss also whether and how a binary source can be distinguished from a three-body system in Lagrange's orbit or others.

### The Shape of an Accretion Disc in a Misaligned Black Hole Binary

**Authors:** Martin, Rebecca G.; Pringle, J. E.; Tout, Christopher A.

**Eprint:** http://arxiv.org/abs/0907.5142

**Keywords:** accretion discs; astro-ph.HE; astrophysics; massive binaries of black holes; spin

**Abstract:** We model the overall shape of an accretion disc in a semi-detached binary system in which mass is transfered on to a spinning black hole the spin axis of which is misaligned with the orbital rotation axis. We assume the disc is in a steady state. Its outer regions are subject to differential precession caused by tidal torques of the companion star. These tend to align the outer parts of the disc with the orbital plane. Its inner regions are subject to differential precession caused by the Lense-Thirring effect. These tend to align the inner parts of the disc with the spin of the black hole. We give full numerical solutions for the shape of the disc for some particular disc parameters. We then show how an analytic approximation to these solutions can be obtained for the case when the disc surface density varies as a power law with radius. These analytic solutions for the shape of the disc are reasonably accurate even for large misalignments and can be simply applied for general disc parameters. They are particularly useful when the numerical solutions would be slow.

### X-ray and Radio Variability of M31*, The Andromeda Galaxy Nuclear Supermassive Black Hole

**Authors:** Garcia, Michael R.; Hextall, Richard; Baganoff, Frederick K.; Galache, Jose; Melia, Fulvio; Murray, Stephen S.; Primini, Frank A.; Sjouwerman, Lorant O.; Williams, Ben

**Eprint:** http://arxiv.org/abs/0907.4977

**Keywords:** astro-ph.HE; astrophysics; observations; supermassive black holes

**Abstract:** We confirm our earlier tentative detection of M31* in X-rays and measure its light-curve and spectrum. Observations in 2004-2005 find M31* rather quiescent in the X-ray and radio. However, X-ray observations in 2006-2007 and radio observations in 2002 show M31* to be highly variable at times. A separate variable X-ray source is found near P1, the brighter of the two optical nuclei. The apparent angular Bondi radius of M31* is the largest of any black hole, and large enough to be well resolved with Chandra. The diffuse emission within this Bondi radius is found to have an X-ray temperature ~ 0.3 keV and density 0.1 cm-3, indistinguishable from the hot gas in the surrounding regions of the bulge given the statistics allowed by the current observations. The X-ray source at the location of M31* is consistent with a point source and a power law spectrum with energy slope 0.9+/-0.2. Our identification of this X-ray source with M31* is based solely on positional coincidence.

### Evolution of the Binary Fraction in Dense Stellar Systems

**Authors:** Fregeau, J. M.; Ivanova, N.; Rasio, F. A.

**Eprint:** http://arxiv.org/abs/0907.4196

**Keywords:** astro-ph.GA; astrophysics; N-body; stellar dynamics

**Abstract:** Using our recently improved Monte Carlo evolution code, we study the evolution of the binary fraction in globular clusters. In agreement with previous N-body simulations, we find generally that the hard binary fraction in the core tends to increase with time over a range of initial cluster central densities for initial binary fractions <~ 90%. The dominant processes driving the evolution of the core binary fraction are mass segregation of binaries into the cluster core and preferential destruction of binaries there. On a global scale, these effects and the preferential tidal stripping of single stars tend to roughly balance, leading to overall cluster binary fractions that are roughly constant with time. Our findings suggest that the current hard binary fraction near the half-mass radius is a good indicator of the hard primordial binary fraction. However, the relationship between the true binary fraction and the fraction of main-sequence stars in binaries (which is typically what observers measure) is non-linear and rather complicated. We also consider the importance of soft binaries, which not only modify the evolution of the binary fraction, but can drastically change the evolution of the cluster as a whole. Finally, we describe in some detail the recent addition of single and binary stellar evolution to our cluster evolution code.

## Probing seed black holes using future gravitational-wave detectors

**Authors:** Gair, Jonathan R; Mandel, Ilya; Sesana, Alberto; Vecchio, Alberto

**Eprint:** http://arxiv.org/abs/0907.3292

**Keywords:** astro-ph.CO; cosmology; detectors; gr-qc

**Abstract:** Identifying the properties of the first generation of seeds of massive black holes is key to understanding the merger history and growth of galaxies. Mergers between ~ 100 solar mass seed black holes generate gravitational waves in the 0.1-10Hz band that lies between the sensitivity bands of existing ground-based detectors and the planned space-based gravitational wave detector, the Laser Interferometer Space Antenna (LISA). However, there are proposals for more advanced detectors that will bridge this gap, including the third generation ground-based Einstein Telescope and the space-based detector DECIGO. In this paper we demonstrate that such future detectors should be able to detect gravitational waves produced by the coalescence of the first generation of light seed black-hole binaries and provide information on the evolution of structure in that era. These observations will be complementary to those that LISA will make of subsequent mergers between more massive black holes. We compute the sensitivity of various future detectors to seed black-hole mergers, and use this to explore the number and properties of the events that each detector might see in three years of observation. For this calculation, we make use of galaxy merger trees and two different seed black hole mass distributions in order to construct the astrophysical population of events. We also consider the accuracy with which networks of future ground-based detectors will be able to measure the parameters of seed black hole mergers, in particular the luminosity distance to the source. We show that distance precisions of ∼ 30% are achievable, which should be sufficient for us to say with confidence that the sources are at high redshift.

## Mass and spin coevolution during the alignment of a black hole in a warped accretion disc

**Authors:** Perego, A.; Dotti, M.; Colpi, M.; Volonteri, M.

**Eprint:** http://arxiv.org/abs/0907.3742

**Keywords:** accretion discs; astro-ph.CO; astro-ph.HE; astrophysics; spin; stellar dynamics; supermassive black holes

**Abstract:** In this paper, we explore the gravitomagnetic interaction of a black hole (BH) with a misaligned accretion disc to study BH spin precession and alignment jointly with BH mass and spin parameter evolution, under the assumption that the disc is continually fed, in its outer region, by matter with angular momentum fixed on a given direction. We develop an iterative scheme based on the adiabatic approximation to study the BH-disc coevolution: in this approach, the accretion disc transits through a sequence of quasi-steady warped states (Bardeen-Petterson effect) and interacts with the BH until the BH spin aligns with the outer angular momentum direction. For a BH aligning with a co-rotating disc, the fractional increase in mass is typically less than a few percent, while the spin modulus can increase up to a few tens of percent. The alignment timescale is between ∼ 100 thousands and ∼ 1 millions years for a maximally rotating BH accreting at the Eddington rate. BH-disc alignment from an initially counter-rotating disc tends to be more efficient compared to the specular co-rotating case due to the asymmetry seeded in the Kerr metric: counter-rotating matter carries a larger and opposite angular momentum when crossing the innermost stable orbit, so that the spin modulus decreases faster and so the relative inclination angle.

## Post-Newtonian theory and the two-body problem

**Authors:** Blanchet, Luc

**Abstract:** Reliable predictions of general relativity theory are extracted using approximation methods. Among these, the powerful post-Newtonian approximation provides us with our best insights into the problems of motion and gravitational radiation of systems of compact objects. This approximation has reached an impressive mature status, because of important progress regarding its theoretical foundations, and the successful construction of templates of gravitational waves emitted by inspiralling compact binaries. The post-Newtonian predictions are routinely used for searching and analyzing the very weak signals of gravitational waves in current generations of detectors. High-accuracy comparisons with the results of numerical simulations for the merger and ring-down of binary black holes are going on. In this article we give an overview on the general formulation of the post-Newtonian approximation and present up-to-date results for the templates of compact binary inspiral.

## Delensing Gravitational Wave Standard Sirens with Shear and Flexion Maps

**Authors:** Shapiro, Charles; Bacon, David; Hendry, Martin; Hoyle, Ben

**Abstract:** Supermassive black hole binary systems (SMBHB) are standard sirens – the gravitational wave analogue of standard candles – and if discovered by gravitational wave detectors, they could be used as precise distance indicators. Unfortunately, gravitational lensing will randomly magnify SMBHB signals, seriously degrading any distance measurements. Using a weak lensing map of the SMBHB line of sight, we can estimate its magnification and thereby remove some uncertainty in its distance, a procedure we call "delensing." We find that delensing is significantly improved when galaxy shears are combined with flexion measurements, which reduce small-scale noise in reconstructed magnification maps. Under a Gaussian approximation, we estimate that delensing with a 2D mosaic image from an Extremely Large Telescope (ELT) could reduce distance errors by about 30-40% for a SMBHB at z=2. Including an additional wide shear map from a space survey telescope could reduce distance errors by 50%. Such improvement would make SMBHBs considerably more valuable as cosmological distance probes or as a fully independent check on existing probes.

### Parameter estimation for coalescing massive binary black holes with LISA using the full 2PN gravitational waveform and spin-orbit precession

**Authors:** Klein, Antoine; Jetzer, Philippe; Sereno, Mauro

**Eprint:** <http://arxiv.org/abs/0907.3318>

**Keywords:** astro-ph.CO; data analysis; massive binaries of black holes; parameter estimation; spin; waveforms

**Abstract:** Gravitational waves emitted by binary systems in the inspiral phase carry a complicated structure, consisting in a superposition of different harmonics of the orbital frequency, the amplitude of each of them taking the form of a Post-Newtonian series. In addition to that, spinning binaries experience couplings which induce a precession of the orbital angular momentum and of the individual spins. So far, all studies of the measurement accuracy of gravitational wave experiments for comparable-mass binary systems have considered either spinless binaries, or spinning binaries without subdominant harmonics in the waveform as well as no amplitude modulations. In this paper, we consider supermassive black hole binaries as expected to be observed with the planned space-based interferometer LISA, and study the measurement accuracy for several astrophysically interesting parameters obtainable taking into account the full 2PN waveform for spinning bodies, as well as spin-precession effects. We find that for binaries with a total mass in the range $10^5 M_\odot < M10$ for $M < 10^7 M_\odot$, 1.5 - 5 times higher than with the restricted waveform. We computed that the full waveform allows to use supermassive black hole binaries as standard sirens up to a redshift of z = 1.6, about 0.4 larger than what previous studies allowed.

### The Quasar SDSS J105041.35+345631.3: Black Hole Recoil or Extreme Double-Peaked Emitter?

**Authors:** Shields, G. A.; Rosario, D. J.; Smith, K. L.; Bonning, E. W.; Salviander, S.; Kalirai, J. S.; Strickler, R.; Ramirez-Ruiz, E.; Dutton, A. A.; Treu, T.; Marshall, P. J.

**Eprint:** <http://arxiv.org/abs/0907.3470>

**Keywords:** astro-ph.CO; astrophysics; kicks/recoil; massive binaries of black holes; observations; supermassive black holes

**Abstract:** The quasar SDSS J105041.35+345631.3 (z = 0.272) has broad emission lines blueshifted by 3500 km/s relative to the narrow lines and the host galaxy. Such an object may be a candidate for a recoiling supermassive black hole, a binary black hole, a superposition of two objects, or an unusual geometry for the broad emission-line region (BLR). The absence of narrow lines at the broad line redshift argues against superposition. New Keck spectra of J1050+3456 place tight constraints on the binary model. The combination of large velocity shift and symmetrical H-beta profile, as well as aspects of the narrow line spectrum, make J1050+3456 an interesting candidate for black hole recoil. Other aspects of the spectrum suggest an extreme case of a double-peaked emitter. We discuss possible observational tests to determine the true nature of this exceptional object.

## Measuring Black Hole Spin via the X-ray Continuum Fitting Method: Beyond the Thermal Dominant State

**Authors:** Steiner, James F.; McClintock, Jeffrey E.; Remillard, Ronald A.; Narayan, Ramesh; Gou, Lijun

**Eprint:** http://arxiv.org/abs/0907.2920

**Keywords:** astro-ph.HE; astrophysics; observations; spin; supermassive black holes

**Abstract:** All prior work on measuring the spins of stellar-mass black holes via the X-ray continuum-fitting method has relied on the use of weakly-Comptonized spectra obtained in the thermal dominant state. Using a self-consistent Comptonization model, we show that one can analyze spectra that exhibit strong power-law components and obtain values of the inner disk radius, and hence spin, that are consistent with those obtained in the thermal dominant state. Specifically, we analyze many RXTE spectra of two black hole transients, 1743-322 and XTE J1550-564, and we demonstrate that the radius of the inner edge of the accretion disk remains constant to within a few percent as the strength of the Comptonized component increases by an order of magnitude, i.e., as the fraction of the thermal seed photons that are scattered approaches 25%. We conclude that the continuum-fitting method can be applied to a much wider body of data than previously thought possible, and to sources that have never been observed to enter the thermal dominant state (e.g., Cyg X-1).

## Unambiguous determination of gravitational waveforms from binary black hole mergers

**Authors:** Reisswig, C.; Bishop, N. T.; Pollney, D.; Szilagyi, B.

**Eprint:** http://arxiv.org/abs/0907.2637

**Keywords:** gr-qc; massive binaries of black holes; numerical relativity; waveforms

**Abstract:** The accurate calculation of gravitational radiation emitted during black hole merger events is important for both improving the chances of detecting such events, as well as for making an astrophysical interpretation of the data once an event has been detected. Gravitational radiation is properly defined only at future null infinity (denoted by Scri), but in practice it is estimated from data calculated at a finite radius. We have used characteristic extraction to calculate gravitational radiation at Scri for the real astrophysical problem of the inspiral and merger of two equal mass non-spinning black holes (this problem has become a standard benchmark case in numerical relativity). Thus we have determined the first unambiguous merger waveforms for this problem. The implementation that has been developed is general purpose, and can be applied to calculate the gravitational radiation, at Scri, given data at a finite radius calculated in another computation.

## Targeted search for continuous gravitational waves: Bayesian versus maximum-likelihood statistics

**Authors:** Prix, Reinhard; Krishnan, Badri

**Eprint:** http://arxiv.org/abs/0907.2569

**Keywords:** data analysis; gr-qc; parameter estimation

**Abstract:** We investigate the Bayesian framework for detection of continuous gravitational waves (GWs) in the context of targeted searches, where the phase evolution of the GW signal is assumed to be known, while the four amplitude parameters are unknown. We show that the orthodox maximum-likelihood statistic (known as F-statistic) can be rediscovered as a Bayes factor with an unphysical prior in amplitude parameter space. We introduce an alternative detection statistic ("B-statistic") using the Bayes factor with a more natural amplitude prior, namely an isotropic probability distribution for the orientation of GW sources. Monte-Carlo simulations of targeted searches show that the resulting Bayesian B-statistic is more powerful in the Neyman-Pearson sense (i.e. has a higher expected detection probability at equal false-alarm probability) than the frequentist F-statistic.

## A Bayesian approach to the study of white dwarf binaries in LISA data: The application of a reversible jump Markov chain Monte Carlo method

**Authors:** Stroeer, Alexander; Veitch, John

**Eprint:** http://arxiv.org/abs/0907.2198

**Keywords:** astro-ph.IM; back/foreground; data analysis; MLDC

**Abstract:** The Laser Interferometer Space Antenna (LISA) defines new demands on data analysis efforts in its all-sky gravitational wave survey, recording simultaneously thousands of galactic compact object binary foreground sources and tens to hundreds of background sources like binary black hole mergers and extreme mass ratio inspirals. We approach this problem with an adaptive and fully automatic Reversible Jump Markov Chain Monte Carlo sampler, able to sample from the joint posterior density function (as established by Bayes theorem) for a given mixture of signals "out of the box", handling the total number of signals as an additional unknown parameter beside the unknown parameters of each individual source and the noise floor. We show in examples from the LISA Mock Data Challenge implementing the full response of LISA in its TDI description that this sampler is able to extract monochromatic Double White Dwarf signals out of colored instrumental noise and additional foreground and background noise successfully in a global fitting approach. We introduce 2 examples with fixed number of signals (MCMC sampling), and 1 example with unknown number of signals (RJ-MCMC), the latter further promoting the idea behind an experimental adaptation of the model indicator proposal densities in the main sampling stage. We note that the experienced runtimes and degeneracies in parameter extraction limit the shown examples to the extraction of a low but realistic number of signals.

## An Improved Black Hole Mass - Bulge Luminosity Relationship for AGNs

**Authors:** Gaskell, C. Martin; Kormendy, John

**Eprint:** http://arxiv.org/abs/0907.1652

**Keywords:** astro-ph.CO; astro-ph.GA; astrophysics; cosmology; observations; supermassive black holes

**Abstract:** Two effects have substantially increased the scatter in the AGN black hole mass - host galaxy bulge luminosity relationship derived from SDSS spectra. The first is that at a fixed black hole mass, the SDSS spectrum depends strongly on redshift because an SDSS fiber sees a larger fraction of the total light of more distant galaxies. The second is that at a given redshift, the fraction of host-galaxy light in the fiber increases with decreasing galaxy luminosity. We illustrate the latter effect using the Kormendy et al. (2009) light profiles of Virgo ellipticals. With allowance for the two effects, we obtain a black hole mass - bulge luminosity relationship for AGNs which has a scatter of only +/- 0.23 dex in mass. This is less than the scatter found for inactive galaxies, and is consistent with the measuring errors. We show that there is a correspondingly tight linear relationship between the fraction of host galaxy light in AGN spectra and the Eddington ratio. This linearity implies that at a given black hole mass, the host luminosities of high-accretion-rate AGNs (NLS1s) and low-accretion-rate AGNs are similar. The relationship between the fraction of host galaxy light and the Eddington ratio provides a simple means of estimating the fraction of host galaxy light in AGN spectra. This means that the real amplitude of variability of low-accretion-rate AGNs is increased relative to NLS1s.

### The role of black holes in galaxy formation and evolution

**Authors:** Cattaneo, A.; Faber, S. M.; Binney, J.; Dekel, A.; Kormendy, J.; Mushotzky, R.; Babul, A.; Best, P. N.; Brueggen, M.; Fabian, A. C.; Frenk, C. S.; Khalatyan, A.; Netzer, H.; Mahdavi, A.; Silk, J.; Steinmetz, M.; Wisotzki, L.

**Eprint:** http://arxiv.org/abs/0907.1608

**Keywords:** astro-ph.CO; astro-ph.GA; astrophysics; cosmology; stellar dynamics; supermassive black holes

**Abstract:** Virtually all massive galaxies, including our own, host central black holes ranging in mass from millions to billions of solar masses. The growth of these black holes releases vast amounts of energy that powers quasars and other weaker active galactic nuclei. A tiny fraction of this energy, if absorbed by the host galaxy, could halt star formation by heating and ejecting ambient gas. A central question in galaxy evolution is the degree to which this process has caused the decline of star formation in large elliptical galaxies, which typically have little cold gas and few young stars, unlike spiral galaxies.

### Cross section, final spin and zoom-whirl behavior in high-energy black hole collisions

**Authors:** Sperhake, U.; Cardoso, V.; Pretorius, F.; Berti, E.; Hinderer, T.; Yunes, N.

**Eprint:** http://arxiv.org/abs/0907.1252

**Keywords:** gr-qc; massive binaries of black holes; numerical relativity; spin; supermassive black holes

**Abstract:** We study the collision of two highly boosted equal mass, nonrotating black holes with generic impact parameter. We find such systems to exhibit zoom-whirl behavior when fine tuning the impact parameter. Near the threshold of immediate merger, these systems can produce black holes rotating close to the Kerr limit and generate radiated energies as large as ∼ 35% of the center of mass energy.

## The Growth of Black Holes: Insights From Obscured Active Galaxies

**Authors:** Greene, Jenny E.; Zakamska, Nadia L.; Liu, Xin; Barth, Aaron J.; Ho, Luis C.

**Eprint:** http://arxiv.org/abs/0907.1086

**Keywords:** astro-ph.CO; astro-ph.GA; stellar dynamics; supermassive black holes

**Abstract:** Obscured or narrow-line active galaxies offer an unobstructed view of the quasar environment in the presence of a luminous and vigorously accreting black hole. We exploit the large new sample of optically selected luminous narrow-line active galaxies from the Sloan Digital Sky Survey at redshifts $0.1 < z < 0.45$, in conjunction with follow-up observations with the Low Dispersion Survey Spectrograph (LDSS3) at Magellan, to study the distributions of black hole mass and host galaxy properties in these extreme objects. We find a narrow range in black hole mass ( = 8.0 +/- 0.7) and Eddington ratio ( = -0.7 +/- 0.7) for the sample as a whole, surprisingly similar to comparable broad-line systems. In contrast, we infer a wide range in star formation properties and host morphologies for the sample, from disk-dominated to elliptical galaxies. Nearly one-quarter have highly disturbed morphologies indicative of ongoing mergers. Unlike the black holes, which are apparently experiencing significant growth, the galaxies appear to have formed the bulk of their stars at a previous epoch. On the other hand, it is clear from the lack of correlation between gaseous and stellar velocity dispersions in these systems that the host galaxy interstellar medium is far from being in virial equilibrium with the stars. While our findings cast strong doubt on the reliability of substituting gas for stellar dispersions in high luminosity active galaxies, they do provide direct evidence that luminous accreting black holes influence their surroundings on a galaxy-wide scale.

## Conformally curved binary black hole initial data including tidal deformations and outgoing radiation

**Authors:** Johnson-McDaniel, Nathan K.; Yunes, Nicolas; Tichy, Wolfgang; Owen, Benjamin J.

**Eprint:** http://arxiv.org/abs/0907.0891

**Keywords:** gr-qc; massive binaries of black holes; post-Newtonian theory; waveforms

**Abstract:** (Abridged) By asymptotically matching a post-Newtonian (PN) metric to two tidally perturbed Schwarzschild metrics, we generate approximate initial data (in the form of a 4-metric) for a nonspinning black hole binary in a circular orbit. We carry out this matching through $O(v^4)$ in the binary's orbital velocity v, so the resulting data are conformally curved. Far from the holes, we use the appropriate PN metric that accounts for retardation, which we construct using the highest-order PN expressions available to compute the binary's past history. The data set's uncontrolled remainders are thus $O(v^5)$ throughout the timeslice; we also generate an extension to the data set that has uncontrolled remainders of $O(v^6)$ in the purely PN portion of the timeslice (i.e., not too close to the holes). The resulting data are smooth, since we join all the metrics together by smoothly interpolating between them. We perform this interpolation using transition functions constructed to avoid introducing excessive additional constraint violations. Due to their inclusion of tidal deformations and outgoing radiation, these data should substantially reduce the initial spurious ("junk") radiation observed in current simulations that use conformally flat initial data. Such reductions in the nonphysical components of the initial data will be necessary for simulations to achieve the accuracy required to supply Advanced LIGO and LISA with the templates necessary for parameter estimation.

## Momentum flow in black-hole binaries: II. Numerical simulations of equal-mass, head-on mergers with antiparallel spins

**Authors:** Lovelace, Geoffrey; Chen, Yanbei; Cohen, Michael; Kaplan, Jeffrey D.; Keppel, Drew; Matthews, Keith D.; Nichols, David A.; Scheel, Mark A.; Sperhake, Ulrich

**Eprint:** http://arxiv.org/abs/0907.0869

**Keywords:** gr-qc; massive binaries of black holes; spin; supermassive black holes

**Abstract:** Research on extracting science from binary-black-hole (BBH) simulations has often adopted a "scattering matrix" perspective: given the binary's initial parameters, what are the final hole's parameters and the emitted gravitational waveform? In contrast, we are using BBH simulations to explore the nonlinear dynamics of curved spacetime. Focusing on the head-on plunge, merger, and ringdown of a BBH with transverse, antiparallel spins, we explore numerically the momentum flow between the holes and the surrounding spacetime. We use the Landau-Lifshitz field-theory-in-flat-spacetime formulation of general relativity to define and compute the density of field energy and field momentum outside horizons and the energy and momentum contained within horizons, and we define the effective velocity of each apparent and event horizon as the ratio of its enclosed momentum to its enclosed mass-energy. We find surprisingly good agreement between the horizons' effective and coordinate velocities. To investigate the gauge dependence of our results, we compare pseudospectral and moving-puncture evolutions of physically similar initial data; although spectral and puncture simulations use different gauge conditions, we find remarkably good agreement for our results in these two cases. We also compare our simulations with the post-Newtonian trajectories and near-field energy-momentum. [Abstract abbreviated; full abstract also mentions additional results.]

## Simulations of Recoiling Massive Black Holes in the Via Lactea Halo

**Authors:** Guedes, Javiera; Madau, Piero; Kuhlen, Micheal; Diemand, Jürg; Zemp, Marcel

**Abstract:** The coalescence of a massive black hole (MBH) binary leads to the gravitational-wave recoil of the system and its ejection from the galaxy core. We have carried out N-body simulations of the motion of a MBH = $3.7x10^6 M_\odot$ MBH remnant in the Via Lactea I simulation, a Milky Way sized dark matter halo. The black hole receives a recoil velocity of Vkick = 80, 120, 200, 300, and 400 km/s at redshift 1.5, and its orbit is followed for over 1 Gyr within a live host halo, subject only to gravity and dynamical friction against the dark matter background. We show that, owing to asphericities in the dark matter potential, the orbit of the MBH is hightly non-radial, resulting in a significantly increased decay timescale compared to a spherical halo. The simulations are used to construct a semi-analytic model of the motion of the MBH in a time-varying triaxial Navarro-Frenk-White dark matter halo plus a spherical stellar bulge, where the dynamical friction force is calculated directly from the velocity dispersion tensor. Such a model should offer a realistic picture of the dynamics of kicked MBHs in situations where gas drag, friction by disk stars, and the flattening of the central cusp by the returning black hole are all negligible effects. We find that MBHs ejected with initial recoil velocities Vkick > 500 km/s do not return to the host center within Hubble time. In a Milky Way-sized galaxy, a recoiling hole carrying a gaseous disk of initial mass ~ MBH may shine as a quasar for a substantial fraction of its wandering phase. The long decay timescales of kicked MBHs predicted by this study may thus be favorable to the detection of off-nuclear quasar activity.

## Competitive feedback in galaxy formation

**Authors:** Nayakshin, Sergei; Wilkinson, Mark I.; King, Andrew

**Eprint:** http://arxiv.org/abs/0907.1002

**Keywords:** astro-ph.CO; astro-ph.GA; astrophysics; cosmology; stellar dynamics; supermassive black holes

**Abstract:** It is now well established that many galaxies have nuclear star clusters (NCs) whose total masses correlate with the velocity dispersion (sigma) of the galaxy spheroid in a very similar way to the well–known supermassive black hole (SMBH) M - sigma relation. Previous theoretical work suggested that both correlations can be explained by a momentum feedback argument. Observations further show that most known NCs have masses $10^8 M_\odot$, which remained unexplained in previous work. We suggest here that this changeover reflects a competition between the SMBH and nuclear clusters in the feedback they produce. When one of the massive objects reaches its limiting M-sigma value, it drives the gas away and hence cuts off its own mass and also the mass of the "competitor". The latter is then underweight with respect to the expected M-sigma mass (abridged).

## Zoom-Whirl Orbits in Black Hole Binaries

**Authors:** Healy, James; Levin, Janna; Shoemaker, Deirdre

**Eprint:** http://arxiv.org/abs/0907.0671

**Keywords:** general relativity; gr-qc; massive binaries of black holes; numerical relativity

**Abstract:** Zoom-whirl behavior has the reputation of being a rare phenomenon in comparable mass binaries. The concern has been that gravitational radiation would drain angular momentum so rapidly that generic orbits would circularize before zoom-whirl behavior could play out, and only rare highly tuned orbits would retain their imprint. Using full numerical relativity, we catch zoom-whirl behavior despite dissipation for a range of orbits. The larger the mass ratio, the longer the pair can spend in orbit before merging and therefore the more zooms and whirls that can be seen. Larger spins also enhance zoom-whirliness. An important implication is that these eccentric orbits can merge during a whirl phase, before enough angular momentum has been lost to truly circularize the orbit. In other words, although the whirl phase is nearly circular, merger of eccentric orbits occurs through a separatrix other than the isco. Gravitational waveforms from eccentric binaries will be modulated by the harmonics of zoom-whirl orbits, showing quiet phases during a zoom and louder glitches during whirls.

## Comparison of post-Newtonian templates for compact binary inspiral signals in gravitational-wave detectors

**Authors:** Buonanno, Alessandra; Iyer, Bala; Ochsner, Evan; Pan, Yi; Sathyaprakash, B. S.

**Eprint:** http://arxiv.org/abs/0907.0700

**Keywords:** Effective one body; gr-qc; massive binaries of black holes; post-Newtonian theory; waveforms

**Abstract:** The two-body dynamics in general relativity has been solved perturbatively using the post-Newtonian (PN) approximation. The evolution of the orbital phase and the emitted gravitational radiation are now known to a rather high order up to $O(v^8)$, v being the characteristic velocity of the binary. The orbital evolution, however, cannot be specified uniquely due to the inherent freedom in the choice of parameter used in the PN expansion as well as the method pursued in solving the relevant differential equations. The goal of this paper is to determine the (dis)agreement between different PN waveform families in the context of initial and advanced gravitational-wave detectors. The waveforms employed in our analysis are those that are currently used by Initial LIGO/Virgo, that is the time-domain PN models TaylorT1, TaylorT2, TaylorT3, TaylorT4 and TaylorEt, the effective one-body (EOB) model, and the Fourier-domain representation TaylorF2. We examine the overlaps of these models with one another and with the prototype effective one-body model (calibrated to numerical relativity simulations, as currently used by initial LIGO) for a number of different binaries at 2PN, 3PN and 3.5PN orders to quantify their differences and to help us decide whether there exist preferred families that are the most appropriate as search templates. We conclude that as long as the total mass remains less than a certain upper limit $M_{crit}$, all template families at 3.5PN order (except TaylorT3 and TaylorEt) are equally good for the purpose of detection. The value of $M_{crit}$ is found to be $\sim 12 M_\odot$ for Initial, Enhanced and Advanced LIGO. From a purely computational point of view we recommend that 3.5PN TaylorF2 be used below Mcrit and EOB calibrated to numerical relativity simulations be used for total binary mass M > Mcrit.

## Gravitational-wave detectability of equal-mass black-hole binaries with aligned spins

**Authors:** Reisswig, Christian; Husa, Sascha; Rezzolla, Luciano; Dorband, Ernst Nils; Pollney, Denis; Seiler, Jennifer

**Eprint:** http://arxiv.org/abs/0907.0462

**Keywords:** data analysis; gr-qc; massive binaries of black holes; waveforms

**Abstract:** Binary black-hole systems with spins aligned or anti-aligned to the orbital angular momentum provide the natural ground to start detailed studies of the influence of strong-field spin effects on gravitational wave observations of coalescing binaries. Furthermore, such systems may be the preferred end-state of the inspiral of generic supermassive binary black-hole systems. In view of this, we have computed the inspiral and merger of a large set of binary systems of equal-mass black holes with spins parallel to the orbital angular momentum but otherwise arbitrary. Our attention is particularly focused on the gravitational-wave emission so as to quantify how much spin effects contribute to the signal-to-noise ratio, to the horizon distances, and to the relative event rates for the representative ranges in masses and detectors. As expected, the signal-to-noise ratio increases with the projection of the total black hole spin in the direction of the orbital momentum. We find that equal-spin binaries with maximum spin aligned with the orbital angular momentum are more than "three times as loud" as the corresponding binaries with anti-aligned spins, thus corresponding to event rates up to 30 times larger. We also consider the waveform mismatch between the different spinning configurations and find that, within our numerical accuracy, binaries with opposite spins $S_1 = -S_2$ cannot be distinguished whereas binaries with spin $S_1 = S_2$ have clearly distinct gravitational-wave emissions. Finally, we derive a simple expression for the energy radiated in gravitational waves and find that the binaries always have efficiencies $E_{rad}/M > 3.6\%$, which can become as large as $E_{rad}/M = 10\%$ for maximally spinning binaries with spins aligned with the orbital angular momentum.

## Use and Abuse of the Model Waveform Accuracy Standards

**Authors:** Lindblom, Lee

**Eprint:** http://arxiv.org/abs/0907.0457

**Keywords:** data analysis; gr-qc; waveforms

**Abstract:** Accuracy standards have been developed to ensure that the waveforms used for gravitational-wave data analysis are good enough to serve their intended purposes. These standards place constraints on certain norms of the frequency-domain representations of the waveform errors. Examples are given here of possible misinterpretations and misapplications of these standards, whose effect could be to vitiate the quality control they were intended to enforce. Suggestions are given for ways to avoid these problems.

## The Final Remnant of Binary Black Hole Mergers: Multipolar Analysis

**Authors:** Owen, Robert

**Eprint:** http://arxiv.org/abs/0907.0280

**Keywords:** gr-qc; horizon; massive binaries of black holes; numerical relativity

**Abstract:** Methods are presented to define and compute source multipoles of dynamical horizons in numerical relativity codes, extending previous work from the isolated and dynamical horizon formalisms in a manner that allows for the consideration of horizons that are not axisymmetric. These methods are then applied to a binary black hole merger simulation, providing evidence that the final remnant is a Kerr black hole, both through the (spatially) gauge-invariant recovery of the geometry of the apparent horizon, and through a detailed extraction of quasinormal ringing modes directly from the strong-field region.

## Ultra-high precision cosmology from gravitational waves

**Authors:** Cutler, Curt; Holz, Daniel E.

**Eprint:** http://arxiv.org/abs/0906.3752

**Keywords:** astro-ph.CO; cosmology; gr-qc; intermediate-mass black holes

**Abstract:** We show that the Big Bang Observer (BBO), a proposed space-based gravitational-wave (GW) detector, would provide ultra-precise measurements of cosmological parameters. By detecting ∼ 300,000 compact-star binaries, and utilizing them as standard sirens, BBO would determine the Hubble constant to 0.1%, and the dark energy parameters $w_0$ and $w_a$ to ∼ 0.01 and 0.1, resp. BBO's dark-energy figure-of-merit would be approximately an order of magnitude better than all other proposed dark energy missions. To date, BBO has been designed with the primary goal of searching for gravitational waves from inflation. To observe this inflationary background, BBO would first have to detect and subtract out ∼ 300,000 merging compact-star binaries, out to $z$ ∼ 5. It is precisely this foreground which would enable high-precision cosmology. BBO would determine the luminosity distance to each binary to ∼ percent accuracy. BBO's angular resolution would be sufficient to uniquely identify the host galaxy for most binaries; a coordinated optical/infrared observing campaign could obtain the redshifts. Combining the GW-derived distances and EM-derived redshifts for such a large sample of objects leads to extraordinarily tight constraints on cosmological parameters. Such "standard siren" measurements of cosmology avoid many of the systematic errors associated with other techniques. We also show that BBO would be an exceptionally powerful gravitational lensing mission, and we briefly discuss other astronomical uses of BBO.

## Supermassive Black Hole Mass Regulated by Host Galaxy Morphology

**Authors:** Watabe, Y.; Kawakatu, N.; Imanishi, M.; Takeuchi, T. T.

**Eprint:** http://arxiv.org/abs/0907.0142

**Keywords:** astro-ph.CO; astrophysics; observations; supermassive black holes

**Abstract:** We investigated the relationship between supermassive black hole (SMBH) mass and host starburst luminosity in Seyfert galaxies and Palomar-Green QSOs, focusing on the host galaxy morphology. Host starburst luminosity was derived from the 11.3 micron polycyclic aromatic hydrocarbon luminosity. We found that the SMBH masses of elliptical-dominated host galaxies are more massive than those of disk-dominated host galaxies statistically. We also found that the SMBH masses of disk-dominated host galaxies seem to be suppressed even under increasing starburst luminosity. These findings imply that final SMBH mass is strongly regulated by host galaxy morphology. This can be understood by considering the radiation drag model as the SMBH growth mechanism, taking into account the radiation efficiency of the host galaxy.

## Relativistic orbits and Gravitational Waves from gravitomagnetic corrections

**Authors:** Capozziello, Salvatore; De Laurentis, Mariafelicia; Forte, Luca; Garufi, Fabio; Milano, Leopoldo

**Eprint:** http://arxiv.org/abs/0906.5530

**Keywords:** EMRI; general relativity; gr-qc; waveforms

**Abstract:** Corrections to the relativistic theory of orbits are discussed considering higher order approximations induced by gravitomagnetic effects. Beside the standard periastron effect of General Relativity (GR), a new nutation effect was found due to the $c^{-3}$ orbital correction. According to the presence of that new nutation effect we studied the gravitational waveforms emitted through the capture in a gravitational field of a massive black hole (MBH) of a compact object (neutron star (NS) or BH) via the quadrupole approximation. We made a numerical study to obtain the emitted gravitational wave (GW) amplitudes. We conclude that the effects we studied could be of interest for the future space laser interferometric GW antenna LISA.

## Oscillation Phenomena in the disk around the massive black hole Sagittarius A*

**Authors:** Miyoshi, M.; Shen, Zhi-Qiang; Oyama, T.; Takahashi, R.; Kato, Y.

**Eprint:** http://arxiv.org/abs/0906.5511

**Keywords:** accretion discs; astro-ph.HE; astrophysics; observations; Sagittarius A*

**Abstract:**

We report the detection of radio QPOs with structure changes using the Very Long Baseline Array (VLBA) at 43 GHz.

We found conspicuous patterned changes of the structure with P = 16.8, 22.2, 31.4, ~ 56.4 min, roughly in a 3:4:6:10 ratio. The first two periods show a rotating one-arm structure, while the P = 31.4 min shows a rotating 3-arm structure, as if viewed edge-on. At the central 50 microasec the P = 56.4 min period shows a double amplitude variation of those in its surroundings.

Spatial distributions of the oscillation periods suggest that the disk of SgrA* is roughly edge-on, rotating around an axis with PA = -10 degree. Presumably, the observed VLBI images of SgrA* remain several features of the black hole accretion disk of SgrA* in spite of being obscured and broadened by scattering of surrounding plasma.

### Shrinking the Braneworld: Black Hole in a Globular Cluster

**Authors:** Gnedin, Oleg Y.; Maccarone, Thomas J.; Psaltis, Dimitrios; Zepf, Stephen E.

**Eprint:** http://arxiv.org/abs/0906.5351

**Keywords:** astro-ph.CO; astro-ph.GA; astrophysics; globular clusters; IMRI; intermediate-mass black holes

**Abstract:** Large extra dimensions have been proposed as a possible solution to the hierarchy problem in physics. One of the suggested models, the RS2 braneworld model, makes a prediction that black holes evaporate by Hawking radiation on a short timescale that depends on the black hole mass and on the asymptotic radius of curvature of the extra dimensions. Thus the size of the extra dimensions can be constrained by astrophysical observations. Here we point out that the black hole, recently discovered in a globular cluster in galaxy NGC 4472, places the strongest constraint on the maximum size of the extra dimensions, L < 0.003 mm. This black hole has the virtues of old age and relatively small mass. The derived upper limit is within an order of magnitude of the absolute limit afforded by astrophysical observations of black holes.

### Globular Clusters and Satellite Galaxies: Companions to the Milky Way

**Authors:** Forbes, Duncan A.; Kroupa, Pavel; Metz, Manuel; Spitler, Lee

**Eprint:** http://arxiv.org/abs/0906.5370

**Keywords:** astro-ph.CO; astro-ph.GA; astrophysics; globular clusters; IMRI; stellar dynamics

**Abstract:** Our Milky Way galaxy is host to a number of companions. These companions are gravitationally bound to the Milky Way and are stellar systems in their own right. They include a population of some 30 dwarf satellite galaxies (DSGs) and about 150 globular clusters (GCs). Here we discuss the relationship between GCs and DSGs using an interactive 3D model of the Milky Way.

## Measuring spin of a supermassive black hole at the Galactic centre – Implications for a unique spin

**Authors:** Kato, Y.; Miyoshi, M.; Takahashi, R.; Negoro, H.; Matsumoto, R.

**Eprint:** http://arxiv.org/abs/0906.5423

**Keywords:** astro-ph.CO; astro-ph.GA; astrophysics; Sagittarius A*; spin; supermassive black holes

**Abstract:** We determine the spin of a supermassive black hole in the context of discseismology by comparing newly detected quasi-periodic oscillations (QPOs) of radio emission in the Galactic centre, Sagittarius A* (Sgr A*), as well as infrared and X-ray emissions with those of the Galactic black holes. We find that the spin parameters of black holes in Sgr A* and in Galactic X-ray sources have a unique value of $\approx 0.44$ which is smaller than the generally accepted value for supermassive black holes, suggesting evidence for the angular momentum extraction of black holes during the growth of supermassive black holes. Our results demonstrate that the spin parameter approaches the equilibrium value where spin-up via accretion is balanced by spin-down via the Blandford-Znajek mechanism regardless of its initial spin. We anticipate that measuring the spin of black holes by using QPOs will open a new window for exploring the evolution of black holes in the Universe.

*Intention and purpose of GW Notes*

A succinct explanation

The electronic publishing service **arXiv** is a dynamic, well-respected source of news of recent work and is updated daily. But, perhaps due to the large volume of new work submitted, it is probable that a member of our community might easily overlook relevant material. This new e-journal and its blog, **The LISA Brownbag (http://www.lisa-science.org/brownbag)**, both produced by the AEI, propose to offer scientist of the Gravitational Wave community the opportunity to more easily follow advances in the three areas mentioned: Astrophysics, General Relativity and Data Analysis. We hope to achieve this by selecting the most significant e-prints and list them in abstract form with a link to the full paper in both a single e-journal (GW Newsletter) and a blog (The LISA Brownbag). Of course, *this also implies that the paper will have its impact increased, since it will reach a broader public*, so that we encourage you to not forget submitting your own work

In addition to the abstracts, in each PDF issue of GW Notes, we will offer you a previously unpublished article written by a senior researcher in one of these three domains, which addresses the interests of all readers.

Thus the aim of The LISA Brownbag and GW Notes is twofold:

- Whenever you see an interesting paper on GWs science and LISA, you can submit the **arXiv** number to our **submission page (http://brownbag.lisascience.org)**. This is straightforward: No registration is required (although recommended) to simply type in the number in the entry field of the page, indicate some keywords and that's it

- We will publish a new full article in each issue, if available. This "feature article" will be from the fields of Astrophysics, General Relativity or the Data Analysis of gravitational waves and LISA. We will prepare a more detailed guide for authors, but for now would like to simply remind submitters that they are writing for colleagues in closely related but not identical fields, and that cross-fertilization and collaboration is an important goal of our concept

Subscribers get the issue distributed in PDF form. Additionally, they will be able to submit special announcements, such as meetings, workshops and jobs openings, to the list of registered people. For this, please register at the **registration page (http://lists.aei.mpg.de/cgi-bin/mailman/listinfo/lisa_brownbag)** by filling in your e-mail address and choosing a password.

### The Astro-GR meetings

*Past, present and future*

Sixty two scientists attended the **Astro-GR@AEI** meeting, which took place September 18-22 2006 at the **Max-Planck Institut für Gravitationsphysik (Albert Einstein-Institut)** in Golm, Germany. The meeting was the brainchild of an AEI postdoc, who had the vision of bringing together Astrophysicists and experts in General Relativity and gravitational-wave Data Analysis to discuss sources for **LISA**, the planned Laser Interferometer Space Antenna. More specifically, the main topics were EMRIs and IMRIs (Extreme and Intermediate Mass-Ratio Inspiral events), i.e. captures of stellar-mass compact objects by supermassive black holes and coalescence of intermediate-mass black holes with supermassive black holes.

The general consensus was that the meeting was both interesting and quite stimulating. It was generally agreed that someone should step up and host a second round of this meeting. Monica Colpi kindly did so and this led to **Astro-GR@Como**, which was very similar in its informal format, though with a focus on all sources, meant to trigger new ideas, as a kind of brainstroming meeting.

Also, in the same year, in the two first weeks of September, we had another workshop in the Astro-GR series with a new "flavour", namely, the **Two Weeks At The AEI (2W@AEI)**, in which the interaction between the attendees was be even higher than what was reached in the previous meetings. To this end, we reduced the number of talks, allowing participants more opportunity to collaborate. Moreover, participants got office facilities and we combined the regular talks with the so-called "powerpointless" seminar, which will were totally informal and open-ended, on a blackboard. The next one was held in Barcelona in 2009 at the beginning of September, **Astro-GR@BCN** and next 2010 it will be the turn of Paris, at the APC.

If you are interested in hosting in the future an Astro-GR meeting, please contact us. We are open to new formats, as long as the *Five Golden Rules* are respected.

A proper Astro-GR meeting **MUST** closely follow the *Five Golden Rules*:

I.    Bring together Astrophysicists, Cosmologists, Relativists and Data Analysts

II.   Motivate new collaborations and projects

III.  Be run in the style of Aspen, ITP, Newton Institute and Modest meetings, with plenty of time for discussions

IIII. Grant access to the slides in a cross-platform format, such as PDF and, within reason, to the recorded movies of the talks in a free format which everybody can play like **Theora**, for those who could not attend, following the good principles of **Open Access**

卌. Keep It Simple and... Spontaneous

A probe of spacetime and astrophysics: EMRIs